\documentclass{jfm}

\usepackage[dvipsnames]{xcolor}
\usepackage{graphicx}
\usepackage{newtxtext}
\usepackage{newtxmath}
\usepackage{natbib}
\usepackage{hyperref}
\hypersetup{
    colorlinks = true,
    urlcolor   = blue,
    citecolor  = nord0,
}

\newcommand{\RomanNumeralCaps}[1]
\linenumbers
\usepackage{amsmath,amssymb,epstopdf}
\usepackage{multirow}
\usepackage{mytools}

\definecolor{nord0}{HTML}{5e81ac}
\definecolor{nord1}{HTML}{d08770}
\definecolor{nord2}{HTML}{a3be8c}
\definecolor{nord3}{HTML}{bf616a}
\definecolor{nord4}{HTML}{b48ead}
\definecolor{nord5}{HTML}{ebcb8b}
\definecolor{nord6}{HTML}{88c0d0}
\definecolor{nord7}{HTML}{eceff4}

\newcommand{\change}[1]{{#1}}

\newcommand{\argmax}{\ensuremath{\mbox{argmax}}}
\newcommand{\vect}[1]{\boldsymbol{#1}}
\newcommand{\pert}{{\psi}}
\newcommand{\norm}[1]{\left\lVert #1 \right\rVert}
\newcommand{\dd}{\mathrm{~d}}

\newcommand{\fulltriangleright}{\raisebox{6pt}{\rotatebox{270}{\mbox{$\blacktriangle$}}}}
\newcommand{\su}{\ensuremath{\langle \vect u \rangle}}
\newcommand{\sud}{\ensuremath{\langle \vect u^2 \rangle}} 
\newcommand{\so}{\ensuremath{\langle \vect \omega \rangle}}
\newcommand{\sod}{\ensuremath{\langle \vect \omega^2 \rangle}} 
\newcommand{\sS}{\ensuremath{\langle \mathsf{S} \rangle}}
\newcommand{\sSd}{\ensuremath{\langle \mathsf{S}^2 \rangle}} 
\newcommand{\ssp}{\ensuremath{\langle s_+ \rangle}}
\newcommand{\ssi}{\ensuremath{\langle s_m \rangle}}
\newcommand{\ssm}{\ensuremath{\langle s_- \rangle}}
\newcommand{\sspd}{\ensuremath{\langle s_+^2 \rangle}}
\newcommand{\ssid}{\ensuremath{\langle s_m^2 \rangle}}
\newcommand{\ssmd}{\ensuremath{\langle s_-^2 \rangle}}
\newcommand{\sA}{\ensuremath{\langle \mathsf{A} \rangle}}
\newcommand{\sAd}{\ensuremath{\langle \mathsf{A}^2 \rangle}} 
\newcommand{\sap}{\ensuremath{\langle a_+ \rangle}}
\newcommand{\sai}{\ensuremath{\langle a_m \rangle}}
\newcommand{\sam}{\ensuremath{\langle a_- \rangle}}
\newcommand{\sapd}{\ensuremath{\langle a_+^2 \rangle}}
\newcommand{\said}{\ensuremath{\langle a_m^2 \rangle}}
\newcommand{\samd}{\ensuremath{\langle a_-^2 \rangle}}
\newcommand{\oso}{\ensuremath{\langle {\vect\omega}_i\mathsf{S}_{i\!j}{\vect\omega}_{\!j} \rangle}}
\newcommand{\osod}{\ensuremath{\langle ({\vect\omega}_i\mathsf{S}_{i\!j}{\vect\omega}_{\!j})^2 \rangle}}
\newcommand{\sss}{\ensuremath{\langle \mathsf{S}_{i\!j}\mathsf{S}_{\!j\!k}\mathsf{S}_{ki} \rangle}}
\newcommand{\sssd}{\ensuremath{\langle (\mathsf{S}_{i\!j}\mathsf{S}_{\!j\!k}\mathsf{S}_{ki})^2 \rangle}}
\newcommand{\usu}{\ensuremath{\langle {\vect u}_i\mathsf{S}_{i\!j}{\vect u}_{\!j} \rangle}}
\newcommand{\usud}{\ensuremath{\langle ({\vect u}_i\mathsf{S}_{i\!j}{\vect u}_{\!j})^2 \rangle}}
\newcommand{\tauij}{\ensuremath{\langle \tau_{ij}S_{ij} \rangle}}
\newcommand{\tauijd}{\ensuremath{\langle (\tau_{ij}S_{ij})^2 \rangle}}
\newcommand{\Q}{\ensuremath{\langle Q \rangle}}
\newcommand{\R}{\ensuremath{\langle R \rangle}}
\newcommand{\Qd}{\ensuremath{\langle Q^2 \rangle}}
\newcommand{\Rd}{\ensuremath{\langle R^2 \rangle}}

\newcounter{alg}
\newenvironment{algorithm}[1]%
        {\begin{list}
            {(#1\arabic{alg})~~~}
            {\usecounter{alg}
             \setlength{\labelwidth}{1.2cm}
             \setlength{\labelsep}{0em}
             \setlength{\itemsep}{0.2ex}
             \setlength{\topsep}{0.6ex}
             \setlength{\leftmargin}{1.2cm}
             \setlength{\rightmargin}{0cm}
             \setlength{\itemindent}{0em}
            }
        }
        {\end{list}}

\title{Identifying causal significance in three-dimensional isotropic turbulence}

\author{Miguel P. Encinar\aff{1}
  \corresp{\email{mencinar@torroja.dmt.upm.es}} \and
  Javier Jim\'enez\aff{1}}

  \affiliation{\aff{1}School of Aeronautics, Universidad Polit\'ecnica de Madrid, 28040 Madrid, Spain}

\begin{document}
\maketitle

\begin{abstract}
    Flow patterns of causal significance to three-dimensional isotropic turbulence are identified through the recently introduced algorithm of \cite{jim:18a}. Localised perturbations are introduced at arbitrary regions of a triple-periodic decaying flow at $\Rey_\lambda \approx 190$, and their evolution is used as a marker of the significance of said regions to the flow. Their dimensions are found to be an important parameter, with sizes of the order of the integral scale being controlled by the kinetic energy content, and sizes within the dissipative range, by the enstrophy and dissipation. The three quantities are found to be important at intermediate (inertial) scales. Prominent differences emerge between regions of high and low significance. The former typically contain strong gradients and/or kinetic energy and the latter are weak. An analysis of the structure of significant and insignificant flow patterns reveals that strain is more efficient than vorticity at propagating the contents of the perturbation to other regions of the flow. Moreover, the flow patterns of significant regions are found to be more complex, typically containing vortex clusters, while simpler vortex sheets are found in insignificant regions. The present results suggest that strategies aiming to manipulate the flow should focus on strain-dominated vortex clusters, avoiding enstrophy-dominated vortex sheets. This is confirmed through an assimilation experiment, in which greater synchronisation between two simulations is achieved when simulations share significant regions rather than insignificant ones. These conclusions have implications for both the control of turbulent flows and the making of predictions based on limited or noisy measurements. 
\end{abstract}



\section{Introduction} \label{sec:intro}
This paper originates from the recent Monte-Carlo simulations of \cite{jim:18a, jim:20b}. In the first of these papers, the author explores whether structures significant to the turbulent flow can be discovered semi-automatically by machine. The second one studies the properties of dipoles in two-dimensional turbulence, which are one of the suggestions of significant pattern obtained from the former. Overall, the campaign was successful at uncovering a flow pattern that had not received much attention until now. The premise of these works is that it is possible to leverage state-of-the-art hardware for turbulence research in a novel approach. From the earliest direct numerical simulations  \citep[DNSs,][]{or:72, sig:81, rog:81, kim:moi:mos:87} to the most recent ones \citep{lee:mos:15,iye:sre:20,vel:enc:21}, the most powerful supercomputers available at the time have been used to compute the largest simulations to date (in terms of degrees of freedom). Alternatively, some lines of research exploit faster hardware to minimise expensive cost functions iteratively, either to find fixed points \citep{nag:90}, orbits \citep{kaw:kid:01}, optimal transients \citep{pri:ker:10}, optimal assimilations \citep{wan:zak:19} or optimal states \citep{mot:kaw:18}. The optimisation of these cost functions involve iterating an expensive Newton-like optimiser, and cannot be applied to simulations at the state-of-the-art complexity. Instead, they need to settle for a combination of low Reynolds number, small boxes, and moderate simulation times, which we can summarise as `small' simulations. These simulations were state of the art about 30 years ago \citep{jim:20a}, and are still relevant today. \cite{jim:18a} explores a different usage scenario for small simulations, one in which modern hardware is used to run massive ensembles of them at an affordable cost. These randomised ensembles, and the analysis produced from them, are reffered by the author as Monte-Carlo Science \citep[MCS,][]{jim:20c}.

MCS is central to this work and we recall its premises here, although we refer the reader to the original works for further details.
The objective of the method is to study causality in dynamical systems through intervention.
The definition of causality is not unique, but a sensible one is that event $A$ is the cause of event $B$ if $B$ happens if and only if $A$ does \citep{pea:09}.
Two key aspects are introduced in the previous statement: time and precedence.
Causality requires time, so instances or `snapshots' of a dynamical systems are agnostic to causality.
It also requires $A$ preceding $B$, as past must be a cause for the future. Finally, there is a time horizon for causality and the delay between $A$ and $B$ is an important parameter. For example, while it is likely that rain in the morning is a cause for wet ground at noon, it is unlikely that rain on March is a cause for wet ground in August. The time between $A$ and $B$ that maximises the probability of the former being causal to the latter is usually known as the `causality horizon'
Thus, elucidating causality in a dynamical system requires occurrences of both $A$ and $B$ in a time history, with $A$ preceding $B$, and somewhat consistent time delays between the two events.
However, the time history may fulfil all the previous requirements, with no causation between $A$ and $B$, as it is well-known that correlation does not imply causality \citep{bee:hit:men:09}. 
An archetypical example of the former is the fact that day always precedes night, but night is not a consequence of day. For these reason, several algorithms exist to elucidate causality from time series detecting spurious correlations \citep{gra:69,sug:etal:12,dua:yan:che:12}. They have shown promising results in fluid mechanics \citep{loz:bae:enc:20, wan:chu:loz:21}, although they are not free from problems \citep{jam:bar:cru:16}.

For dynamical systems that can be manipulated by experiments, such as a direct numerical simulation (DNS) of the Navier--Stokes equations, an alternative is modifying the system and studying the consequences. For example, by modifying $A$ with some perturbation and observing the effect it has on $B$.
If $A$ is the sole cause of $B$ it can be expected that at least some actions on $A$ have a large impact on $B$, while manipulations that are `orthogonal' to $A$ leave $B$ unaffected.
In principle, evaluating the causal impact of an action is straightforward in a DNS, albeit with some limitations.
One needs to define a causal norm of interest, e.g.
large reduction of drag for an industrial flow, which takes the role of `$B$', and manipulate some cause `$A$' to test the veracity of the hypothesis that $A$ is the cause of $B$.
A successful example of this is the opposition control of \cite{choi94}, where the near-wall vortices are opposed by blowing/suction from the wall, effectively reducing drag.
In this sense it can be said that the near-wall vortices are a cause of drag, as destroying them affects the latter.
However they are probably not the only cause, as the friccion is reduced but the flow does not relaminarise. Thus, some other structures act as a different cause for drag and were not discovered in the previous experiments, which were based on a preconception of which variables are important for the system.
What the MCS introduces is a methodology to `automatically' deduce which are the causes for a given choice of $B$.
Instead of trying one intervention from \emph{a priori} assumptions on $A$, one tries a massive ensemble of actions, and checks which ones are effective.
It is expected that from the set of effective actions one can deduce which `$A$s' may be causes for the chosen `$B$'.
MCS attempts to extract information about the flow from the information provided by the interventions. In this sense, the intervention (or perturbations) can be interpreted as a probe for the flow dynamics.
As stated before, the procedure has shown promising results in \citep{jim:20b}, where the authors studied the properties of tight dipoles in two-dimensional turbulence, a previous suggestion of cause for a particular choice of consequence. Another example of success, related to the opposition control example above is \cite{pas:flo:20}, where the authors find that opposing streamwise velocity streaks is also an effective way of reducing drag.

In the present paper we explore the potential of the MCS procedure in three-dimensional decaying homogeneous isotropic turbulence (HIT).
Decaying turbulence is an interesting problem for MCS, as ensembles are the only way of computing statistics. The flow is not ergodic, which naturally leads to the computation of independent experiments.
Modern hardware, such as graphical processing units (GPUs), allow for relatively small problems to run fast enough to make massive simulation ensembles of reasonable HIT flows \citep{vel:jim:21}.
For example, in the present case it is possible to run one turnover of HIT in a $256^3$ grid in about $75~\mathrm{GPU\mbox{-}seconds}$.
That implies that a moderate cluster of 8 GPUs can produce an ensemble of $10\mkern2mu000$ experiments in approximately one week and a half.
The amount of data generated by those experiments presents a bigger challenge.
Assuming that 30 snapshots of the state of the flow are taken during one turnover, the previous experiment generates $40~\mathrm{TB/week}$, which is unmanageable for most research labs.
The consequence is that compromises have to be made, and extending MCS to three dimensions implies attempting limited observations.
Three dimensional turbulence is not only more challenging than the two dimensional case in practical considerations, but also on theoretical grounds.
In two-dimensional flows, enstrophy is an inviscid invariant \citep{ons:49},
which results in an inverse cascade confined between the energy-injection scale and the largest scale allowed by the boundaries \citep{kra:67}.
The implication is that obtaining sensible separation of scales in two-dimensional turbulence is hard and computationally expensive.
Thus, the experiments in \citep{jim:18a,jim:20c} have little scale separation and essentially there are few vortex sizes.
In contrast, the present experiments (even at a moderate Reynolds number) have considerable scale separation which makes our results very dependant on the size of the perturbations.

One important limitation of our work is the finite number of types of interventions that we can try. The Monte-Carlo Science problem is akin to looking for needles in a haystack: the possible interventions are infinite, and the number we can try is very limited. For example, one could try to zero the velocity vector within a region, but it is also possible to zero the vorticity vector instead. Several examples of reasonable families of perturbations can be found in \citep{jim:20c}, but the space of the ones not attempted remains infinite. As a result, many potential features of the flow, perhaps more interesting than the ones we show in the present paper, were possibly missed by our search. Nevertheless, it should be acknowledged that this problem is shared by the `traditional science' approach. In the latter, the problem of limited capacity for testing is replaced by personal bias. For example, choosing to study intense vortical regions may hide the energy-containing structures from the researcher. Nevertheless, the identified structures by either method remain relevant despite the fact that many other structural descriptions of the flow are possible. The aim of the present work is to relate the structural description of the flow to its dynamics by experimentation, rather than from preconceived ideas, and to relate them with previously known features of the flow.

Perhaps the limitation on the number of trials suggests the search for an `optimal' perturbation that maximises
some norm. If only a few perturbations could be tried it could be argued that an attempt should
be made to optimise their growth. However, we do not attempt this search for two reasons. In the first place, it is unclear whether
optimal perturbations constitute the best probes for the flow dynamics. In this work, perturbations are
introduced mainly to investigate the dynamics of the flow, not of the perturbations themselves. In
this context the fact that they are optimal is of little relevance. Optimal perturbations would point 
to a particular structure of the flow that may or may not be more interesting than the ones pointed out by
other types of perturbations. The second reason concerns the cost of finding the optimal. For finite size
perturbations, the search for an optimal is computationally expensive, as it relies on iterative methods. Also,
nonlinearity implies that the optimal depends on the amplitude introduced to the flow. Thus,
the cost of finding a single optimal for a particular amplitude is that of trying several suboptimal perturbations.
Since perturbations are used as probes, that computational power is better invested in covering different perturbations (e.g. perturbations of different sizes).

The dynamics of isotropic turbulence is a well researched topic, although structural descriptions of the flow are still limited.
Most of the classical theories of cascades \citep{kol:41b,kol:61,fri:78,men:sre:91} focus on predictions of the velocity structure functions rather than on the instantaneous structure of the flow patterns.
This stands opposite to free shear flows \citep{bro:ros:74}, and boundary layers \citep{lum:67,kli:rey:sch:run:67,wal:eck:bro:72} where the instantaneous structure of the flow patterns has been a topic of research from the beginning.
More recently, some structural aspects of HIT have been investigated. Vorticity is known to concentrate in the form of thin tubes, although is also possible to find it organised in vortex sheets \citep{sig:81,she:jak:ors:91,vin:men:91,jim:wra:saf:rog:93,hor:tak:05}. In contrast to the tubular vorticity structures, the rate-of-strain concentrates in `flake'-like structures \citep{moi:jim:04,leu:swa:dav:12} with shapes and fractal dimensions closer to surfaces than to tubes. Structures can also be extracted from the invariants of the velocity gradient tensor, $Q$ and $R$, which also point to vortical structures \citep{cho:per:can:90,hun:wra:moi:88}. All these structures are deduced from intense gradients, associating them to the dissipative range. To identity structures in the inertial range, other works filter either the kinetic energy field \citep{car:vel:17} or the enstrophy and dissipation fields \citep{hos:yam:97,leu:swa:dav:12,ber:pul:hor:09}. These works show that there are gradient-based structures at larger scales, and that some aspects of their geometry are different from those at the dissipative range, while other characteristics remain similar. We will see how these descriptions are related to the ones obtained from MCS.

Finally, although the main objective of the present paper is studying the flow field, we will also study the evolution of the perturbations used to study it.
The properties of infinitesimal perturbations are well studied, through the research of Lyapunov exponents \citep{eck:rue:85,yam:ohk:88}.
Although in principle these are properties of the evolution of the perturbations, in a practical matter they turn out to be properties of the ergodic attractor \citep{ose:68}, as they are defined from an infinitely long evolution.
There are also adaptations of the theory to shorter time horizons, the so called finite-time Lyapunov exponents \citep{aur:bof:97}, which deal with locality in time but still assume infinitesimal perturbations.
The evolution of finite perturbations has also received attention, mostly for the related problems of predictability \citep{aur:bof:96}, which is related to MCS, and data assimilation \citep{di:maz:bif:20,wan:tam:22}.
The latter focus on which are the conditions that a finite perturbation has to satisfy in order to be eliminated by a data assimilation program, which is opposite in nature to what the present manuscript deals with.
The object of study in many works on predictability are the finite-size Lyapunov exponents, i.e.
the evolution of finite amplitude perturbations.
In MSC, the perturbations we use are both finite-time and finite-size, with the additional property of being localised within the flow, which is necessary in order to be able to detect structures. Characterising the growth and time scales of these perturbations is important for their relevance as dynamical probes.

The remainder of the paper is organised in seven sections. The numerical experiments are described in \S\ref{sec:experiments}, with \S\ref{sec:perturbations} detailing how perturbations are introduced in the flow. Section \ref{sec:significance} follows, which focuses on the properties of the extreme perturbations themselves, while \S\ref{sec:regions} does so on the coarse-grained properties of the perturbed regions. The structural properties of these regions are presented in \S\ref{sec:structures}, both for the average structures and for the instantaneous ones. Finally, \S\ref{sec:assimilation} proves the relevancy of the identified flow patterns with a numerical assimilation experiment and section \S\ref{sec:conclusions} closes.

\section{Numerical experiments}\label{sec:experiments}

We study the temporal evolution of an incompressible turbulent fluid, as given
by the three-dimensional Navier--Stokes equations (NSE),
\begin{gather}
\partial_t u_i + u_j\partial_j u_i = - \partial_i p + \nu\partial_{jj} u_i + f_i,\label{eq:N1}\\
    \partial_i u_i = 0\label{eq:N2},
\end{gather}
where $\vect{f}\equiv f_i$ is a forcing, $\nu$ is the kinematic viscosity and $p$ is the
kinematic pressure. Throughout the paper, repeated indices imply summation, and $i = 1,2,3$ stands for the three spatial directions. Equations (\ref{eq:N1}, \ref{eq:N2})
are supplemented by triply periodic boundary conditions of spatial period $L=2\pi$,
resulting in a turbulent flow which is statistically homogeneous and 
approximately isotropic (HIT). The forcing injects constant
power in the largest wavenumber sphere, $k < 2$, where $k = \left|\vect{k}\right|$ is the wavevector $\vect{k}$ magnitude, and can be switched
off to produce a naturally decaying flow. A statistically steady state ensures that the average dissipation equals the constant power input in forced simulations.

Equations (\ref{eq:N1}--\ref{eq:N2}) are integrated in time using a fully phase-shifting dealiased
pseudospectral method \citep{rog:81}, with a three-step Runge-Kutta serving as
the time stepper \citep{spa:87}. Both the algorithm and the GPU implementation have been validated by previous works \citep{car:vel:17,vel:jim:21}. We use $256^3$ collocation points at $k_{\max{}}\eta_0 = 1$ for a $\Rey_\lambda \approx 190$ and $L_{\mathcal{E}0}/\eta_0 \approx 177$, where $\eta_0$ is the Kolmogorov scale of the initial flow field, $\Rey_\lambda$ is the Reynolds number based on the Taylor microscale, and $L_{\mathcal{E}0}$ is the initial integral scale. The three parameters are computed as in \cite{bat:53},
\begin{equation}
    L_\mathcal{E} = \frac{\pi}{2(q^\prime)^2}\int_0^\infty E_{qq}(k, t)/k\dd k,
\end{equation}
and
\begin{equation}
    (q^\prime)^2 = \frac{2}{3}\int_0^\infty E_{qq}(k, t)\dd k,
\end{equation}
where $E_{qq}$ is the energy spectrum and $q^\prime$ the root mean square (rms) velocity. The Taylor microscale $\lambda$ is
\begin{equation}
\lambda^2 = 15\nu q^{\prime2}/\epsilon,
\end{equation}
where $\epsilon$ is the dissipation, and $\Rey_\lambda = q^\prime\lambda/\nu$.

The initial conditions for the decaying
simulations are taken from the statistically steady state of
simulations with the same parameters as the decaying ones, but forced as described above.
They are spaced from each other by $30$ turnover times of the steady simulation, guaranteeing their statistical independence. Decaying simulations serve as independent experiments to detect structures in an
environment free from the influence of the forcing term. The length and time scales used to normalise the flow fields are computed for each of the initial conditions, and differ slightly from one to another.
The results of this paper use up to 50 initial conditions, probed with $O(100\mbox{--}10000)$ perturbations (depending on their size), for a total of approximately $10^6$ simulations of two turnover times each.

\begin{table}
    \begin{center}
        \begin{tabular}{lccccc} 
                & $\rmDelta_0$ & $\rmDelta_1$ & $\rmDelta_2$ & $\rmDelta_3$ & $\rmDelta_4$\\
                \hline
            $\rmDelta_i/L_{\mathcal{E}0}$ & 0.067 & 0.134 & 0.27 & 0.54 & 1.08\\
            $\rmDelta_i/\eta_0$ & 15 & 30 & 60 & 120 & 240\\
            $k_i\eta_0$ & 0.419 & 0.209 & 0.105 & 0.052 & 0.026\\
            Marker & \textcolor{nord0}{\fulltriangleright} & \textcolor{nord1}{\fullstar}& \textcolor{nord2}{\fullsquare} & \textcolor{nord3}{\fullcirc} & \textcolor{nord4}{\fulltriangle}\\

        \end{tabular}
    \caption{Perturbation sizes and markers used in the plots}%
    \label{tab:perturbations}
    \end{center}
\end{table}

\begin{figure}
    \begin{center}
    \includegraphics{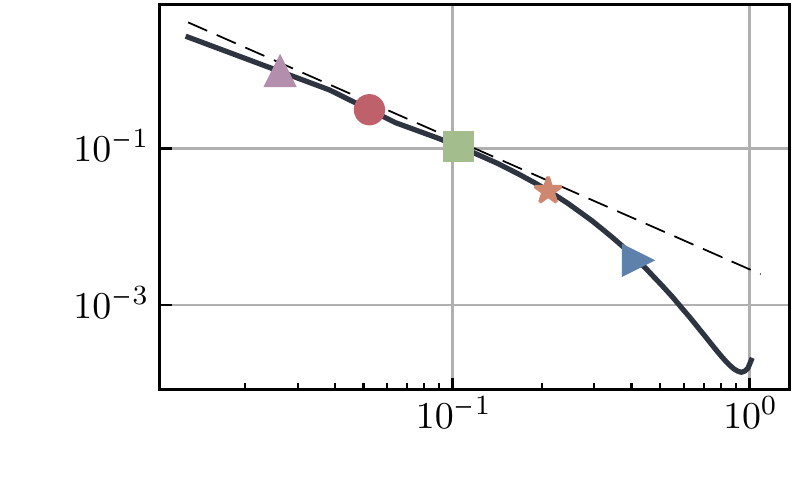}%
    \mylab{-3.5cm}{4pt}{$k\eta_0$}%
    \mylab{-8.2cm}{2.8cm}{$E_{q}$}%
    \end{center}

    \caption{Average energy spectrum of the initial conditions. Symbols as table \ref{tab:perturbations}. The markers represent the approximate size of the perturbations. \label{fig:speqq}}
\end{figure}

\section{Characteristics of the perturbations}\label{sec:perturbations}
Initial conditions $\vect u_0 \equiv [u_{01}, u_{02}, u_{03}]$ of a turbulent steady state are allowed to decay up to 60--70\% of its initial energy by running (\ref{eq:N1}, \ref{eq:N2}) without forcing. From now on, the zero subindex references values at the initial conditions. These decaying simulations, $\vect u_{\mathrm{ref}}(\vect x, t)$, serve as reference cases. Perturbed initial conditions are generated from $\vect u_0$, by adding localized perturbations to it,
\begin{equation}
    u_{0i}^\pert(x_i; \xi_{i}, \rmDelta) = u_{0i}(x_i) + \pert_i(x_i; \xi_{i}, \rmDelta),\label{eq:ploff0}
\end{equation}
where the perturbation $\vect \pert$ is generated from the product of a Gaussian kernel,
\begin{equation}
    g(x_i;\rmDelta) = \exp{(-\norm{x_i}^2/\rmDelta^2)},
\end{equation}
where $\norm{\cdot}$ is the Euclidean norm, and the velocity field is
\begin{equation}
    \pert^\dagger_i(x_i; \xi_{i}, \rmDelta) = -u_{0i}(x_i)g(x_i - \xi_{i}, \rmDelta).
\end{equation}
The perturbation $\pert^\dagger$, mimics the effect of an obstacle of characteristic size $\rmDelta$ at $\vect x = \vect \xi$, which would stop the velocity around and within it. However, incompressibility requires $\vect u_0^\pert$ to be divergence free, and $\pert^\dagger$ is projected to the closest perturbation (under the $L_2$-norm) that satisfies this requirement,

\begin{equation}
    \pert_i = \pert_i^\dagger - \partial_i\Psi,\label{eq:ploffn1}
\end{equation}
where the scalar field $\Psi$ is the solution to the Poisson equation,
\begin{equation}
    \partial_{jj}\Psi = \partial_i \pert_i^\dagger.\label{eq:ploffn}
\end{equation}
The final perturbation $\vect \pert$ is very close to $\vect \pert^\dagger$, although it does not completely stop the flow, acting like a mildly permeable obstacle. 

Figure \ref{fig:speqq} shows the spectrum of the initial conditions, as well as markers indicating the values of $k_\rmDelta = 2\pi/\rmDelta$ used in our experiments, as shown in table \ref{tab:perturbations}. The sizes shown are the radius in terms of the $1/e$ limit of the Gaussian. The five samples are distributed across the spectral range with the intention of sampling the dissipative range with $\rmDelta_0$, the inertial range with $\rmDelta_1$-$\rmDelta_3$, and the integral scale with $\rmDelta_4$.

\begin{figure}
    \includegraphics[width=0.5\textwidth]{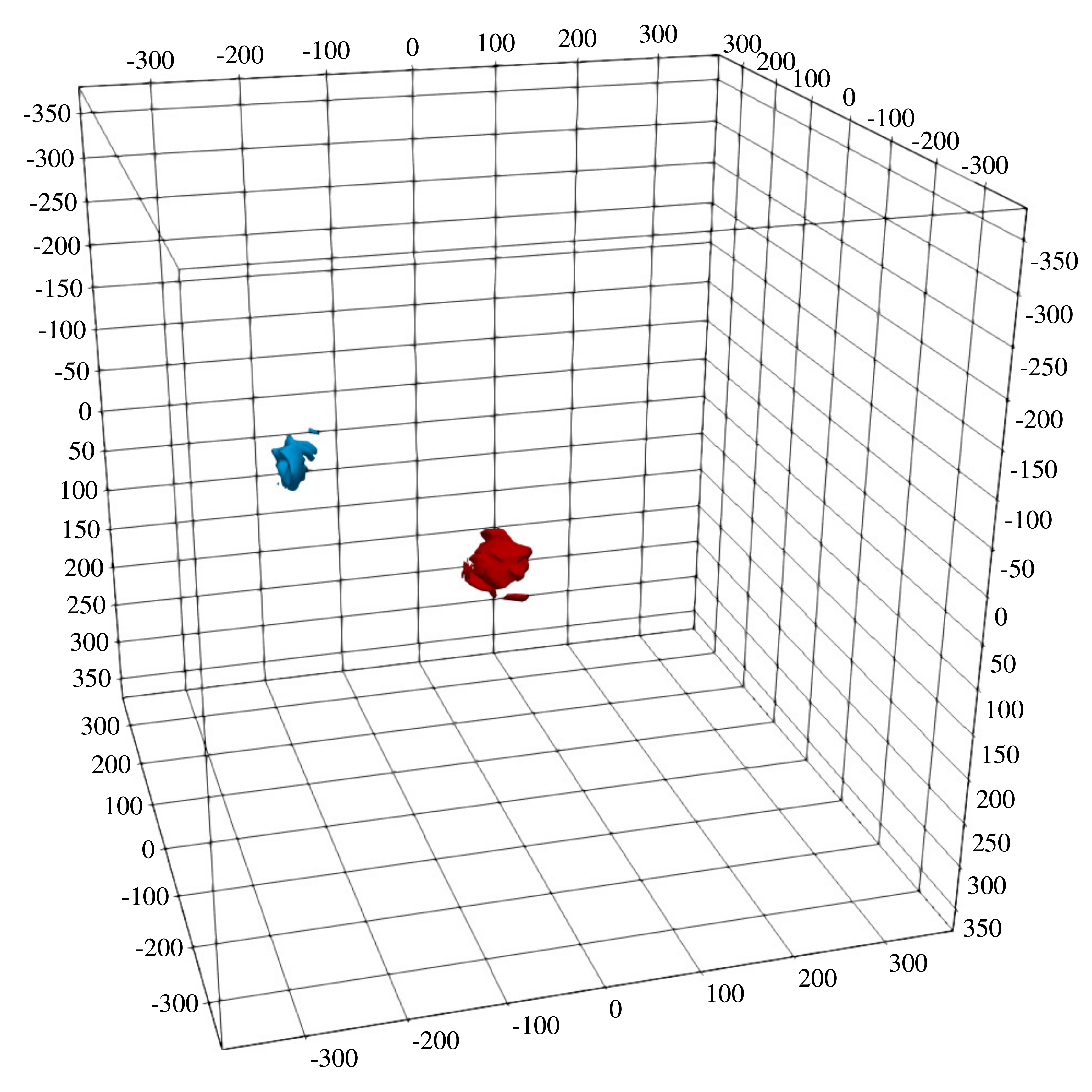}%
    \includegraphics[width=0.5\textwidth]{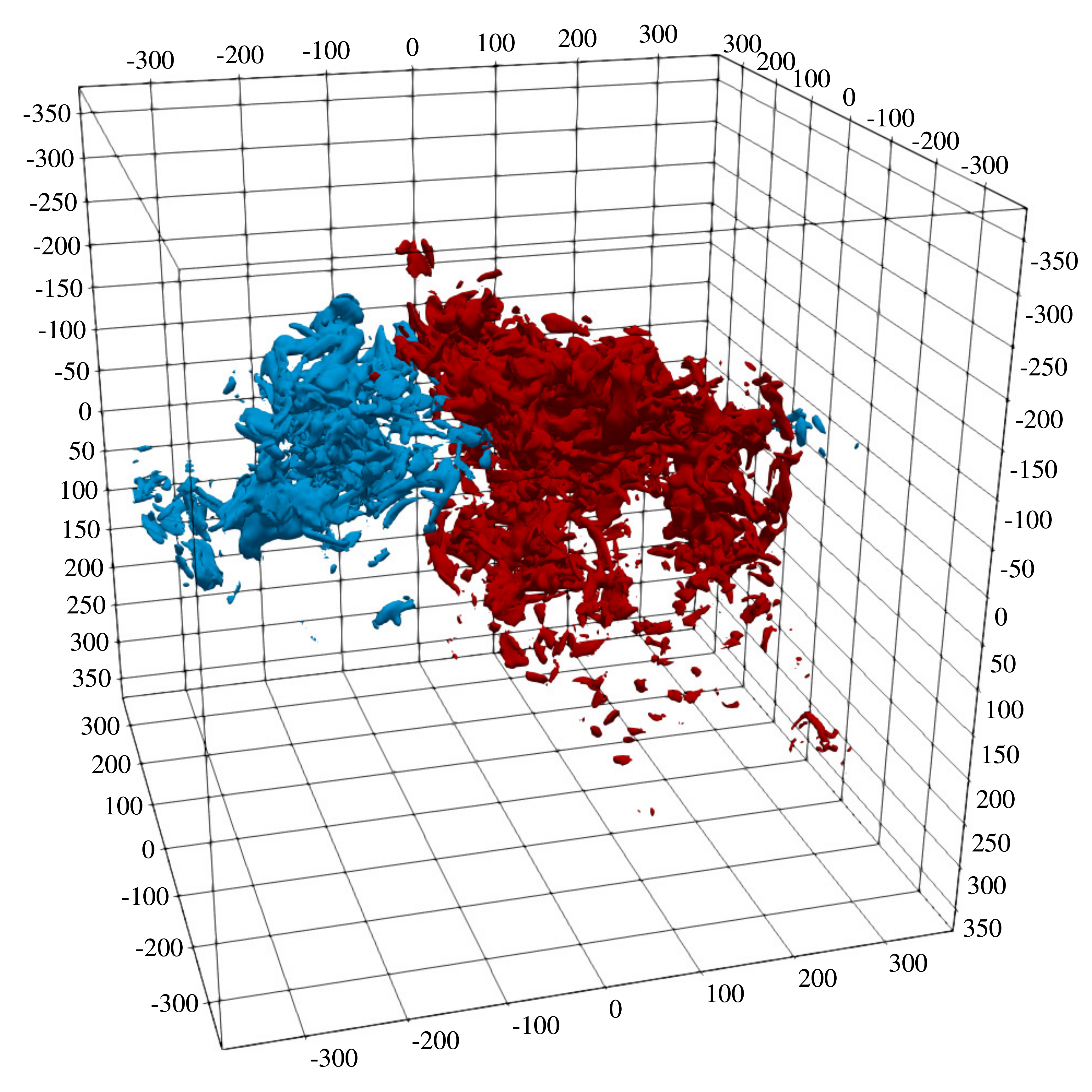}%
    \mylab{-0.98\textwidth}{1.0cm}{(a)}%
    \mylab{-0.49\textwidth}{1.0cm}{(b)}%

    \caption{Example of two different perturbations, one in the 95th percentile of growth (red), and one in the 5th percentile of growth, (blue). (a) $tq^\prime_0/L_{\mathcal{E}0} = 0$. (b) $tq^\prime_0/L_{\mathcal{E}0} \approx 1.2$.  \label{fig:3dper}}
\end{figure}

Perturbed initial conditions are generated for a set of $\vect \xi$ positions collocated in a Cartesian grid, with $\delta\vect \xi \approx 3\rmDelta$, which we found to be sufficient to probe the whole flow field. The perturbed initial conditions are evolved in time, generating a set of time-dependent perturbed solutions, $\vect u_{\mathrm{per}}(\vect x, t; \vect \xi)$, which allow us to define the perturbation norm field,
\begin{equation}
    \pert \equiv \left\|{\vect\pert}\right\|(\vect x, t; \vect \xi) = \norm{ \vect u_{\mathrm{ref}}(\vect x, t) - \vect u_{\mathrm{per}}(\vect x, t; \vect \xi) }.
\end{equation}
The perturbation norm field records the impact of the perturbation on the flow evolution as a function of the position $\vect x$, the position $\vect \xi$ where the perturbation of size $\rmDelta$ was introduced, and the observation time $t$. The spectrum of the scalar field $\pert$ is defined as,
\begin{equation}
    E_{\pert}(k, t; \vect \xi) = \int_{\Sigma_k} \hat\pert\hat\pert^*(\vect k, t; \vect \xi)\dd\sigma
\end{equation}
where $\Sigma_k$ is the surface of the sphere of constant wavevector magnitude, and the hat denotes triply Fourier transformation. The perturbation spectrum can be further integrated over $k$, giving the squared $L_2$-norm of the kinetic energy of the perturbation,
\begin{equation}
    \pert_q^{\prime2}(t; \vect \xi) = \int_0^{k_{\max{}}} E_{\pert}(k, t; \vect \xi)\dd k \equiv \int_\Omega \left\| \vect \psi \right\|^2 \dd {\vect x},
\end{equation}
where $\Omega$ stands for the full domain, and the squared $L_2$ magnitude of the perturbation enstrophy,
\begin{equation}
    \pert_\omega^{\prime2}(t; \vect \xi) = \int_0^{k_{\max{}}} k^2E_{\pert}(k, t; \vect \xi)\dd k,
\end{equation}
which due to incompressibility is proportional to the magnitude of the perturbation strain.
Both magnitudes measure the global impact of the perturbation at a given time, the former measuring the effect on the velocity, and the latter on the gradients of the flow.

Figure  \ref{fig:3dper} shows an example of $\psi$ for two different values of $\vect \xi$ and at two different times. Figure \ref{fig:3dper}(a) shows the initial perturbation and \ref{fig:3dper}(b) after $1.2~tq^\prime_0/L_{\mathcal{E}0}$. Two things should be noted. First, perturbations remain local, despite the presence of global effects affecting them (e.g. the pressure). Second, based on their location on the flow, perturbations may grow very differently. Both perturbations have a radius  of $\approx 30\eta$, their centres are $\approx 200\eta$ apart, and their initial energy is very similar. After a time of the order of one local eddy turnover, the `red' perturbation has grown over nine times more than the `blue' one in terms of $\psi^{\prime2}_q$. 

\begin{figure}
    \centering
    \includegraphics[width=0.99\textwidth]{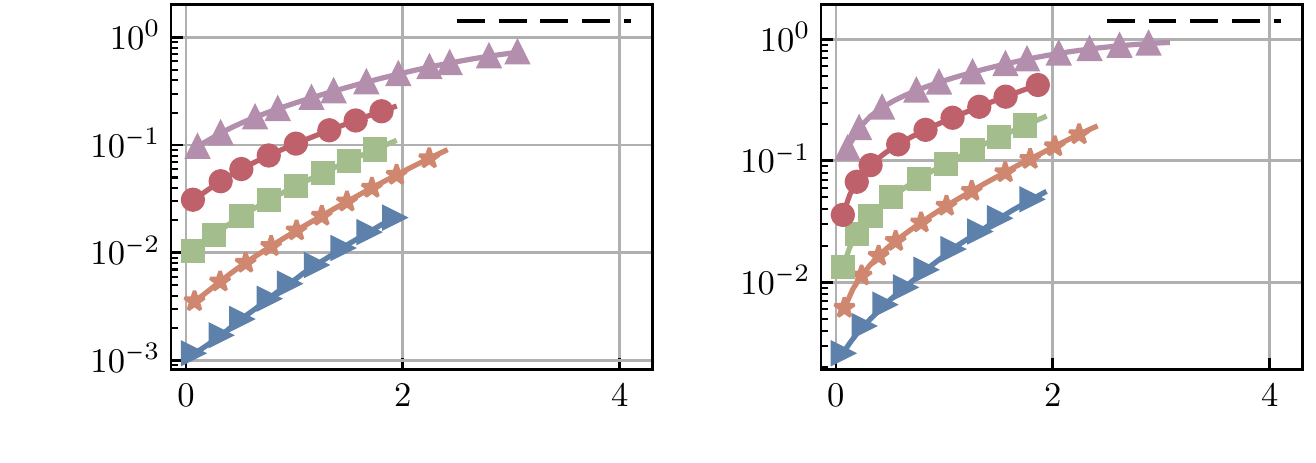}%
    \mylab{-3cm}{0.15cm}{$\tau$}%
    \mylab{-9.6cm}{0.15cm}{$\tau$}%
    \mylab{-6.5cm}{2.45cm}{$\dfrac{\pert_\omega^{\prime}}{\omega^{\prime}}$}%
    \mylab{-13cm}{2.45cm}{$\dfrac{\pert_q^{\prime}}{q^{\prime}}$}%
    \mylab{-13cm}{4.5cm}{(a)}%
    \mylab{-6.5cm}{4.5cm}{(b)}%

    \vspace{0.2cm}
    \includegraphics{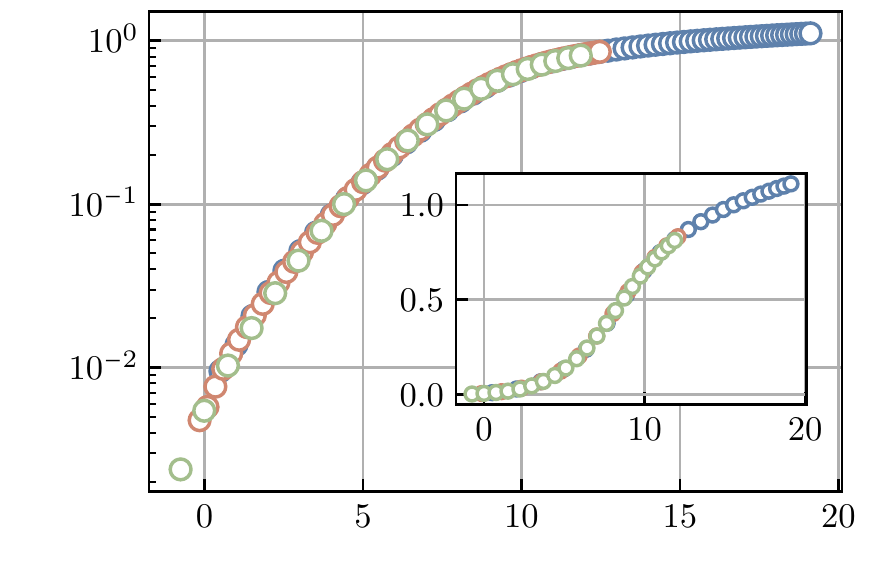}%
    \mylab{-4cm}{0.1cm}{$\tau$}%
    \mylab{-8.5cm}{3cm}{$\dfrac{\pert_q^{\prime}}{q^{\prime}}$}%
    \mylab{-8.4cm}{5.75cm}{(c)}%

    \caption{Mean perturbation $L_2$-squared growth $\pert_q^{\prime}/q^{\prime}_0$ (a) and $\pert_\omega^{\prime}/\omega^{\prime}_0$ (b) as a function of time, for all the perturbation sizes. Symbols as in table \ref{tab:perturbations}. The dashed lines are a exponential-saturation model that saturates at $\pert_q^{\prime}/q^{\prime}_0 = \sqrt{2}$. (c) Perturbation growth for 3 cases with different initial energy (symbols), that run much longer. \label{fig:pergr}}
\end{figure}

Figure \ref{fig:pergr} shows the mean perturbation growth of $\pert_q^\prime$ and $\pert_\omega^\prime$ as a function of time, for the values of $\rmDelta$ in table \ref{tab:perturbations}. They are normalised with the contemporary magnitudes of kinetic energy and enstrophy of the unperturbed field, ensuring a proper representation of the perturbation growth despite the decay of the base flow. As expected for a homogeneous flow, the initial value is proportional to $\rmDelta^3$, which is the volume of the perturbations. Because the time scale changes in a decaying flow, we define the local eddy turnover as $L_\mathcal{E}/q^\prime$, which allows us to define the non-dimensional time,
\begin{equation}
    \tau \equiv \int_0^{\tau} \frac{q^\prime\dd t}{L_{\mathcal{E}}},
\end{equation}
which captures the slowing time scale of the flow.
On average, all the perturbations grow with time, regardless of their size. While this may seem natural for a chaotic system, effects such as synchronisation are known to happen for perturbations in Fourier space at high wavenumbers. For example, in \cite{yos:yam:kan:05}, two simulations are forced to share the low wavenumbers up to a cutoff, leaving the remainder wavenumbers unconstrained. The latter synchronise in both simulations if the cutoff is in the dissipative range. 
In contrast, perturbations here are local in space and have a non-local spectrum in Fourier space, which explains why they do not vanish even in the dissipative range.

In contrast to infinitesimal perturbations, our finite-amplitude perturbations  grow sub-exponentially, as the growth rate $(\partial_t q/q)$ slightly decreases with time. 
Smaller perturbations grow faster, although their growth rate is never such that they overgrow larger perturbations. Regardless of their size, both the kinetic energy and enstrophy of the perturbations seem to approach an asymptote. It is important to recall that both base and perturbed flows are decaying, and by the end of the time window shown in the figure, their turbulent kinetic energy is 60--70\% of the initial one. This implies that the approached asymptotic value must be decaying too, and thus hard to estimate. However, under the scaling shown in figure \ref{fig:pergr}, which compensates for the decay of kinetic energy in the simulations and for the changing time scale, the growth is close to exponential until the magnitude of the perturbation is of the same order than that of the flow field, when it starts to saturate. This is confirmed in figure \ref{fig:pergr}(c), which shows how cases that run for much longer behave like an exponential model with saturation. In this context, a good reference is $\psi'_q/q' = \sqrt{2}$, which would be the constant asymptote for uncorrelated flows with similar energy. The time series in figure \ref{fig:pergr}(c) collapse together with the introduction of an arbitrary time delay that compensates for the difference in initial energy. 
\change{This suggests that the dependence of the average perturbation magnitude with their average initial energy can be modelled with the introduction of a virtual origin.}
The collapse is remarkable, considering that at the end of the evolution of these extended cases, the perturbations contain approximately one fifth of the initial energy of the flow.

Small differences exist between $\psi'_q$ and $\psi'_\omega$. The latter shows two different time scales, a fast one, $\tau < 0.3$, in which new gradients are generated in the perturbed region, and another one comparable to that of $\psi'_q$, which slowly approaches the asymptote. The best indication that the initial growth is a different time scale, is that the growth rate is approximately independent of the perturbation size, except for $\rmDelta_0$ which is a small scale itself.
A reasonable explanation is that the faster initial growth is related to out-of-the-attractor dynamics caused by the abrupt initial perturbation, and a consequence of the flow recreating the abnormally missing small scales. We will refer to this first part of the evolution ($\tau < 0.3$) as the `transient' from now on.

\begin{figure}
    \centering
    \includegraphics[width=0.99\textwidth]{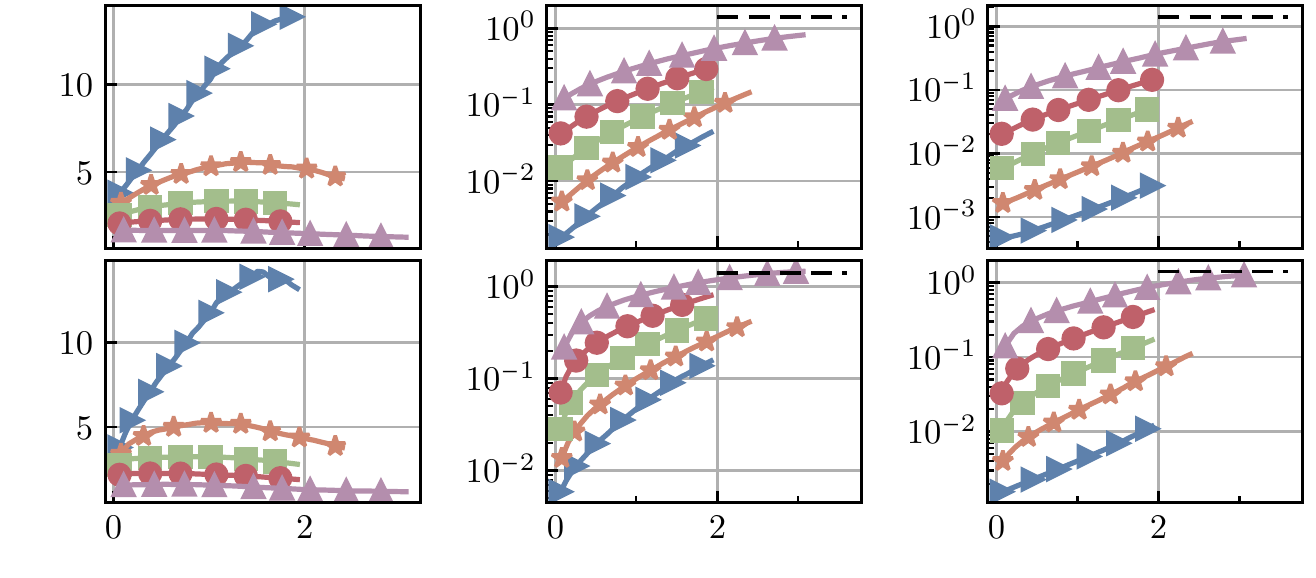}%
    \mylab{-0.145\textwidth}{10pt}{$\tau$}%
    \mylab{-0.478\textwidth}{10pt}{$\tau$}%
    \mylab{-0.811\textwidth}{10pt}{$\tau$}%
    \mylab{-4.6cm}{2.5cm}{$P_\omega^{5}$}%
    \mylab{-4.6cm}{5.0cm}{$P_q^{5}$}%
    \mylab{-9.0cm}{2.5cm}{$P_\omega^{95}$}%
    \mylab{-9.0cm}{5.0cm}{$P_q^{95}$}%
    \mylab{-13.3cm}{2.5cm}{$\mathcal{R}_{\omega}$}%
    \mylab{-13.3cm}{5.0cm}{$\mathcal{R}_{q}$}%
    \mylab{-13.cm}{5.75cm}{(a)}%
    \mylab{-13.cm}{3cm}{(d)}%
    \mylab{-8.9cm}{5.75cm}{(b)}%
    \mylab{-8.9cm}{3cm}{(e)}%
    \mylab{-4.4cm}{5.75cm}{(c)}%
    \mylab{-4.4cm}{3cm}{(f)}%
    \vspace{1em}

    \includegraphics[width=0.99\textwidth]{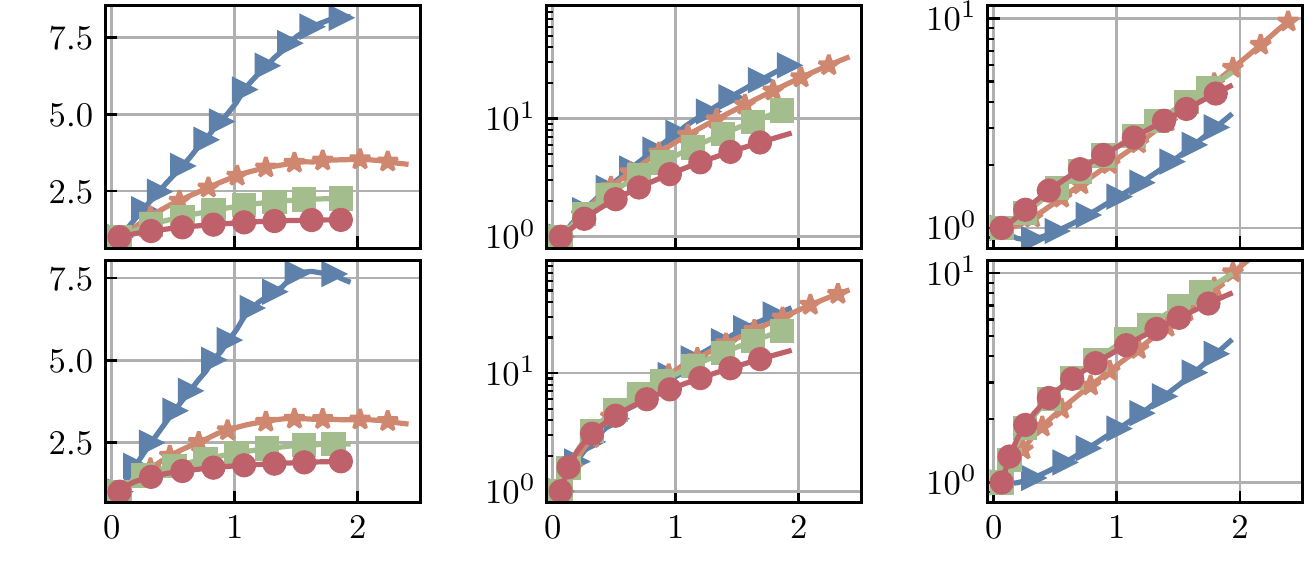}%
    \mylab{-0.145\textwidth}{0pt}{$\tau$}%
    \mylab{-0.478\textwidth}{0pt}{$\tau$}%
    \mylab{-0.811\textwidth}{0pt}{$\tau$}%
    \mylab{-4.1cm}{2.5cm}{$\tilde P_{\omega}^{5}$}%
    \mylab{-4.1cm}{5.0cm}{$\tilde P_{q}^{5}$}%
    \mylab{-8.75cm}{2.5cm}{$\tilde P_{\omega}^{95}$}%
    \mylab{-8.75cm}{5.0cm}{$\tilde P_{q}^{95}$}%
    \mylab{-13.5cm}{2.0cm}{$\mathcal{\tilde R}_{\omega}$}%
    \mylab{-13.5cm}{4.5cm}{$\mathcal{\tilde R}_{q}$}%
    \mylab{-13.4cm}{5.85cm}{(g)}%
    \mylab{-13.4cm}{3cm}{(j)}%
    \mylab{-8.5cm}{5.85cm}{(h)}%
    \mylab{-8.5cm}{3cm}{(k)}%
    \mylab{-4.3cm}{5.85cm}{(i)}%
    \mylab{-4.3cm}{3cm}{(l)}%
    
    \caption{Statistics of the growth of the 95th ($P^{95}$) and the 5th ($P^{5}$) percentiles of  $\psi'_{q}$ in (a,b,c), $\psi'_{\omega}$ in (d,e,f), $\tilde \psi'_{q}$ in (g,h,i) and $\tilde \psi'_{\omega}$ (j, k, l) for perturbations that reduce the kinetic energy. (a,d,g,j) Significance ratio, defined as the quotient between the percentiles. (b,e,h,k) $P^{95}$. (c, f, i, l) $P^{5}$.\label{fig:pertu}\label{fig:pertur}}
\end{figure}

\section{From perturbations to significance}\label{sec:significance}


While figure \ref{fig:pergr} gives an idea of the `typical' growth of a perturbation, figure \ref{fig:pertu} shows the evolution of the extreme ones. We divide the experiments at each time step in percentiles, adopting the notation $P^{\alpha}_a(t)$ for the $\alpha$-th percentile of the variable $\psi_a(t; \xi)$ and considering each $\xi$ and ${\vect{u}}_{\mathrm{ref}}$ as an individual experiment. The percentile is defined as the value of $\psi_a(t)$ below which a percentage $\alpha$ of the samples fall. For small values of $\alpha$, we consider `insignificant' perturbations those below $P^{\alpha}_a(t)$ and `significant' ones those above $P^{100-\alpha}_a(t)$. We also define the $\alpha$-`significance ratio' as,
\begin{equation}
    \mathcal{R}^{\alpha}_a = \frac{P^{(100-\alpha)}_a}{P^{\alpha}_a(t)}.\label{eq:sig}
\end{equation}
The high/low percentiles and the ratio \eqref{eq:sig} serve
as proxies for the maximum and minimum values of each variable, and for the quotient between the two respectively, because these statistics are very hard to converge. For the remainder of the paper we
limit our analysis to the 5\% most and least intense
perturbations ($\alpha = 5$), and omit the $\alpha$ superscript when referring to $\mathcal{R}$. We checked that our results are qualitatively similar
as long as $3 < \alpha < 10$. Lower values of $\alpha$ require considerably more experiments  to obtain converged statistics (which for $\alpha = 5$ are $O(10^5)$ for each size), and higher
values of $\alpha$ start to include perturbations  that
can no longer be considered `extreme'.
These definitions are consistent with causally significant and insignificant events in the context of Monte Carlo science. In turn, the significance ratio measures how different are both sets from each other, and gives an indication of how hard is to tell them apart. The top row of figure \ref{fig:pertu} shows these three measures for the energy norm and the second row for the enstrophy norm.
A slightly different question is which perturbations are amplified the most or the least over the same period.
The last two rows of figure \ref{fig:pertur} show the significance ratio and percentiles for the amplification of the kinetic energy, $\tilde \psi_{q}(t) = \psi_{q}(t)/\psi_{q}(0)$, and of the gradients, $\tilde \psi_{\omega}(t) = \psi_{\omega}(t)/\psi_{\omega}(0)$.

Figures \ref{fig:pertu}(a) and \ref{fig:pertu}(b) show that smaller perturbations reach higher significance ratios than larger ones, especially for the smallest perturbation, $\rmDelta_0$, whose size is in the dissipative range. For the largest case, $\rmDelta_4$, the significance ratio is always lower than 1.5, and the ratio is largely explained by the difference in the initial conditions. The implication is that the ratio has little to do with the dynamics, and is almost entirely traceable to the energy removed from the flow by the perturbation. The initial magnitude of these $\rmDelta_4$ perturbations is about one tenth of the initial field, and considering the similarity in their growths, it can be concluded that their size is too large to identify individual structures. This is not the case with the rest of the perturbations, where the energy subtracted to the field stays small. For example, the $\rmDelta_2$ perturbations initially subtract an average $0.03\%$ of the energy of the flow, and the difference in energy between reference and perturbed simulation stays constant in percentage through the evolution. In contrast, the perturbation energy at $\tau \approx 2$ is about $3\%$ of the energy of the field, which is ten times larger than the energy difference. It can be concluded that $\rmDelta_4$ perturbations essentially contain a piece of the flow that is almost homogeneous on its own. Note that this can be used as an alternative definition of the integral scale, and that $\rmDelta_4/L_\mathcal{E} \approx 1$. Since the objective of this work is the identification of individual structures, we do not pursue experiments at this size. 

Figure \ref{fig:pertu} also shows that the smallest perturbation $\rmDelta_0$ behaves differently than the larger ones. The greater significance ratio of $\rmDelta_0$ in figure \ref{fig:pertu}(a,d,g,j) is not due to the significant perturbations but to the evolution of the insignificant ones.
The growth rate of the perturbations decreases monotonically with time in almost every case except in $P^{5}$ for $\rmDelta_0$, in which it increases as a function of time, both for the perturbation energy and for its gradients. The other exception is the kinetic energy of $\rmDelta_1$ which shows almost constant growth rate in figure \ref{fig:pertur}(c, i). This behaviour of the two smaller perturbations is specially clear in the amplifications, where all perturbations start from unity at $\tau = 0$. Their small size allows some $\rmDelta_0$ perturbations to be dominated both by the larger scales that contain them, and by dissipation \citep{yos:yam:kan:05}. 
Our results show that this behaviour is not possible for velocity perturbations larger than $\rmDelta_1$. 
Comparing figure \ref{fig:pertur}(k) to \ref{fig:pertur}(l) shows that $\rmDelta_0$ collapses with the inertial perturbations in $\tilde P^{95}_\omega$ but departs from them in $\tilde P^{5}_\omega$. 
If we consider an initially linear evolution of the perturbations, we can separate their initial energy into a projection over the unstable manifold and another one over the stable one. In order for the perturbation to experience a transient contraction, the projection over the stable manifold has to dominate the dynamics initially. As the size of the perturbation grows, the probability of injecting energy over the unstable manifold grows in detriment of the stable one, until at approximately $\Delta > \Delta_1$, no contracting perturbation can be found.
From the point of view of the structure of the flow, this is possible because the flow is intermittent, and thus $\rmDelta_0$ can be smaller than the local dissipative range \citep{fri:ver:93}. In turn, significant $\rmDelta_0$ perturbations
are larger than the local dissipative range and thus contain some inertial 
must contain dynamics free from the contraction by the dissipation.

Regardless of the norm and of the perturbation size, all perturbations reach maximum significance ratios within the interval $\tau \in (0.5, 1.5)$, with small differences in the peaking time. The norm based on the gradients peaks slightly faster than the one based on the kinetic energy but far from the initial transient, in which the perturbation gradients grow much faster than its velocities. 

    

The significance ratios defined from the amplification,  $\mathcal{\tilde R}^{5}_{q}$ and $\mathcal{\tilde R}^{5}_{\omega}$ in figure \ref{fig:pertur}(g, j), share most of their statistics with the absolute magnitude ones, $\mathcal{R}_{q}$ and $\mathcal{R}_{\omega}$ in figure \ref{fig:pertu}(a, d). The time of maximum significance ratio is slightly delayed in every case, while remaining of the same order. Their maximum significance ratios are shallow, and their amplitudes tend to plateau after some time. The average values of the significance ratios based on amplification are lower, due mostly from the influence of the magnitude of the initial perturbation being removed. This can be easily seen from their definitions, 
\begin{equation}
    \psi^\prime_{q}(\tau) \approxeq \psi^\prime_{q}(0)\tilde \psi^\prime_{q}(\tau) = \psi^\prime_q(0)\exp\left(\int_0^\tau \Lambda(t) \dd t \right),\label{eq:absrellog}
\end{equation}
where $\Lambda(t) = \mathrm{d}[\log(\psi^\prime_{q}(t))]/\mathrm{d}t$ is the average finite-size finite-time Lyapunov exponent. \change{Note that $\Lambda(t)$ is the local growth rate of the average growth and, in general, it is different from the growth rate of each perturbation. Thus, \eqref{eq:absrellog} is an approximation in which $\tilde \psi^\prime_{q}$ and $\Lambda(t)$ do not depend on $\psi^\prime_{q}(0)$}. It suggests that the difference between absolute norms and amplifications must come from the memory of the initial perturbation in the system. If $\psi^\prime_{q}(0)$ and $\psi^\prime_{q}(\tau_{\max{}})$ are approximately independent, then the significance ratio of the magnitudes is necessarily larger or equal than the ratio of the amplifications. 
The consequence of having smaller values of the significance ratio in the amplifications is that significant and insignificant regions are harder to tell apart, and thus individual structures are harder to identify in this case.

Finally, figure \ref{fig:pertur}(k) shows excellent collapse of the maximum enstrophy growth during the transient. This is in agreement with our hypothesis, that the growth of enstrophy in the initial transient depends on the regeneration of the dissipative scales and thus is roughly universal.


The analysis so far hinges on a single definition of perturbation. In order to confirm our results we run similar experiments, with perturbations that target the vorticity inside a compact region instead of the velocity. The procedure is analogous to eqs. (\ref{eq:ploff0}--\ref{eq:ploffn}), but replacing the velocity field by the vorticity field. These perturbations are harder to associate with a physical mechanism that modifies the local rate of rotation, but serve our purpose of probing the flow for structural differences related to the significance. Because they reinforce our previous conclusions but do not provide any new strong one, we discuss them in the appendix \ref{app:op}.

\begin{figure}
    \centering{\vspace{1em}}
    \includegraphics[width=0.99\textwidth]{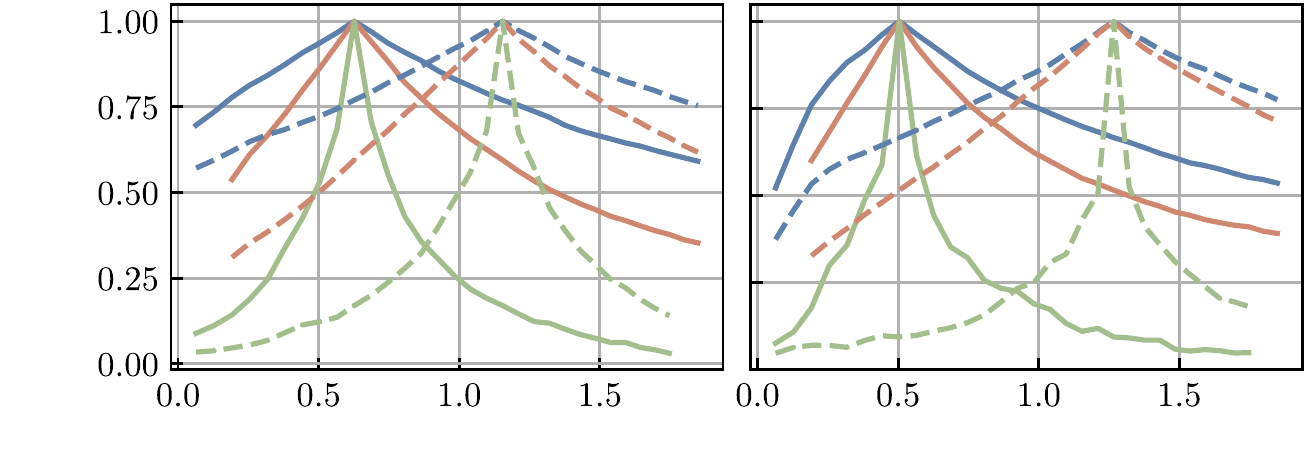}%
    \mylab{-3cm}{4pt}{$\tau$}%
    \mylab{-8.9cm}{4pt}{$\tau$}%
    \mylab{-13cm}{2.75cm}{$\mathcal{P}$}%
    \mylab{-11.5cm}{4.25cm}{$(a)$}%
    \mylab{-5.5cm}{4.25cm}{$(b)$}%
    \caption{Persistence of significance for velocity perturbation of size $\rmDelta_2$, as a function of time. Blue lines are (a) $\mathcal{P}(\tau, q)$ and (b) $\mathcal{P}(\tau, \omega)$. Orange lines are (a) $\mathcal{P}(\tau, {\tilde q})$ and (b) $\mathcal{P}(\tau, {\tilde \omega})$. Green lines are (a) $\mathcal{P}(\tau, \Lambda)$ and (b) $\mathcal{P}(\tau, \Lambda_\omega)$. (a, b) Solid lines are $\tau_{\max} = \argmax(\mathcal{R}_\omega)$ and dashed lines are $\tau_{\max} = \argmax(\mathcal{R}_q)$\label{fig:persis}}
\end{figure}

Our objective is characterising regions that show extreme behaviour when perturbed. For this reason, being able to distinguish significant regions from insignificant ones is fundamental for the rest of the analysis. Under this premise, the most obvious time delay to define causation is the time of maximum significance ratio, $\tau_{\max}$. Moreover, the consistency of time delays across very different perturbation sizes, norms and types of perturbations indicates that the interval $\tau \in (0.5, 1.5)$ is most relevant in our experiments. Before committing ourselves to one time delay, we explore how critical this choice is for the identification of significant regions. 
Figure \ref{fig:persis} shows the `persistence' of significances defined as,

\begin{equation}
    \mathcal{P}(\tau; a) = \frac{\mathrm{prob.}\left[\left(\psi^\prime_a(\tau)>P^{95}_a\right) \& \left(\psi^\prime_a(\tau_{\max})>P^{95}_a\right)\right]}{\mathrm{prob.}\left[\psi^\prime_a(\tau_{\max})>P^{95}_a\right]},\label{eq:perse}
\end{equation}
i.e. the probability that a perturbation that is considered significant at $\tau=\tau_{\max}$ is also significant at other times.
We show the results for $\rmDelta_2$, but they are qualitatively similar for the other inertial sizes. 
Blue lines show $\mathcal{P}(\tau, q)$, revealing how persistent is the absolute
norm. Taking the time of maximum significance according to $\mathcal{R}^{5}_q$ ($\tau_{\max} \approx 1.15$) as a reference, the minimum persistence in the interval $\tau \in (0.5, 1.5)$ is 70\%, which justifies considering $\tau_{\max}$ representative of the whole interval. Taking $\tau_{\max} \approx 0.6$, which is the maximum for $\mathcal{R}_\omega$ maintains 75\% of the significant regions at the previous choice of $\tau_{\max}$, and more that 50\% elsewhere. The persistence of the amplification, shown in orange, behaves qualitatively similar but is everywhere worse. Still, it gives a minimum of 50\% of persistence within the $\tau \in (0.5, 1.5)$ interval. The difference between the absolute and the relative persistences should come from the effect of the norm of the initial perturbation (as hypothesised in \eqref{eq:absrellog}), which persists for a very long time. The persistence of $\Lambda(\tau)$ is represented by green lines, showing that is a very local quantity, where the persistence decays to less than 10\% within the $(0.5, 1.5)$ interval. In contrast, the persistence of the initial norm of the perturbation $\psi^\prime_{q}(0)$, decays initially but never goes below 30\% (not shown). \change{These results suggest that the approximations introduced in \eqref{eq:absrellog} are justified}. The local nature of $\Lambda(\tau)$ explains well the different behaviour of $\mathcal{R}_{q}^{5}$ and $\mathcal{\tilde R}_{{q}}$. The latter plateaus when the growth rates of initially different perturbations are equal, as seen in figure \ref{fig:pertur}(g, j). After some time, all sufficiently weak perturbations approach the growth rate given by the largest Lyapunov exponent of system \citep{ose:68}. The beginning of the plateau (or a very weak maximum) indicates that the growth rate of every perturbation becomes approximately equal, making it a sensible choice for the causality horizon. In $\mathcal{R}_q$, the added factor $\psi^\prime_{q}(0)$ gives an initial significance that decays in importance respect to the growth of the integral term at latter times. The product of a decaying significance in $\psi^\prime_{q}(0)$ and a growing one in $\tilde \psi^\prime_{q}(\tau)$ results in a maximum of the significance ratio. At this time the memory from the initial perturbation is of the same order than the growth of the integral of $\Lambda(\tau)$. 
Based on our analysis, it is reasonable to use the time of maximum significance ratio as the causality horizon for each norm.
For the remainder of the manuscript we show results for significant and insignificant perturbations at $\tau=\tau_{\max}$, which is different for the different sizes and norms, but they can be thought of representative of the interval $\tau \in (0.5, 1.5)$ in every case. 

It should be noted that \eqref{eq:perse} is based on the intersection of significant sets, but it could have been formulated equivalently for the intersection of insignificant sets. The persistence defined that way is always larger than the one shown in figure \ref{fig:persis}, so the analysis we have performed is the most restrictive of the two. More details about the relations among the different norms are in the appendix \ref{app:b}.

\begin{figure}
    \centering
    \includegraphics[width=0.99\textwidth]{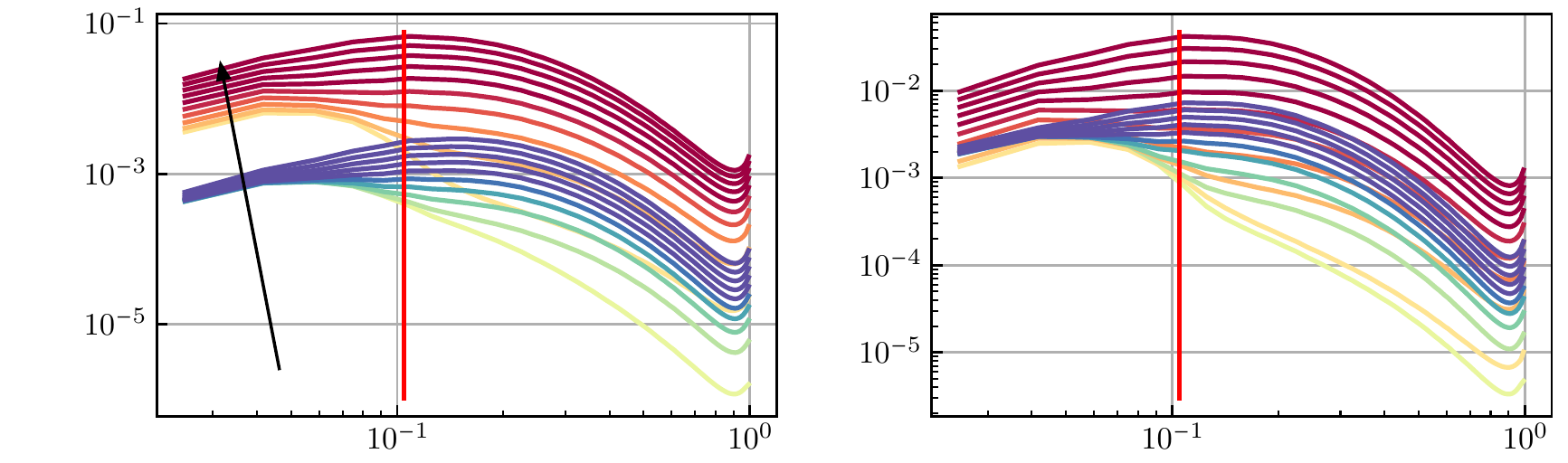}%
    \mylab{-3cm}{0pt}{$k\eta_0$}%
    \mylab{-9.25cm}{0pt}{$k\eta_0$}%
    \mylab{-13.2cm}{2.0cm}{$\dfrac{E_{\psi}}{E_q}$}%
    \mylab{-6.55cm}{2.0cm}{${\tilde E}_{\psi}$}%
    \mylab{-13.05cm}{3.6cm}{(a)}%
    \mylab{-6.5cm}{3.6cm}{(b)}%

    \caption{Average spectra of the significant (red) and insignificant (blue) velocity perturbations of $\rmDelta_2$, classified according to (a) $\psi'_q$ and (b) $\tilde \psi'_{q}$ as a function of time, from light to dark. The darkest lines corresponds to $\tau_{\max}$ in each case. The vertical red line is $k_2\eta_0$.\label{fig:speu}}
\end{figure}

Fixing the causality horizon $\tau_{\max}$ allows us to define sets of significant and insignificant perturbations. We can explore the differences between both sets by conditioning statistics of the perturbations to either of them. Figure \ref{fig:speu} shows the conditionally averaged spectra $E_{\psi}$ for significant/insignificant perturbations as a function of time and for two perturbation sizes,  $\rmDelta_0$ and $\rmDelta_2$. The top row shows the spectra of the sets based on $\mathcal{R}_q$ and the bottom one of the ones based on  $\mathcal{\tilde R}_q$.
It is remarkable that the difference between both norms amounts mostly to a vertical translation of the significant spectra, removing the offset of $\psi_q(0)$ in \eqref{eq:absrellog}.
Initially, the spectral mass is centred around the perturbation size, shifting towards smaller scales as the perturbation starts to grow.
This phenomenon can be attributed to the initial transient, and thus applies both to significant and insignificant perturbations.
The difference between significant regions and insignificant ones lies mostly on the scales larger than the perturbation, which grow substantially in the significant perturbations, and remain inactive in the insignificant ones.
This difference is more acute in the inertial sizes such as $\rmDelta_2$, where small scales grow comparably for both sets, but insignificant perturbations develop no large scales.
The evolution is less clear for the smallest perturbations, $\rmDelta_0$, because most wavenumbers are larger than the perturbation size, and the difference in growth rate among sets is distributed across most of the spectrum (not shown).
Caution should be taken when interpreting the higher energy growth of the perturbation in the significant case.
Because $E_\psi$ is the spectra of the difference between two fields, larger values are not necessarily associated to larger energies, but to large spatial energy differences among the two.
Thus, the significant perturbations are capable of displacing or deforming large scales in space, generating large energy differences that show in the difference spectra. This is confirmed by visual inspection of the perturbed fields.

\begin{figure}
    \centering
    \includegraphics[width=0.99\textwidth]{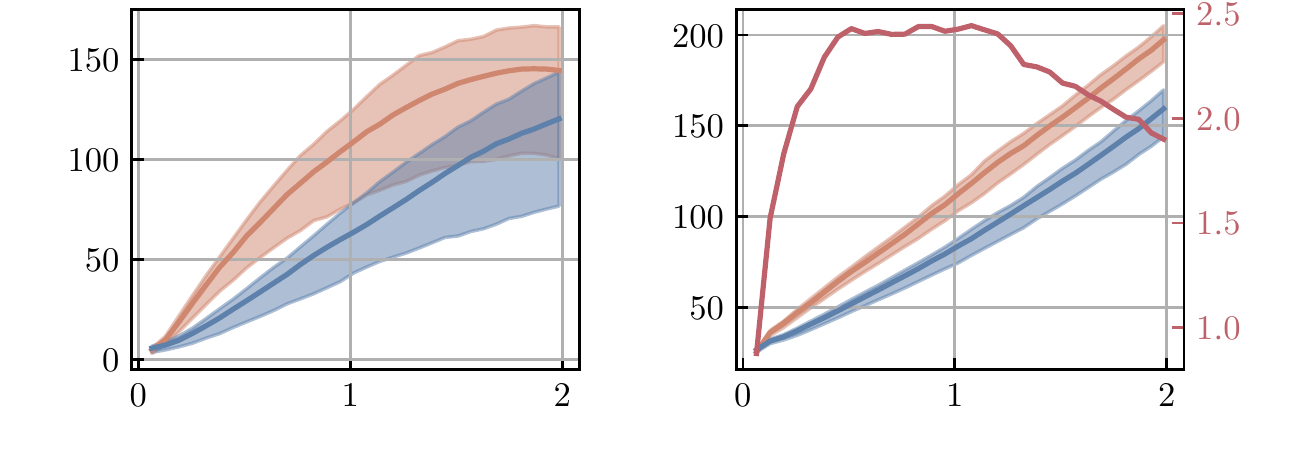}%
    \mylab{-4cm}{4pt}{$\tau$}%
    \mylab{-10cm}{4pt}{$\tau$}%
    \mylab{-13.25cm}{2.85cm}{${d/\eta}$}%
    \mylab{-7cm}{2.85cm}{${D/\eta}$}%
    \mylab{-0.8cm}{2.85cm}{\textcolor{nord3}{$\left(\dfrac{D_{\mkern-2mu\text{si}}}{D_\text{in}}\right)^3$}}%
    \mylab{-13.25cm}{4.5cm}{(a)}%
    \mylab{-7.25cm}{4.5cm}{(b)}%

    \caption{(a) Average distance of the centroid of the $q$-perturbations of $\rmDelta_2$ as a function of time. (b) Average growth in size of the perturbations in the same case. The red line and red ordinate are the cubed ratio of the significant size over the insignificant. (a, b) Blue lines are the significant perturbations and orange lines are the insignificant ones. The shaded contours contain 90\% of the probability mass.\label{fig:physper}}
\end{figure}

Finally, we explore the geometric aspects of the evolution of absolutely significant and insignificant $q$-perturbations. We first define the centroid $\vect{x}_{\pert}$ of the perturbation energy distribution,
\begin{equation}
    \vect{x}_{\pert}(\tau; \vect{x}_0) = \dfrac{\int_\Pi \vect{x}\left\|\psi_q\right\|^2(\vect{x}, t; \vect{x}_0) \dd \vect{x}}{\psi_q^{\prime2}(\tau; \vect{x}_0)}, \label{eq:dd}
\end{equation}
where $\Pi$ stands for the full domain, and
we construct the correlation tensor of $\psi_q$,
\begin{equation}
    \mathsf{D}_{i\!j}(\tau; \vect{x}_0) = \dfrac{\int_\Pi \big(x_i - x_{\pert i}\big)\big(x_{\!j} - x_{\pert j}\big)\left\|\psi_q\right\|^2(\vect{x}, t; \vect{x}_0)\dd\vect{x}}{\psi_q^{\prime2}(\tau; \vect{x}_0)},
\end{equation}
whose eigenvalues are $D_i^2$, and a characteristic dimension is $D = {\mkern6mu^{^3}\mkern-11mu}\sqrt{D_0D_1D_2}$.
This value can be interpreted as the radius of the equivalent sphere of constant energy density that contains the same perturbation energy.

Figure \ref{fig:physper}(a) shows the averaged evolution of the displacement of the centroid $d = \| {\vect x}_\pert - {\vect x}_0\|$ for $\rmDelta_2$. The initial value is about $10\eta$, one third of the radius of the perturbation. It initially grows for both sets of perturbations but is significantly faster for the significant. This could be expected as significant perturbations have higher local velocity magnitudes and are thus more likely to be immersed in a region of faster velocity. After the first turnover, the total displacement becomes comparable for both families of perturbations as the significant ones decrease their speed. At this point, the centroid has moved between three and four times its initial displacement, and the effect of the initial displacement is negligible.

The magnitude $D$ is shown in figure \ref{fig:physper}(b).  It's initial value is very close to $30\eta$ for both significant and insignificant perturbations, which is the radius of the Gaussian almost exactly. This validates using $D$ as a measure of the typical radius of the perturbation at other instants. Significant perturbations initially grow faster than insignificant ones, until $\tau\approx 1$, where the two growth rates become equal. The red line shows the ratio of the volume of the averaged significant perturbations over the insignificant ones $(D_{\mkern-2mu\text{si}}/D_\text{in})^3$. In the interval of maximum significance, $\tau\approx (0.5, 1.5)$, it stays approximately constant and approximately equal to the ratio of perturbation energies (see figure \ref{fig:pertu}(a)). This shows that the difference between significant and insignificant perturbations in the inertial range is not one of intensities, which are comparable, but of size. Significant perturbations spread faster until they are approximately $2.5$ times bigger than insignificant ones. During the transient, the growth rate of the perturbation sizes is almost linear with time, and much faster than a passive scalar would grown if seeded in the same place \citep{ric:26}. The faster growth rate implies that the production term of the perturbation or the pressure term (see appendix \ref{app:a}) must dominate their spatial growth. More details about the differences between a passive scalar, a passive vector which only has pressure and a perturbation can be found in \cite{tsi:01}.

\section{From significance to flow markers}\label{sec:regions}

In the previous section we have explored the properties of the perturbation field
and extracted significant and insignificant perturbations. In this section we explore which features of the flow field locations where the perturbations are introduced, are different in each set. 

First, we construct a library of coarse-grained averages over the regions where the perturbations are introduced, e.g,
\begin{equation}
    \langle a \rangle(\vect \xi; \rmDelta) = \int_\Pi g(\vect x - \vect \xi; \rmDelta) a(\vect x, t=0) \dd \vect x,
\end{equation}
where $a$ stands for any scalar, vector or tensor field (e. g. vorticity field). For scalar variables we compute the averaged magnitude $\langle a \rangle$ and the square $\langle a^2 \rangle$.
For vector and tensor fields, we define the magnitude as the $L_2$-norm of the averaged vector $|\langle \vect a \rangle|$, and the averaged square stands for the averaged squared magnitude $\langle |\vect a |^2\rangle$, and for the squared Frobenius norm of tensors. For simplicity we abuse our notation, omitting the magnitude operator from now on when refferencing the averages (e.g. $\su$ instead of $\langle |{\boldsymbol u}|\rangle$). 
Similarly, $\oso$ stands for $\langle{\vect\omega}_i\rangle\langle \mathsf{S}_{i\!j}\rangle\langle {\vect\omega}_{\!j}\rangle$, i.e. the product of the mean fields, etc. Precise definitions are gathered in table \ref{tab:class}. Some elements of the library are self-explanatory, but others require a brief description. 
The tensor $\mathsf{A}_{i\!j} = \mathsf{\Omega}_{i\!j}\mathsf{S}_{\!j\!k} + \mathsf{S}_{i\!j}\mathsf{\Omega}_{\!j\!k}$, is indicative of the presence of vortex sheets \citep{hor:tak:05}.
Since $\mathsf{A}$ is a second-order tensor, we compute the same coarse-grained quantities as for $\mathsf{S}$. We also track the invariants of the velocity gradient tensor \citep{cho:per:can:90}, $Q$ and $R$, both their mean and the mean of their squares. \change{They are defined through the second $\mathrm{I}_2()$ and third $\mathrm{I}_3()$ invariant operators respectively, applied to the velocity gradient tensor (see table \ref{tab:class})}.
The vortex stretching,
${\vect\omega}_i\mathsf{S}_{i\!j}{\vect\omega}_{\!j}$ and the rate-of-strain self-amplification $\mathsf{S}_{i\!j}\mathsf{S}_{\!j\!k}\mathsf{S}_{ki}$, are important quantities related to the evolution of the velocity gradients \citep{tsi:19}. Another important quantity is the source term of the perturbation energy at the initial condition, ${\vect u}_i\mathsf{S}_{i\!j}{\vect u}_{\!j}$ \citep[][see also appendix \ref{app:a}]{tsi:01}.
Similarly, $\tau_{i\!j}\mathsf{S}_{i\!j}$ is an indicator of the energy flux in the cascade \citep{ger:pio:moi:cab:91} across a given scale.  We use $g(\vect x; \rmDelta)$ as a filter for the computation of $\tau_{i\!j}$, with the same width as the perturbation. This leaves us with 32 classifiers in total.

\begin{table}
    \begin{center}
        \renewcommand{\arraystretch}{1.4}
        \scalebox{0.9}{\begin{tabular}{cccp{6.5cm}c} 
                $i$ & Symbol & Definition & \multicolumn{1}{c}{Meaning} & $\big(\mathcal{M}_q, \mathcal{\tilde M}_q\big)$\\
            \hline
            0 & $\su$ & $\left|\int\! gu_i\right|$ & Magnitude of the average velocity& (0.90, 0.66)\\ 
            1 & $\sud$ & $\int\! gu_iu_i$ & Kinetic energy& (0.98, 0.50)\\ 
            2 & $\so$ & $\left|\int\! g\omega_i\right|$ & Magnitude of the average vorticity& (0.61, 0.59)\\ 
            3 & $\sod$ & $\int\! g\omega_i\omega_i$ & Enstrophy& (0.74, 0.67)\\ 
            4 & $\sS$ & $\left|\int\! g\mathsf{S}_{i\!j}\right|$ & Magnitude of the average strain& (0.72, 0.76)\\
            5 & $\sSd$ & $\int\! g\mathsf{S}_{i\!j}\mathsf{S}_{i\!j}$ & Normalised dissipation& (0.75, 0.73)\\
            6 & $\ssp$ & $(\int\! g\mathsf{S}_{i\!j})_+$ & Most stretching eigenvalue of the average strain& (0.73, 0.76)\\ 
            7 & $\ssi$ & $(\int\! g\mathsf{S}_{i\!j})_m$ & Intermediate eigenvalue of the average strain& (0.63, 0.62)\\ 
            8 & $\ssm$ & $(\int\! g\mathsf{S}_{i\!j})_-$ & Most compressive eigenvalue of the average strain& (0.72, 0.74)\\ 
            9 & $\sspd$ & $\int\! g(\mathsf{S}_{i\!j})_+^2$ & Average squared magnitude of the most stretching eigenvalue of the strain& (0.75, 0.73)\\ 
            10 & $\ssid$ & $\int\! g(\mathsf{S}_{i\!j})_m^2$ & Intermediate eigenvalue of the average strain& (0.66, 0.55)\\ 
            11 & $\ssmd$ & $\int\! g(\mathsf{S}_{i\!j})_-^2$ & Most compressive eigenvalue of the average strain& (0.75, 0.72)\\ 
            12 & $\sA$ & $\left|\int\! g\mathsf{A}_{i\!j}\right|$ & Magnitude of the average vortex sheet tensor (VST)& (0.75, 0.70)\\
            13 & $\sAd$ & $\int\! g\mathsf{A}_{i\!j}\mathsf{A}_{i\!j}$ & Magnitude of the squared norm of the VST.& (0.75, 0.71)\\
            14 & $\sap$ & $(\int\! g\mathsf{A}_{i\!j})_+$ & Most stretching eigenvalue of the average VST& (0.74, 0.69)\\ 
            15 & $\sai$ & $(\int\! g\mathsf{A}_{i\!j})_m$ & Intermediate eigenvalue of the average VST& (0.66, 0.55)\\ 
            16 & $\sam$ & $(\int\! g\mathsf{A}_{i\!j})_-$ & Most compressive eigenvalue of the average VST& (0.74, 0.69)\\ 
            17 & $\sapd$ & $\int\! g(\mathsf{A}_{i\!j})_+^2$ & Average squared magnitude of the most stretching eigenvalue of the VST& (0.75, 0.71)\\ 
            18 & $\said$ & $\int\! g(\mathsf{A}_{i\!j})_m^2$ & Intermediate eigenvalue of the average VST& (0.64, 0.64)\\ 
            19 & $\samd$ & $\int\! g(\mathsf{A}_{i\!j})_-^2$ & Most compressive eigenvalue of the average VST& (0.75, 0.69)\\ 
            20 & $\Q$ & $\mathrm{I}_2\big(\partial_j \int\! gu_i\big)$  & Second invariant of the averaged velocity gradient tensor & (0.65, 0.6)\\ 
            21 & $\Qd$ & $\int\! g \mathrm{I}_2(\partial_ju_i)^2$ & Average squared magnitude of the second invariant of the velocity gradient tensor & (0.75, 0.69)\\ 
            22 & $\R$ & $\mathrm{I}_3\big(\partial_j \int\! gu_i\big)$  & Third invariant of the averaged velocity gradient tensor & (0.67, 0.57)\\ 
            23 & $\Rd$ & $\int\! g \mathrm{I}_3(\partial_ju_i)^2$ & Average squared magnitude of the third invariant of the velocity gradient tensor& (0.75, 0.69)\\ 
            24 & $\oso$ & $\int\! g\omega_i \int\! g\mathsf{S}_{i\!j} \int\! g\omega_j $ & Vortex stretching of the average magnitudes& (0.78, 0.74)\\ 
            25 & $\osod$ & $\int\! g(\omega_i\mathsf{S}_{i\!j}\omega_j)^2$ & Average squared magnitude of the vortex stretching & (0.75, 0.7)\\ 
            26 & $\sss$ & $\int\! g\mathsf{S}_{i\!j} \int\! g\mathsf{S}_{\!j\!k} \int\! g\mathsf{S}_{\!ki}$ & Strain self amplification of the average magnitudes& (0.78, 0.75)\\ 
            27 & $\sssd$ & $\int\! g(\mathsf{S}_{i\!j}\mathsf{S}_{\!j\!k}\mathsf{S}_{\!ki})^2$ & Average squared magnitude of the strain self amplification & (0.76, 0.71)\\ 
            28 & $\usu$ & $\int\! g u_i \int\! g\mathsf{S}_{i\!j} \int\! gu_j $ & Perturbation production of the averaged magnitudes & (0.83, 0.67)\\ 
            29 & $\usud$ & $\int\! g(u_i\mathsf{S}_{i\!j}u_j)^2$ & Average squared magnitude of the perturbation production& (0.97, 0.53)\\ 
            30 & $\tauij$ & $\int\! g\tau_{i\!j}\mathsf{S}_{i\!j}$ & Average subgrid energy transfer& (0.70, 0.60)\\ 
            31 & $\tauijd$ & $\int\! g(\tau_{i\!j}\mathsf{S}_{i\!j})^2$ & Average squared subgrid energy transfer& (0.73, 0.66)
    \end{tabular}}
    \caption{Library of classifiers. $\mathrm{I}_i$ is the operator that computes the $i$-th invariant of a tensor. $\mathcal{M}_q$ and $\mathcal{\tilde M}_q$ are the magnitude and amplification classification scores for $q$-perturbations of size $\rmDelta_2$. See the text for more information.}%
    \label{tab:class}
    \end{center}
\end{table}

\begin{figure}
    \centering\vspace{1em}
    \includegraphics[width=0.99\textwidth]{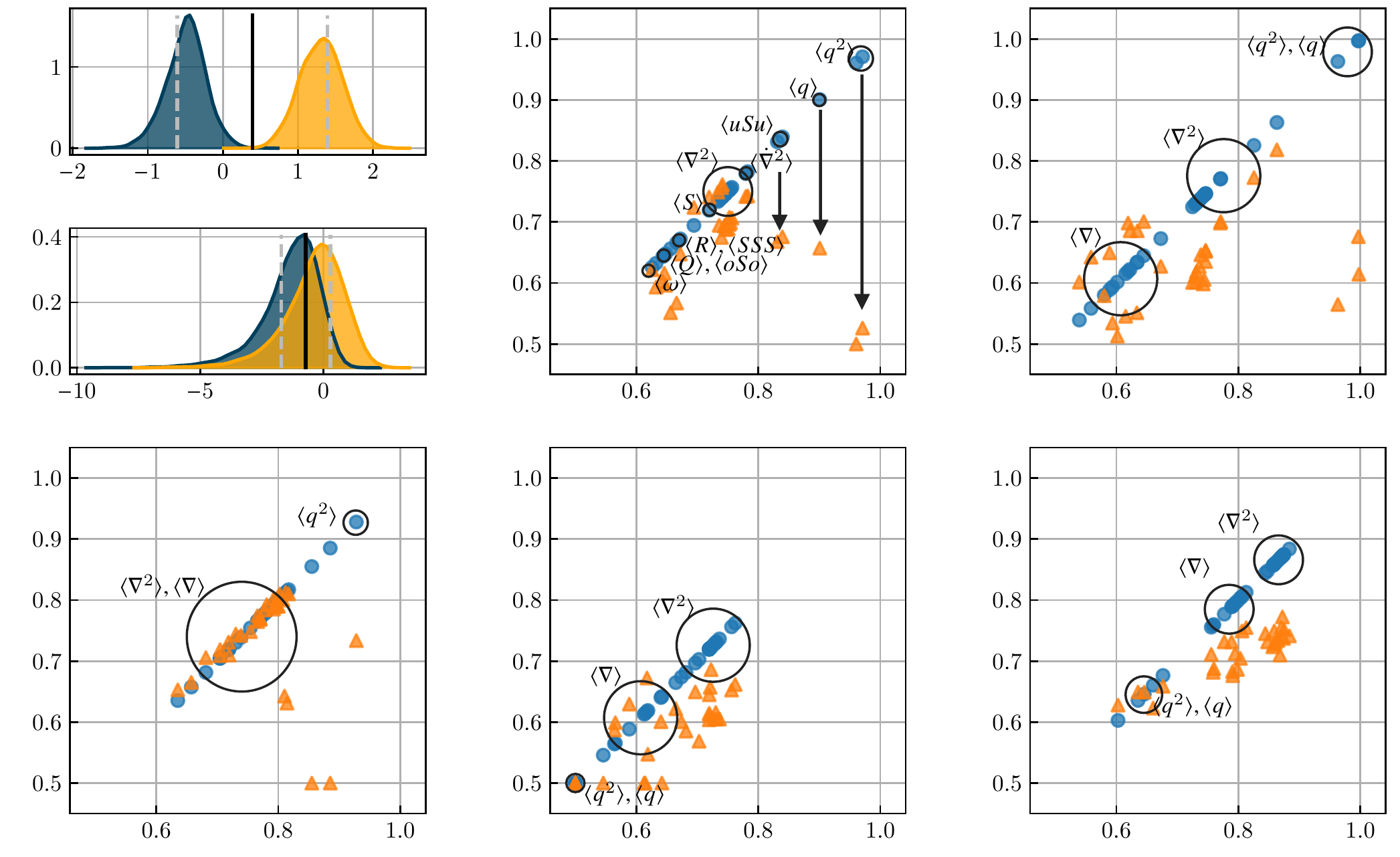}%
    \mylab{-2.25cm}{0pt}{$\mathcal{M}_q$}%
    \mylab{-6.75cm}{0pt}{$\mathcal{M}_q$}%
    \mylab{-11.5cm}{0pt}{$\mathcal{M}_q$}%
    \mylab{-4.4cm}{2.3cm}{$\tilde {\mathcal{M}}_q$}%
    \mylab{-9.0cm}{2.3cm}{$\tilde {\mathcal{M}}_q$}%
    \mylab{-13.6cm}{2.3cm}{$\tilde {\mathcal{M}}_q$}%
    \mylab{-2.25cm}{4.2cm}{$\mathcal{M}_q$}%
    \mylab{-6.75cm}{4.2cm}{$\mathcal{M}_q$}%
    \mylab{-4.4cm}{6.5cm}{$\tilde {\mathcal{M}}_q$}%
    \mylab{-9.0cm}{6.5cm}{$\tilde {\mathcal{M}}_q$}%
    \mylab{-11.0cm}{4.05cm}{${\log\so}$}%
    \mylab{-11cm}{6.2cm}{${\log\sud}$}%
    \mylab{-13.55cm}{7.25cm}{{pdf}}%
    \mylab{-13.55cm}{5.0cm}{{pdf}}%
    \mylab{-13.7cm}{7.9cm}{(a)}%
    \mylab{-13.7cm}{5.75cm}{(b)}%
    \mylab{-9.15cm}{7.9cm}{(c)}%
    \mylab{-4.5cm}{7.9cm}{(d)}%
    \mylab{-13.7cm}{3.85cm}{(e)}%
    \mylab{-9.15cm}{3.85cm}{(f)}%
    \mylab{-4.5cm}{3.85cm}{(g)}%
    
    \caption{Classification of significant/insignificant regions. (a) Example of a good classifier ($\log\langle{q^2}\rangle$ for $q$-perturbations at size $\rmDelta_3$) and the norm $\mathcal{R}_q$. (b) Example of a bad classifier for the same case ($\log\so$). (c-g) Classification score for different perturbations and significance criteria. The abscissa is the score associated to $\mathcal{R}_q$. Colours are the classifier of the ordinate, (blue) $\mathcal{R}_q$, and (orange) $\mathcal{\tilde R}_q$. (c) $q$-perturbations at $\rmDelta_2$. (d) Classification of $q$-perturbations at $\rmDelta_3$. (e) Classification of $q$-perturbations at $\rmDelta_0$. (f) Classification of $\omega$-perturbations at $\rmDelta_2$. (g) Classification of $\omega$-perturbations at $\rmDelta_0$.\label{fig:class1d}}
\end{figure}

To generate a figure of merit of which elements of the library are good markers of significance, we use linear support vector machines \citep[SVM,][]{bos:guy:vap:92}, which are generalisations of one-dimensional thresholds. 
A thresholds separates two variables if their probability mass does not overlap. SVMs assign a score $\mathcal{M}$ of $1.0$ if the separation is complete and a score of $0.5$ if no separation is achieved, and thus the best guess is a random one. 
This is illustrated in figure \ref{fig:class1d}(a), where the conditional probability density functions (pdfs) for the significant and insignificant sets are shown for $(\log{\sud})$, and a black line shows the optimum threshold. In contrast to figure \ref{fig:class1d}(a), which shows an example of a good classifier with score $\mathcal{M} \approx 0.97$, figure \ref{fig:class1d}(b) is an example of a bad one $\mathcal{M} \approx 0.605$, in which the value of $(\log\so)$ marginally separates the pdfs. Thus, knowledge of $\so$ marginally improves the odds of choosing an element from a desired set of the two, while knowledge of $\sud$ almost guarantees that an element is assigned to the desired set. As seen in figure \ref{fig:class1d}(b), no classifier with score below $\mathcal{M} \approx 0.7$ should be taken too seriously. For example, two sets represented by identical normal distributions with their means separated by one standard deviation have an approximate score of $0.7$; and two, three and four standard-deviation separations give scores of $0.84$, $0.93$ and $0.98$ respectively. As anticipated by the example in figure \ref{fig:class1d}(a, b), we use the natural logarithm of the variables as a classifier, because we found that it makes the significant and insignificant pdfs similar and close to normals in most cases. Because some variables are not positive definite, we use the logarithm of the absolute value, and we also use a linear classifier which preserves the information about the sign. We only keep the best score of the two classifiers in those cases.

The remainder of figure \ref{fig:class1d} is devoted the classification of different perturbations using one variable at a time.
Figure \ref{fig:class1d}(c) shows the scores of the classification of $q$-perturbations for $\rmDelta_2$. Each marker represents the score of one classifier. The horizontal axis shows the score for the significance defined using the absolute $q$-norm, $\mathcal{R}_q$, whereas the vertical one shows the absolute $q$-norm for the blue dots, and the relative $q$-norm for the orange triangles. The dots do not add information but help to compare both norms visually. Several aspects of the figure stand out, but perhaps the most immediate one is that scores tend to be clustered. Within these clusters, classifiers have negligible differences in their classification scores, so they can be considered equivalent to each other.

For the absolute $q$-norm, $\sud$ is the best classifier, with a score of almost 1. The magnitude of the perturbation source term, $\usud$ is equally good, but the reason is that it is a quadratic form similar to $\sud$. This is tested using the classifier $\usud/\sud$, which measures the initial alignment of the perturbation with the rate-of-strain, and has units of strain. This classifier does not contain the initial energy, only its amplification, and it is a very poor classifier $(\mathcal{M}_q\approx 0.55)$. The mean values of both quantities, $\su$ and $\usu$ follow the score of $\sud$. They are indicators of the kinetic energy at scales of the order of the region size, and the amplification of the perturbation at the same scale, respectively.  At a score of $\mathcal{M}_q \approx 0.75$, we have a cluster of classifiers labelled in figure \ref{fig:class1d}(c) as `$\langle\nabla^2\rangle$', representing the following square gradients and related quantities: $\sod$, $\sSd$, $\sAd$, $\Qd$, $\Rd$, $\langle a_\pm ^2 \rangle$, $\langle s_\pm ^2\rangle$. All of them contain information associated to strong vortices or strong vortex sheets and their associated rate-of-strain. 

Two classifiers sit at $\mathcal{M}_q \approx 0.78$; $\sssd$ and $\osod$, grouped together as $\langle\dot\nabla^2\rangle$. These are the total magnitude of the production terms of the velocity gradients.
Because they are slightly better classifiers than the gradient magnitudes, it could be argued that strong gradient production is related to significance. For this reason, we also tested the classifiers $\sssd/\sSd^3$ and $\osod/(\sSd\sod^2)$, which contain little information about the magnitude of the gradients, and focus on gradient amplification. They can be thought of the average rate-of-strain self alignment and vorticity-strain alignment. Their classification scores are low $\approx 0.6$, which shows that the relevancy of the production terms as classifiers is mostly tied to the intensity of the gradients, and only secondarily to their amplification. A similar argument can be made for $\sAd$, which contains information about vortex sheets but also about the gradients magnitude. The classifier $\sAd/(\sSd\sod)$ only contains information about the vorticity-strain structure but not about its magnitude. It completely fails, with $\mathcal{M}_q \approx 0.5$. The same is true for the squared invariants of the velocity gradient tensor, as the classifiers $\Qd/\sSd^2$ and $\Rd/\sSd^3$ give a score of $\mathcal{M}_q\approx 0.55$.

The ranking of classifiers continues with the magnitude of the mean rate-of strain, $\sS$, scoring $\mathcal{M}_q \approx 0.72$. Aside from $\su$, it is the only mean magnitude that provides some classification. The remaining one, $\so$, is a poor classifier $\mathcal{M}_q \approx 0.605$. 
Although unlabelled, the next classifier is $\tauij$ with $\mathcal{M}_q \approx 0.73$. If the influence of $\mathcal{S}_{i\!j}^2$ is removed from the classifier, the score drops to the level of a random guess $\mathcal{M}_q \approx 0.5$. This suggests that the cascading process is irrelevant for classification, which is associated instead to the local rate of strain. All other quantities in table \ref{tab:class} fail to classify significance.

We can now compare these numbers with the classification of the amplification of kinetic energy $\tilde q$. The classification score of gradients is comparable to the absolute norm, either slightly better or slightly worse, but the score of the kinetic energy markers plummets, making them useless as classifiers. This is shown explicitly in the plot by the vertical black arrows. 

Figures \ref{fig:class1d}(d, e) are equivalent to figure \ref{fig:class1d}(c) for a larger ($\rmDelta_3$) and a smaller ($\rmDelta_0$) size, respectively. The general structure of the of the classifiers is the same, although some differences exist. First, classifiers that rely on kinetic energy are better for larger perturbation sizes and classifiers relying on gradients are better for smaller ones. This is consistent with common turbulence knowledge that gradients are associated to small scales and kinetic energy to larger ones, but arises here through the introduction of perturbations. Second, the amplification of kinetic energy (and enstrophy) is classified better the smaller the perturbation. Regardless of the size, the kinetic energy is a bad classifier of the amplification and the ability of the gradients to classify it improves for small perturbations. This is  separate from the general improvement of gradients as classifiers for the absolute norm. For $\rmDelta_2$ the scores of the gradients for amplifications and absolute norms are almost equal, whereas the former are lower than the latter in $\rmDelta_3$, and the converse applies to $\rmDelta_0$.
The most likely explanation of this behaviour comes from the effect of the initial perturbation. In larger perturbations, the magnitude of the initial perturbation is more relevant, and estimating their norm after a given time correlates well with estimating their initial norm. This is different for small perturbations, where the initial norm is not as important, as most of their magnitude at the time of classification comes from the amplification. 

Finally, \ref{fig:class1d}(f) and \ref{fig:class1d}(g) show the classification scores for $\omega$-perturbations for $\rmDelta_2$ and $\rmDelta_0$ respectively.
The main difference between these perturbations and those of kinetic energy is that kinetic energy markers are never good classifiers. They are unable to classify significance for any norm, $\mathcal{M} \in (0.5, 0.6)$, and their quality lowers for smaller perturbations. In contrast to that, the markers of intense gradients have slightly better scores than for $q$-perturbations of the same size. The scores of the amplification of the kinetic energy are very similar in $\omega$-perturbations and in $q$-perturbations, and improve for smaller perturbations. 

The overall picture is that (i) classifiers related to the kinetic energy are best for the absolute norm, (ii) classifiers that are related to the squared gradients or squared production of gradients are reasonable for all norms and (iii) complex classifiers add little value over simpler ones. Averages act as a filter when they operate before squaring the magnitude, but as accumulators when they operate after squaring. That square gradients are better classifiers than averages suggests that significance is related to the scales contained in the cell, i.e. of the cell size and smaller, rather than exclusively on scales of the order of the cell. This also applies to the kinetic energy, where squared magnitudes are almost perfect classifiers, and averages are not. 
Despite having worse scores, scales at the order of the perturbation are not irrelevant, as classifiers such as $\sS$ and $\su$ are able to provide some classification.

\begin{figure}
    \centering\vspace{1em}
    \includegraphics[width=0.99\textwidth]{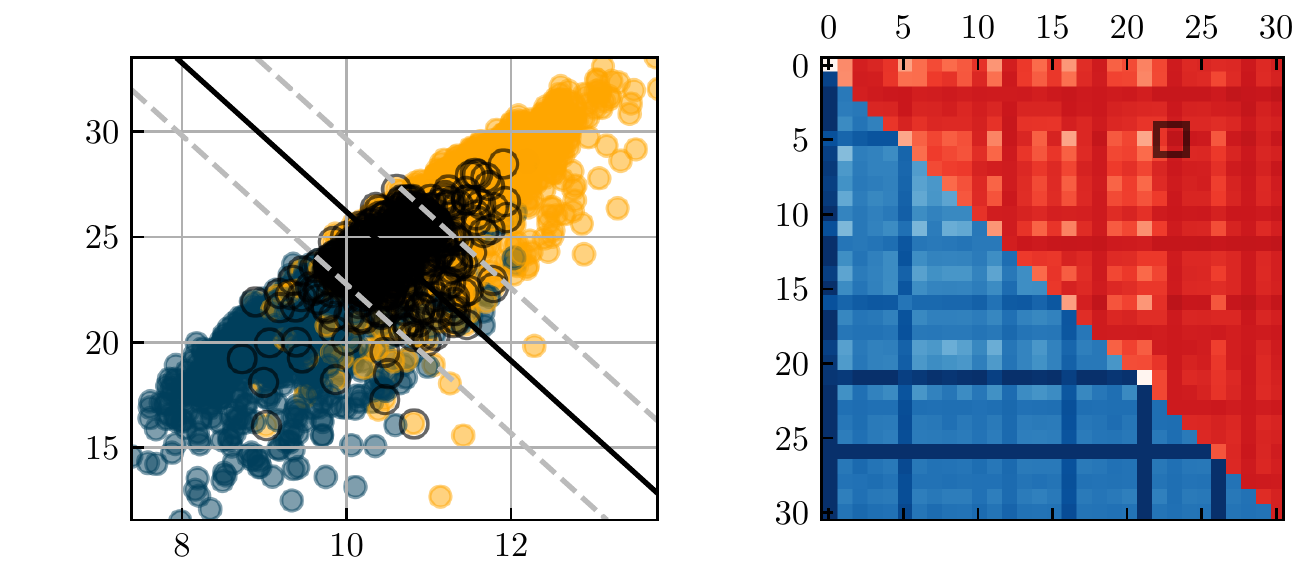}%
    \mylab{-9.25cm}{0.0cm}{${\log(a)}$}%
    \mylab{-13.15cm}{2.75cm}{${\log(b)}$}%
    \mylab{-5.75cm}{3.0cm}{$i$}%
    \mylab{-3.0cm}{5.75cm}{$i$}%
    \mylab{-12.5cm}{5.5cm}{(a)}%
    \mylab{-5.75cm}{5.5cm}{(b)}%
    \caption{Classification of significant/insignificant regions using two classifiers simultaneously. (a) Example of a case where two classifiers contribute to a better classification. (b) Classification score for pairs of classifiers from (light - 0.5) to (dark - 1.0) for $q$-perturbations at size $\rmDelta_3$. (Blue) $\mathcal{R}_q$ and (red) $\mathcal{\tilde R}_q$. The dark bracketed square corresponds to (a). \label{fig:class2d}}
\end{figure}

Figure \ref{fig:class2d} explores how much the classification improves using more than one classifier at the same time. For multiple classifiers, the support vector machine generalises the threshold
to an hyperplane separating both sets. An example of the procedure for two
classifiers is shown in figure \ref{fig:class2d}(a). A useful feature of the log-log transformation in two dimensions is that hyperplanes correspond to products of powers of the
classifiers, e.g. $b^na^m= c$ for two classifiers $(a, b)$, which may hint 
relevant groups.
For example, a notable combination is $(\su, \sS)$, which works well for the classification of $\mathcal{R}_q$ for $q$-perturbations at $\rmDelta_2$. The resulting hyperplane in log-log space approximately suggests the group $\su^2\sS$, which has the same units than the source term of the perturbations $\usu$ and is an excellent classifier. Figure 
\ref{fig:class2d}(b) shows a map with the scores for every pair of predictors in the library both for the absolute norm and the amplification of kinetic energy. Whenever an off-diagonal point is dark (high score), one of the two predictors has essentially the same score by itself. The result is that dark lines `radiate' vertically and horizontally from the diagonal, indicating that only marginal gain is achieved by using multiple predictors at the same time. Most non-dimensional groups found are worse versions of better classifiers already contained in the library. \change{It should be noted that we `discovered' $\usu$ through this procedure. We realised that it was the production term at the initial condition, and it was added to the library \emph{a posteriori}, based on this suggestion.} Similar conclusions apply to using three classifiers at the same time, where the gains in classification are negligible (not shown).

\begin{figure}
    \centering\vspace{1em}
    \includegraphics[width=0.99\textwidth]{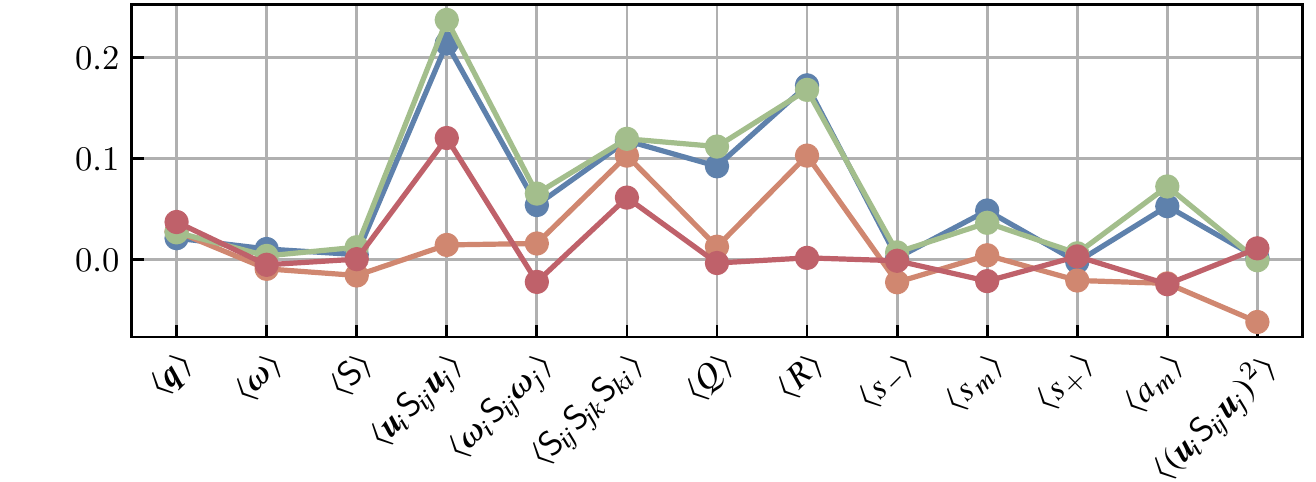}%
    \mylab{-13.25cm}{1.75cm}{\rotatebox{90}{$\mathcal{M}_{p\rightarrow f} - \mathcal{M}_{f\rightarrow p} $}}%
    \caption{Difference between the prediction and postdiction score of the different classifiers for $q$-perturbations of size $\rmDelta_2$. Blue is $\mathcal{R}^5_q$, orange is $\mathcal{\tilde R}^5_q$, green is $\mathcal{R}_\omega$ and red is $\mathcal{\tilde R}_{\omega}$. Each norm is taken at the time of maximum significance. Only classifiers with differences larger than $|\Delta\mathcal{M}| > 0.03$ are shown. \label{fig:cross}}
\end{figure}

Finally, figure \ref{fig:cross} shows the difference between the `prediction' classification scores shown in figure \ref{fig:class1d}(c), and the `postdiction' scores, which evaluate how easy is to separate strong from weak values of a classifier with knowledge of the significance. Thus, they are a measure of how the past state of the flow can be estimated with knowledge of the future significances. Although in a strict sense causality does not require postdiction, it is useful to know which causes are necessary ones, in the sense that they can be postdicted, and which ones are not. The general behaviour of the classifiers is that those based on squared magnitudes can be postdicted while the ones based on averages cannot. Most average classifiers whose postdiction does not worsen have scores close to $0.5$ (e.g $\Q$ for norms based on the gradients). The solely exception is $\sS$ which is a reasonable predictor and postdictor.
These results add to the evidence that the relevant structures are small scales contained within the perturbation and the mean rate of strain, while other classifiers are less important.

\section{From flow markers to structures}\label{sec:structures}

The analysis above shows that the local intensity of gradients is the most reliable predictor of the amplification of perturbations and their absolute magnitude.
 However, the predictions are not perfect, and it is likely that other properties characterise these regions.
 In this section, we extract the significant and insignificant regions from the flow and study their structure.
 We work with the significant and insignificant sets of $\rmDelta_2$, using $\mathcal{\tilde R}_q$ for $q$-perturbations, which are representative of the inertial range.
 Our findings indicate that other perturbation sizes are less informative, and the absolute norms of these perturbations are not particularly noteworthy. In contrast, the relative norms are more relevant, as the absolute norms can be largely explained by the magnitude of the perturbation introduced to the flow, as observed in the larger cases such as $\rmDelta_4$.

\begin{figure}
    \centering
    \includegraphics[width=.49\textwidth]{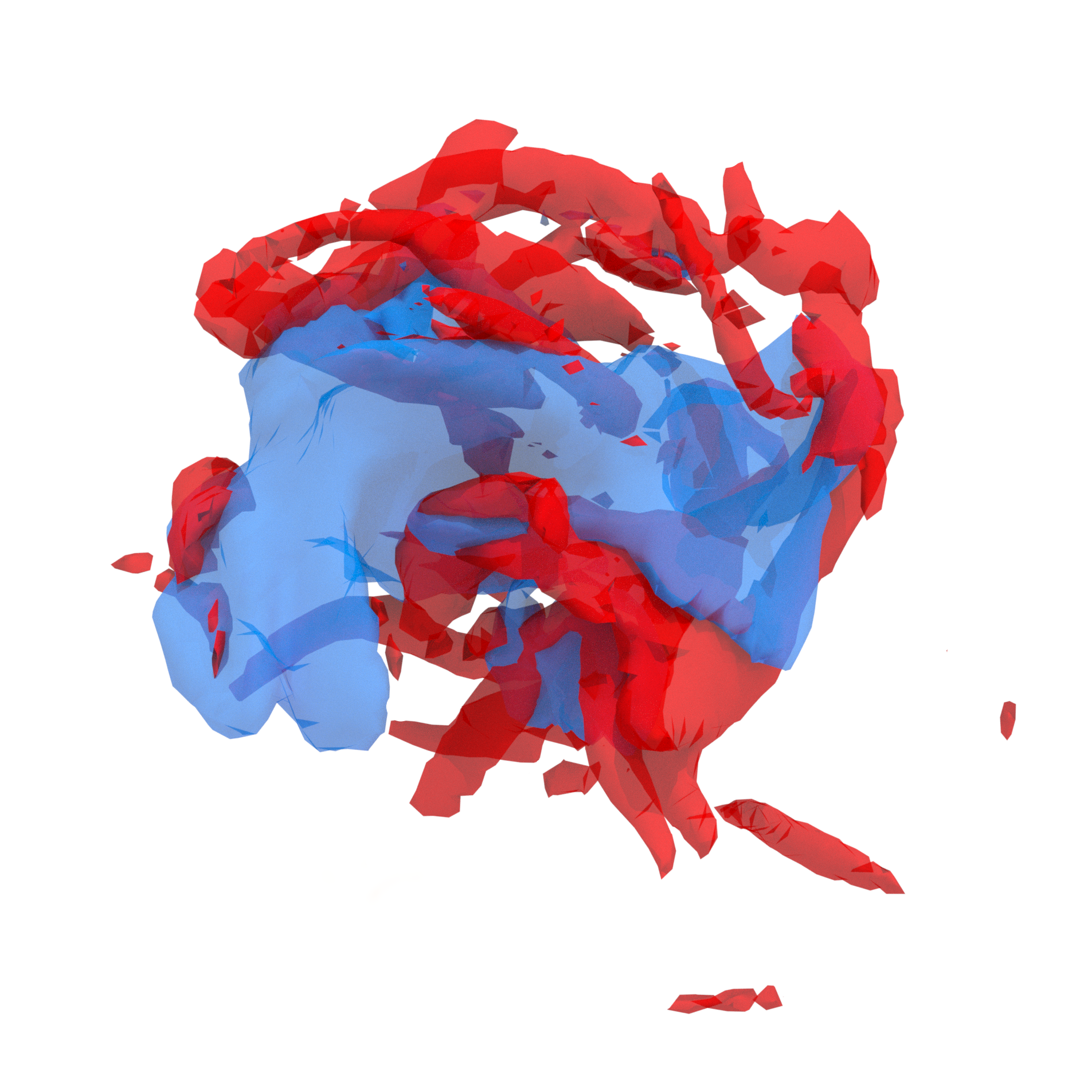}
    \includegraphics[width=.49\textwidth]{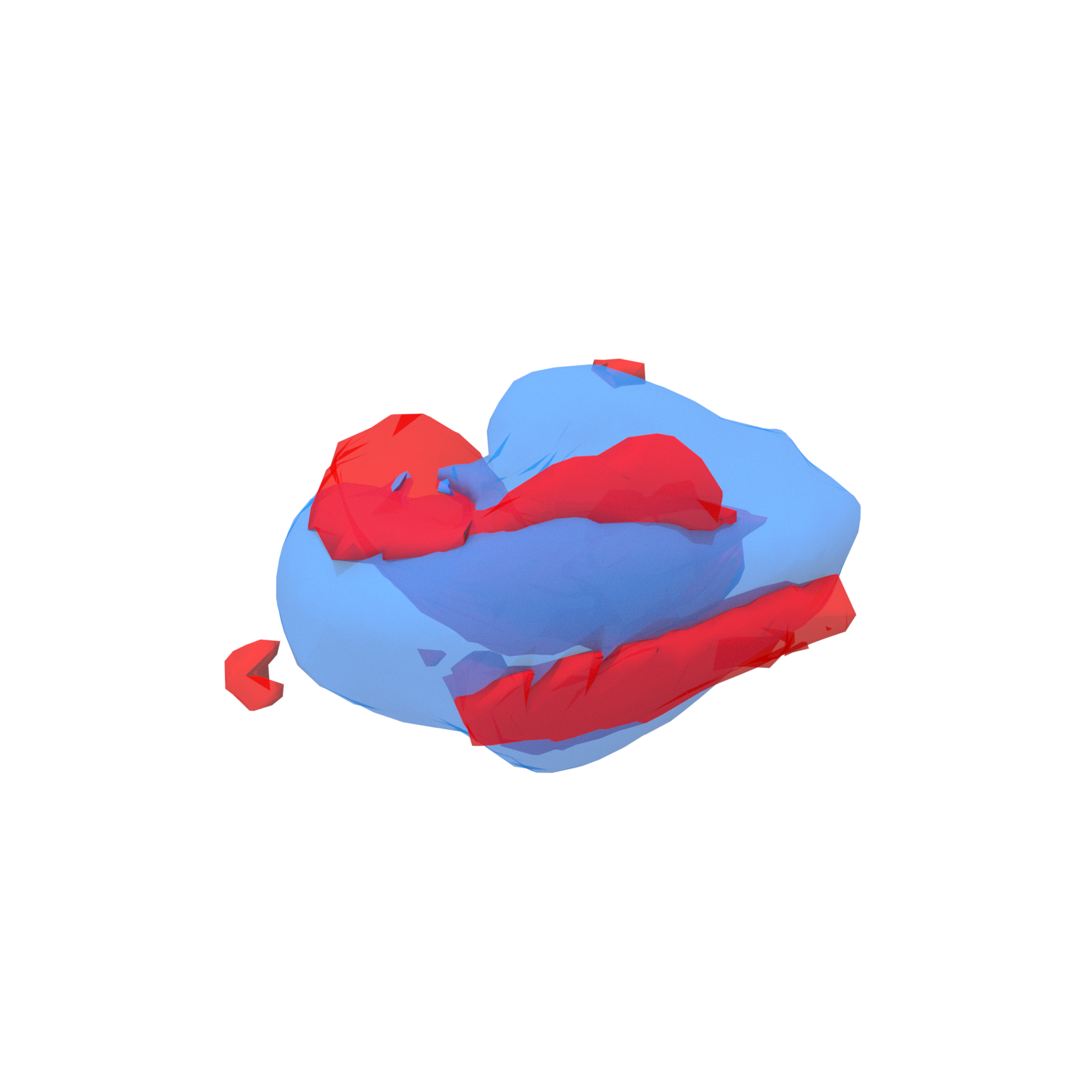}%
    \mylab{-13cm}{6cm}{(a)}%
    \mylab{-6cm}{6cm}{(b)}%

    \includegraphics[width=.49\textwidth]{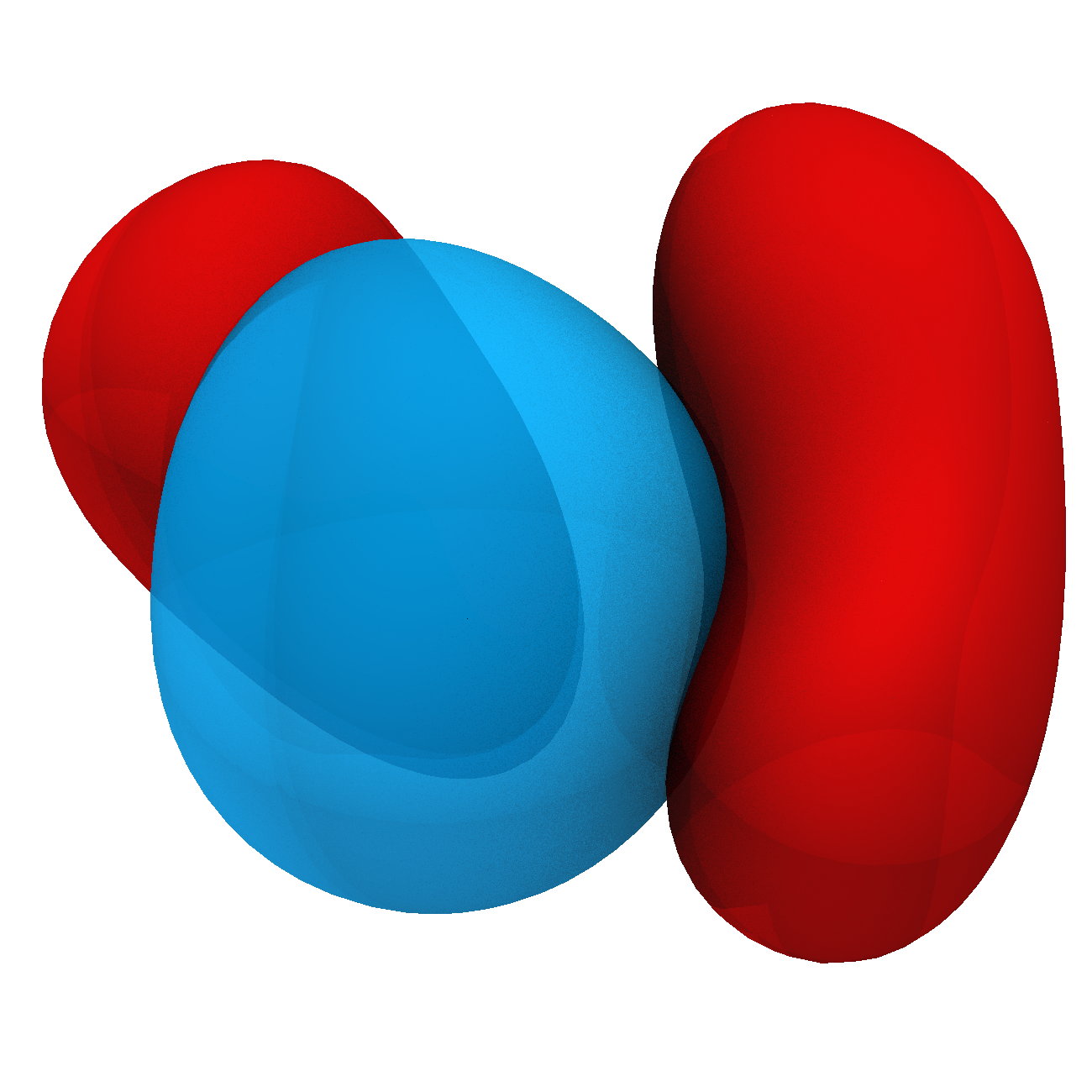}
    \includegraphics[width=.49\textwidth]{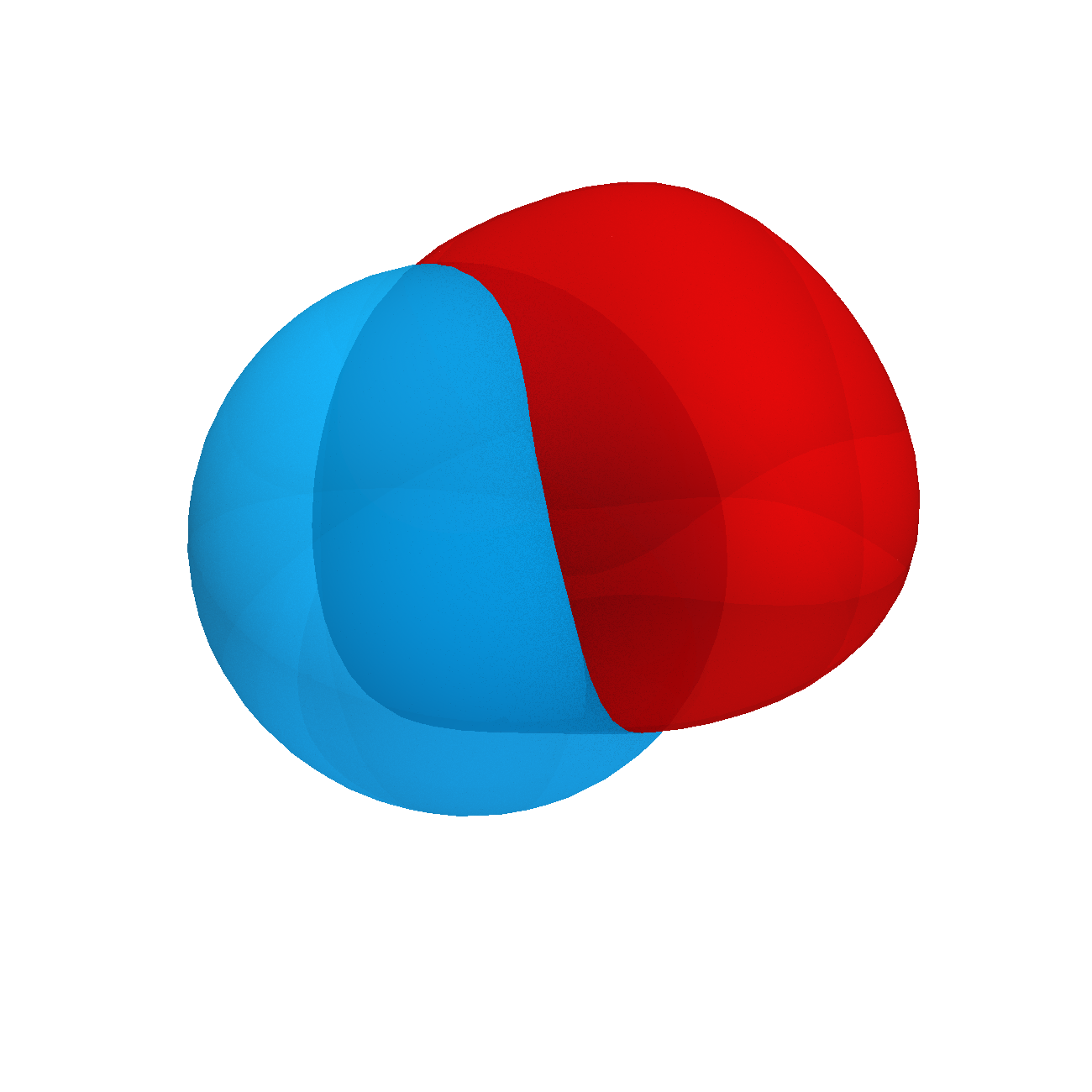}%
    \mylab{-13cm}{6cm}{(c)}%
    \mylab{-6cm}{6cm}{(d)}%
    \caption{Example of a significant (a, c) and insignificant (b, d) region for $\mathcal{R}_q$ and $\mathcal{\tilde R}_{q}$ for $\rmDelta_2$. Red isocontours are enstrophy and blue are kinetic energy. Both isocontour levels are chosen using a percolation analysis.\label{fig:3dexample}}
\end{figure}

Figure \ref{fig:3dexample} shows an example of a significant (both absolutely and relatively) region of $\rmDelta_2$, and of an insignificant region. The structure is extracted by multiplying the flow field by a wider window $g(\vect{x}-\vect{\xi}; 3\rmDelta_2)$, capturing both the perturbed region and its surroundings. We show isosurfaces of the enstrophy (red) and the turbulent kinetic energy (blue). The latter is made Galilean invariant by subtracting from the velocity field the mean velocity of the region. For the remainder of the section we refer to this quantity as `kinetic energy'. We apply a percolation analysis \citep{moi:jim:04} independently to the significant and insignificant sets to obtain adequate values for the thresholds for each quantity. If a constant threshold was used for both sets, most members of the insignificant set would be completely below the threshold.
Both structures involve an almost centred kinetic energy object surrounded by enstrophy structures. The remainder of this section studies both sets of perturbations by computing several statistics about them.

Perhaps the simplest statistics are two conditional averages, one representative of significant structures and another one representative of insignificant ones. Obtaining a conditional average of an isotropic flow requires choosing consistent reference axes for each region.

For example, aligning the mean velocity vector, we obtain a conditional average reminiscent of a velocity jet for the significant structures, but retain little information of the velocity gradients.
\change{On average, this average is able to represent 80\% of the energy of the significant regions, but only 0.02\% of their enstrophy.}
The information this provides is marginal, as the fact that strong kinetic energy is present in significant regions was already known. Changing the rotation criteria to align the most intense vorticity (mean vorticity is a bad marker of significance), the rate-of-strain eigenframe, or the covariance of the gradient distribution results in an average projection of the vorticity that is never better than 10\%. The conclusion is that, as hinted in figure \ref{fig:3dexample}(a), the local structure of the gradients is complex enough that a conditional average is never representative of the individual elements.

Alternatively, it is possible to obtain a local average representative of the filtered velocity gradients, which are simpler objects. Figures \ref{fig:3dexample}(c, d) show examples of significant and insignificant regions for $\rmDelta_2$, extracted from the filtered velocity field. We use the filter $g(\boldsymbol{x};\rmDelta_2)$ for the velocity field, which avoids the introduction of another length scale to the flow. The filtered fields are much simpler and typically contain one large region of intense kinetic energy, one to three connected regions of strong enstrophy, and one to two regions of strong rate-of strain. The procedure we found to produce the best results for the filtered case is similar to the one used by \cite{els:mar:2010}, which relies on orienting the flow to the local eigenframe of the rate-of-strain. In our case, the rate-of-strain is a local average. While the rate-of-strain eigenframe sets the angles of the coordinates, the directions are undetermined, and can be manipulated through a reflection. The algorithm we used to compute the conditional flow goes as follows,
\begin{algorithm}{CF}
    \item Take the filtered flow field centred on $\boldsymbol{x}_0$, for a significant/insignificant region. In the following, the `fields' stands for the fields of velocity, vorticity, and rate of strain.
    \item Multiply the fields by $g(\boldsymbol{x}; 3\rmDelta)$, isolating the neighbourhood of the region
    \item Compute the average rate-of-strain of the isolated neighbourhood and rotate the fields. The fields are now aligned with the rate-of-strain eigenframe.
    \item Segment the rotated enstrophy field to identify the strongest enstrophy structure in the neighbourhood. Compute the average vorticity vector of said structure.\label{CF:segment}
    \item Separate the conditional average into three classes, depending on which component of the vorticity vector has the largest magnitude. Note that the magnitude of each component is a measure of alignment with each eigenvector of the rate of strain.\label{CF:branch}
    \item Use the three free reflections to ensure that (i) the strongest component of the vorticity vector is positive and (ii) the centre of mass of the vortex is on the positive semi-space normal to its strongest vorticity component. \label{CF:voral}
\end{algorithm}
The procedure departs from \cite{els:mar:2010} at (CF\ref{CF:branch}), as these authors did not consider different conditionals for different alignments between the vorticity and the strain. Other differences between our conditional and theirs have to come either from CF\ref{CF:voral}, which is a different orientation criterion, or from the condition to significance or insignificance. It should be noted that the flow field belongs to the $\mathrm{O}(3)$ symmetry group, and thus the same result can be achieved with just one reflection. However, this requires computing CF\ref{CF:segment} and CF\ref{CF:branch} before rotating the field, as the directions of the axis need to be known beforehand. Instead, reflections are computationally cheap once the field is aligned with the Cartesian grid, and the algorithm is more efficient when the reflections are performed at the last step.

\begin{table}
    \begin{center}\setlength{\tabcolsep}{6pt}\renewcommand{\arraystretch}{1.1}
        \begin{tabular}{ccccccccccccc}
            $\rmDelta$ & Kind & Total & \multicolumn{3}{c}{Compressive} & \multicolumn{3}{c}{Intermediate} & \multicolumn{3}{c}{Stretching}\\[-5pt]
            \hline\\[-14pt]
                       &&& \% & $q^i$ & $\omega^i$ & \% & $q^i$ & $\omega^i$ & \% & $q^i$ & $\omega^i$\\
                       \cline{3-12}\\[-4pt]
            \multirow{3}{*}{$\rmDelta_2$} &Significant & \multirow{3}{*}{3135} & 3.4 & 0.47 & 0.05 & 31 & 4.57 & 0.43 & 66 & 9.12 & 0.75\\
                                          &Insignificant & & 12.2 & 0.33 & 0.11 & 30.7 & 0.83 & 0.28 & 53.6 & 1.30 & 0.37\\
                                          &Random & & 7.7 & 0.39 & 0.08 & 31.7 & 1.59 & 0.34 & 60.6 & 2.96 & 0.53\\[4pt]
            \multirow{3}{*}{$\rmDelta_4$} & Significant & \multirow{3}{*}{1966} & 10.99 & 0.58 & 0.10 & 44.66 & 2.75 & 0.48 & 44.35 & 2.73 & 0.46\\
                                          & Insignificant & & 15.16 & 0.27 & 0.08 & 37.95 & 0.81 & 0.24 & 46.90 & 0.93 & 0.29\\
                                          & Random & & 13.78 & 0.42 & 0.10 & 42.37 & 1.48 & 0.36 & 43.85 & 1.49 & 0.35\\
        \end{tabular}
    \end{center}
    \caption{Information about the conditional structures of the amplification of kinetic energy for $\rmDelta_2$. `Total' is the number of events found for each category. `\%' is the percentage of the total structures that fall in each of the three categories. $q^i$ and $\omega^i$ are measures of the total energy and enstrophy of the conditional average (see the body of the manuscript for more details).\label{tab:3dconds}}
\end{table}

\begin{figure}
    \centering
    \includegraphics[width=.40\textwidth]{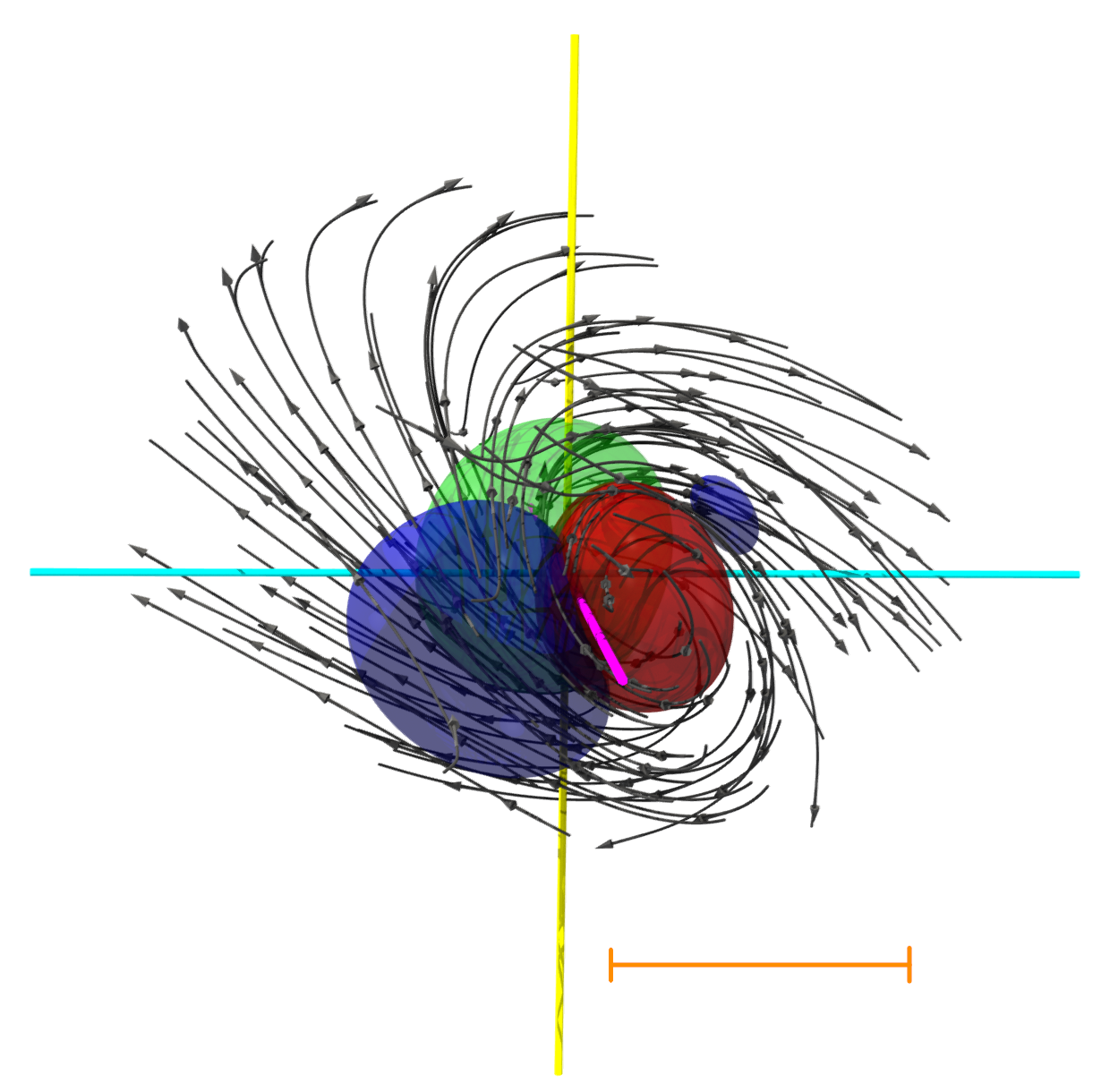}%
    \hspace{20pt}%
    \includegraphics[width=.40\textwidth]{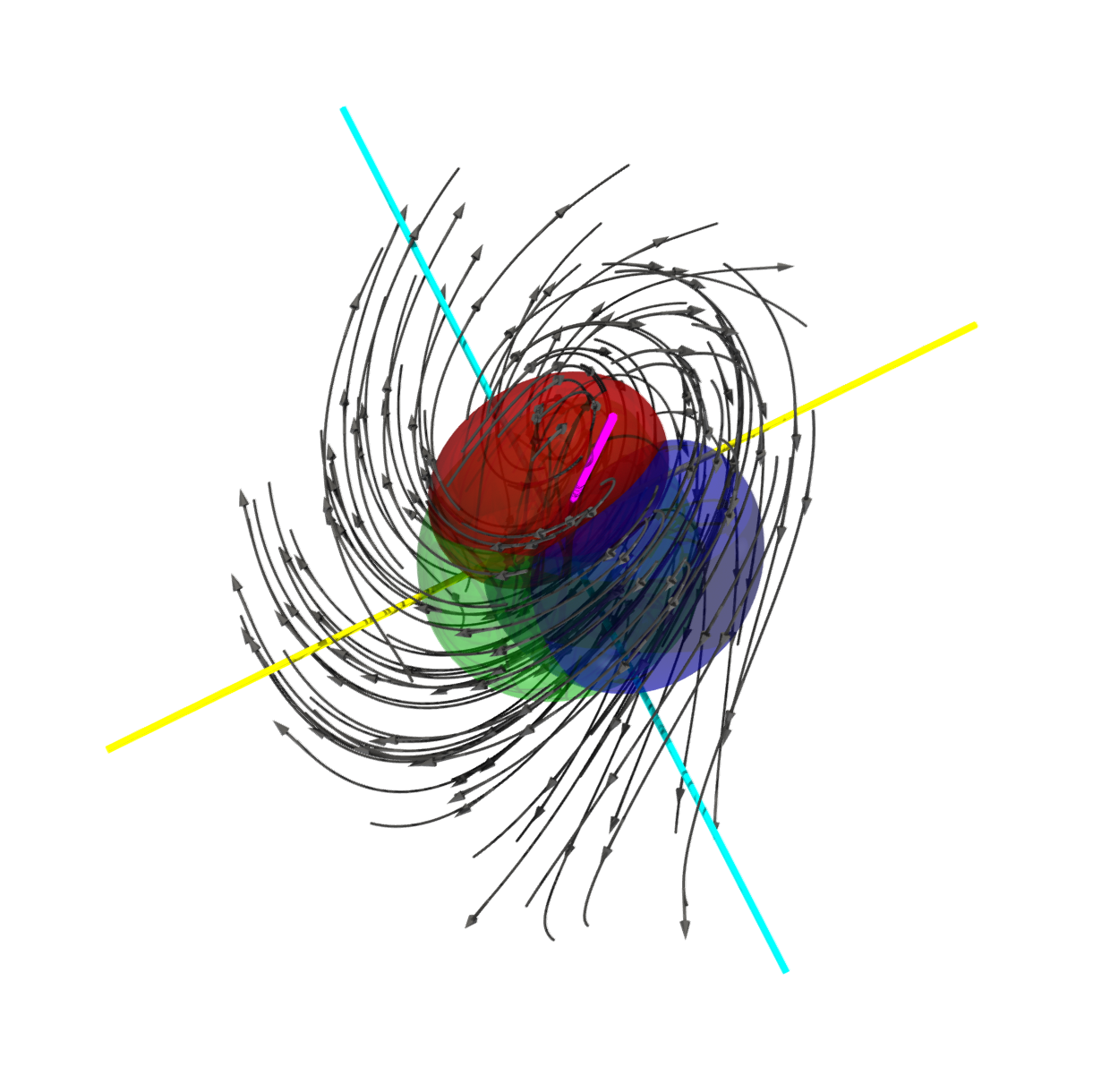}%
    \mylab{-11cm}{4.75cm}{(a)}%
    \mylab{-5cm}{4.75cm}{(b)}

    \includegraphics[width=.40\textwidth]{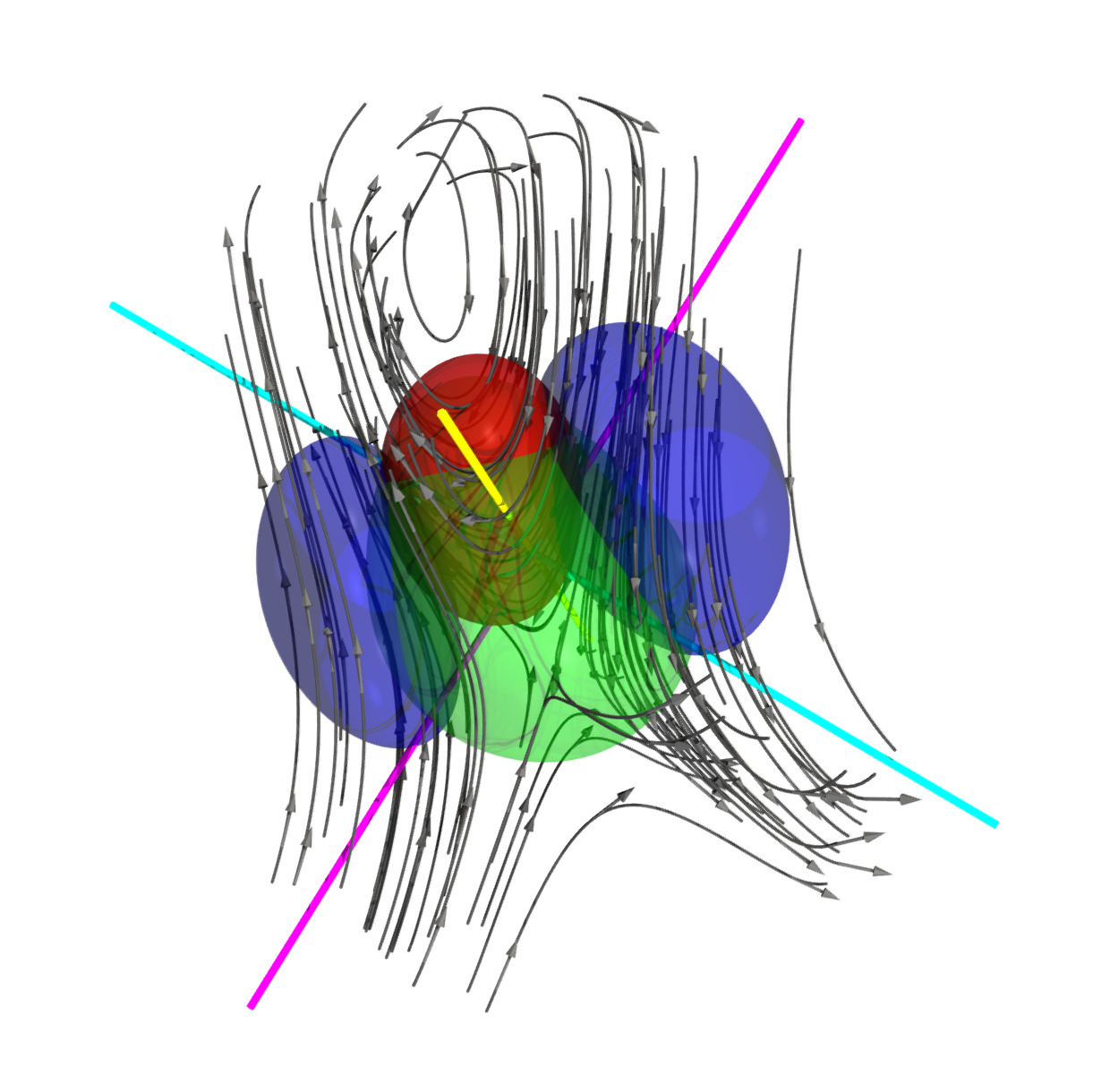}%
    \hspace{20pt}%
    \includegraphics[width=.40\textwidth]{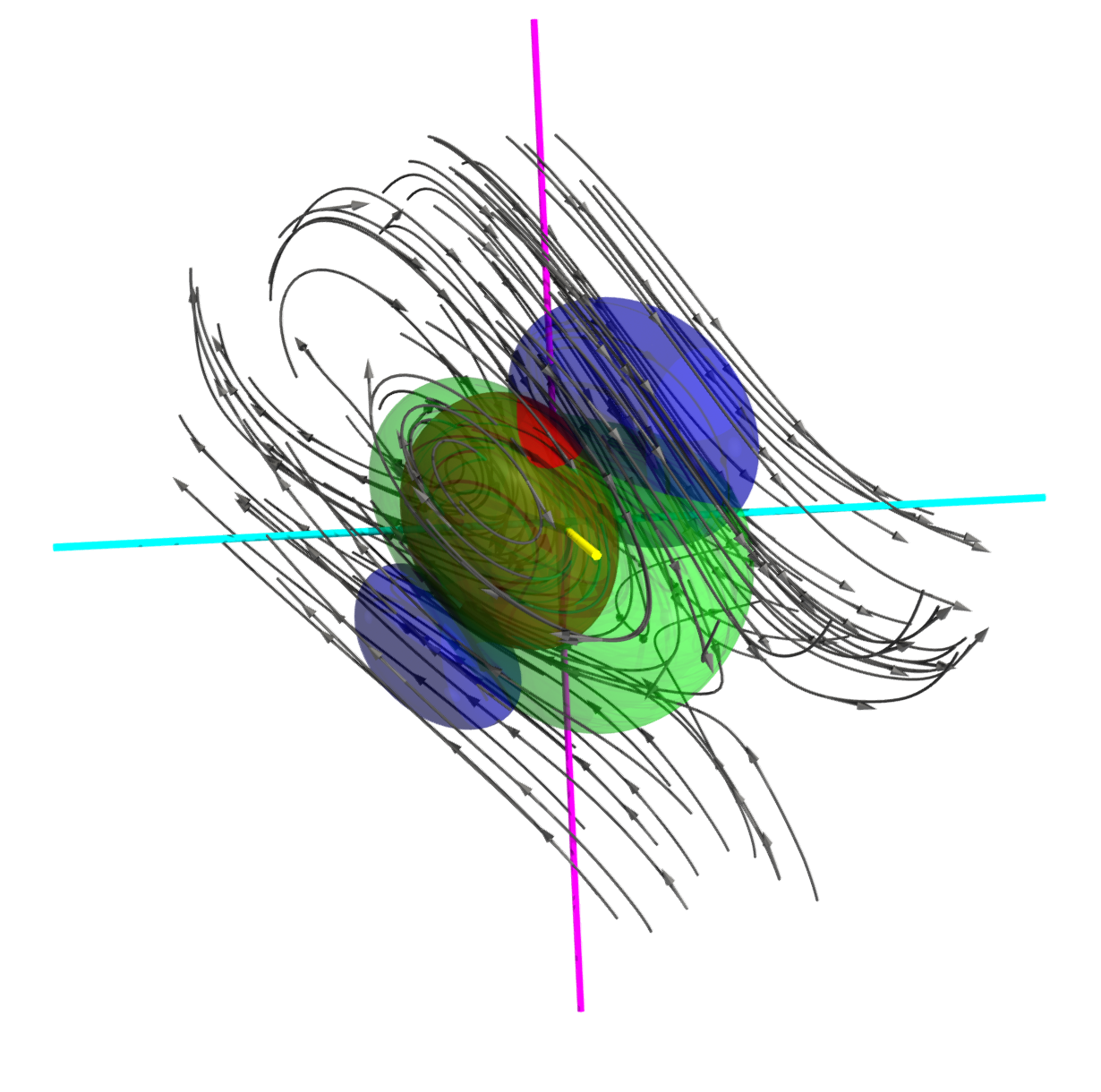}%
    \mylab{-11cm}{4.75cm}{(c)}%
    \mylab{-5cm}{4.75cm}{(d)}

    \includegraphics[width=.40\textwidth]{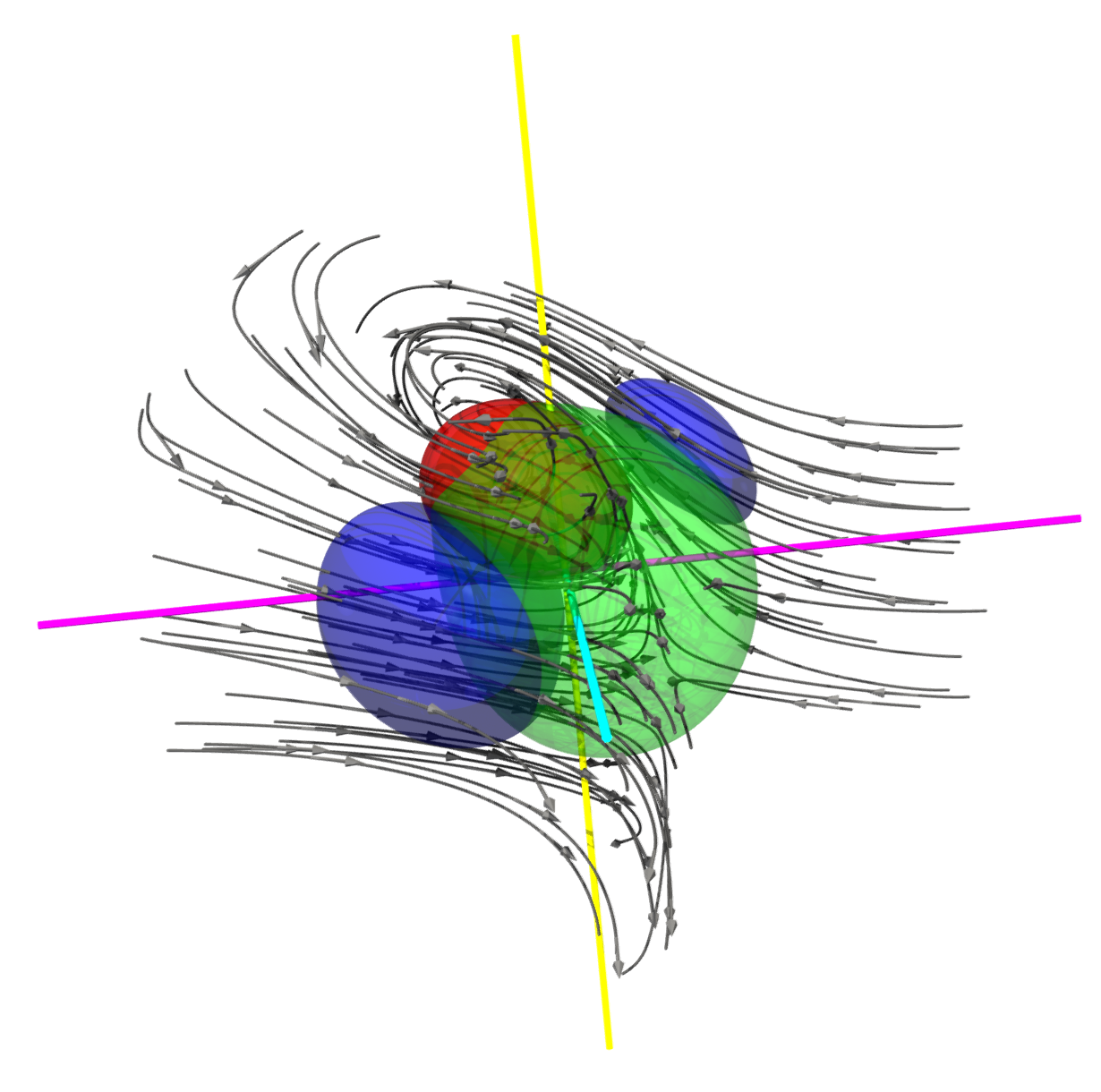}%
    \hspace{20pt}%
    \includegraphics[width=.40\textwidth]{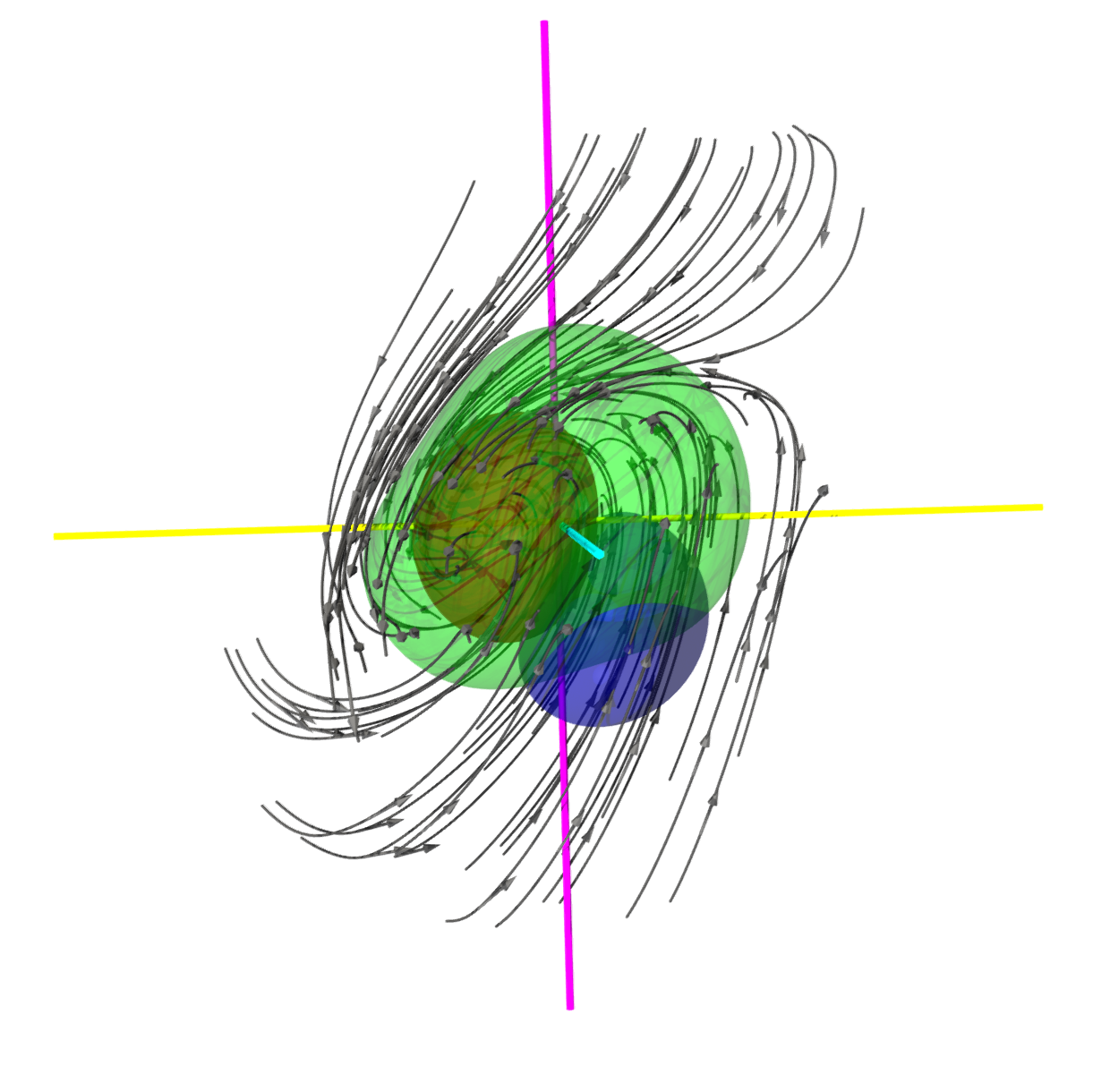}%
    \mylab{-11cm}{4.75cm}{(e)}%
    \mylab{-5cm}{4.75cm}{(f)}
    
    \caption{Flow fields conditioned to the presence of significant (a, c, e) and insignificant (b, d, f) regions, as described by algorithm CF. Computed using the relative $\tilde q$-norm for $q$-perturbations of $\rmDelta_2$. (a-f) Red isosurfaces are enstrophy, green isosurfaces are strain magnitude and blue isosurfaces are local kinetic energy magnitude. The thresholds are proportional to the total kinetic energy/ enstrophy of each conditional structure, but are equal in all cases. Gray lines with arrows are streamlines in a frame of reference that moves with the structure. The coloured axes are the three principal directions of the average rate-of-strain, magenta is the most compressive, yellow the intermediate and cyan the most stretching. (a, b) The strongest vorticity is aligned with the most compressive eigenvector. (c, d) Like (a, b) but for the intermediate eigenvector. (e, f) Like (a, b) but for the most stretching eigenvector. The orange scale bar in (a) corresponds to $60\eta$. \label{fig:3dconds}}
\end{figure}

Table \ref{tab:3dconds} gathers the most relevant averaged information pertaining to the conditional structures. The first important disparity among the two sets is the discrepancy in the number of elements for each alignment. The strong vorticity contained in significant regions is preferentially aligned with the most stretching eigenvalue (66\%) and seldom aligns with the most compressive (3.4\%). These preferences are also observed in the insignificant conditionals, but they are noticeably milder (12.2\% for the compressive and 53.6\% for the stretching). It is important to note that these alignments are based on the average strain of a region of size $3\rmDelta_2$ and the average vorticity vector of the strongest enstrophy object of size $\rmDelta_2$ contained in that region. That vorticity aligns preferentially with the most stretching eigenvalue of the strain at larger (coarse-grained) scales has been previously established \citep{leu:swa:dav:12,loz:hol:jim:16}. The present results 
provide further insight by highlighting that this preference is accentuated in strong and significant regions, while weak and insignificant regions show less preferential alignment. 
Table \ref{tab:3dconds} also shows the quantifiers $q^i$ and $\omega^i$, defined as,
\begin{equation}
    \label{eq:qi}
    \left(q^i\right)^2 = \frac{\int_\Sigma{q_c^2(\vect{x}) \dd \vect{x}}}{\int_\Sigma \left(q_f^\prime\right)^2g^2(\vect{x};3\rmDelta_2)\dd \vect{x}},
\end{equation}
where $(q_f^\prime)^2$ is the average kinetic energy of the filtered velocity field, and $q_c^2(\vect{x})$ is the kinetic energy of the conditional. They represent the strength of the conditional field compared with a hypothetical field of uniform average intensity. The quantifiers reinforce the idea that significant structures have stronger gradients and strong kinetic energy compared with the insignificant ones, even when they are classified with the amplification. The data also indicate that the most prevalent significant structure, which is aligned with the most stretching eigenvector, is the strongest of the three. On the other hand, the most compressive significant structure has relatively weak enstrophy, yet it still falls into the significant category, possibly due to its high kinetic energy. Values for a random set extracted from the flow fall between significant and insignificant, further suggesting that the main characteristic of these sets is their intensity. Finally, we repeated the full procedure using $\rmDelta_0$, which lies on the dissipative range. Here, viscosity ensures that the much smaller perturbed regions are reasonably smooth, and contain one to three vortices, so no filter is needed. The regions are somewhat similar to the filtered structures at size $\rmDelta_2$, although with longer vorticity structures.
Table \ref{tab:3dconds} contains the averages for $\rmDelta_0$, showing similar qualitative characteristics to those of filtered $\rmDelta_2$. The most notable difference is that significant structures contain proportionally more kinetic energy for $\rmDelta_2$. This is not remarkable, as large scales have more kinetic energy than small scales and we have shown that the relevancy of kinetic energy as a significance marker decays with the size of the perturbation.

More insight is gained from figure \ref{fig:3dconds}, where the six conditional flow fields are shown. Structures go in pairs of almost identical structures for each alignment. They show one intense vortex interacting with the background strain, with one or two regions of intense kinetic energy bracketing the vortex. Strain and vorticity are not collocated, and the strong strain is associated to an stagnation point in the stretched (c, e) patterns. The conditional aligned with the intermediate eigenvector is remarkably similar to the unconditional, unfiltered average of \cite{els:mar:2010}, which aligns with the intermediate eigenvector of the unfiltered strain. The dominant pattern (e) shows a vortex that is not completely parallel to the most stretching eigenvector but inclined towards the intermediate one. Unlike (c), which is almost two-dimensional and faithfully represented by the figure, the pattern in (e) is clearly three-dimensional, and the rotation we show represents our best attempt to display both the stagnation point and the vortex. 
Lastly, pattern (a), shows a vortex being compressed by the external strain field, a transient state previously reported \citep{ver:jim:orl:95}, and less likely for strong gradients. 

Aside from the stronger kinetic energy hinted by table \ref{tab:3dconds}, patterns (a,c,e) and (b,d,f) show different topologies, which
\begin{figure}
    \centering\vspace{1em}
    \includegraphics[width=0.99\textwidth]{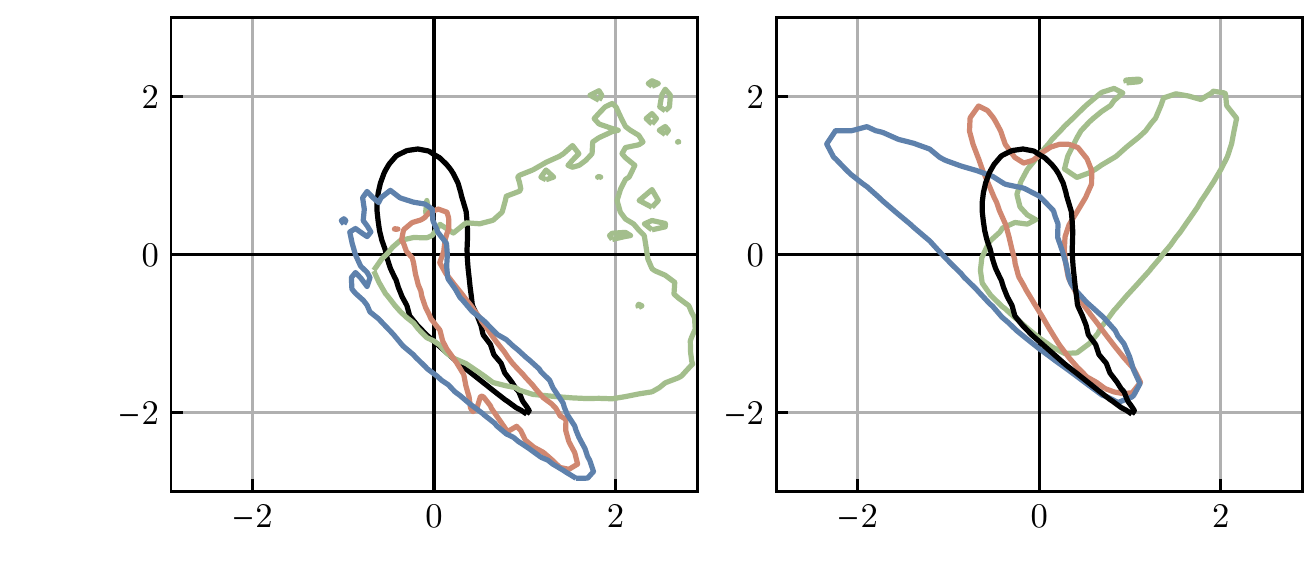}%
    \mylab{-3.05cm}{0.1cm}{$R/(Q^\prime)^{(3/2)}$}%
    \mylab{-9.65cm}{0.1cm}{$R/(Q^\prime)^{(3/2)}$}%
    \mylab{-13cm}{3.2cm}{$Q/Q^\prime$}%
    \mylab{-12.15cm}{5.5cm}{(a)}%
    \mylab{-6cm}{5.5cm}{(b)}%
    \caption{Joint pdf of the invariants of the conditional flow fields in figure \ref{fig:3dconds}. Contours contain 80\% of the probability mass. Green lines are figures \ref{fig:3dconds}(a, b), orange lines are figures \ref{fig:3dconds}(c,d) and blue are \ref{fig:3dconds}(e,f). (a) Significant structures. (b) Insignificant structure. (a, b) The black contour is the unconditional joint pdf for the filtered velocity. \label{fig:qrcond}}
\end{figure}
are further explored in figure \ref{fig:qrcond}.
 It shows the joint probability density function of the invariants of the velocity gradient tensor $(Q, R)$ for the conditionals in figure \ref{fig:3dconds}. Note that the pdfs are computed for the conditionals themselves and not for the elements that form those fields. Therefore, the amount of data is limited, and the results are necessarily noisy. 
 The invariants are useful to quantify the local topology of the flow, and the intensity of the gradients. 
 Figure \ref{fig:qrcond} shows the joint pdfs for the invariants, each normalised with the rms $Q^\prime_s$ of their conditional flow.
 These joint pdfs can be compared with the black contours, which represent the unconditional joint pdf for the filtered velocity field. One key difference between significant and insignificant structures is that insignificant structures have stronger vorticity ($Q > 0$) and significant ones have stronger strain ($Q < 0$).
 This is particularly acute for the intermediate and stretching significant structures, where most of their probability mass falls along the lower right quadrant \citep{vie:84}.
 In contrast to those, the insignificant structures lean towards the
 upper half-plane, which represents vorticity dominated regions.
 Finally, the figure helps clarifying the behaviour of the structures  where vorticity aligns with the compressive eigenvalue. Because significant structures are dominated by strain, the compressive one lies in a region of very low unconditional probability, whereas the insignificant one lies in the more probable upper half plane. 
The overall picture is that intense strain-dominated regions are more sensitive to perturbations than weak vorticity dominated regions.


 \subsection{Statistics of the structure of significant and insignificant regions}

 The classifiers in \S\ref{sec:significance} have shown that the structures within significant regions are just as important for significance, of not more so, than larger scale structures. Instead of averaging out these structures, losing important information, we now provide a statistical description of them. 
 We obtain them by applying a threshold to the enstrophy and kinetic energy fields, and segmenting connected strong enstrophy and kinetic energy regions. 
As before, the threshold values are chosen independently for the significant and insignificant sets of regions, following a percolation analysis. 

\begin{figure}
    \centering
    \includegraphics{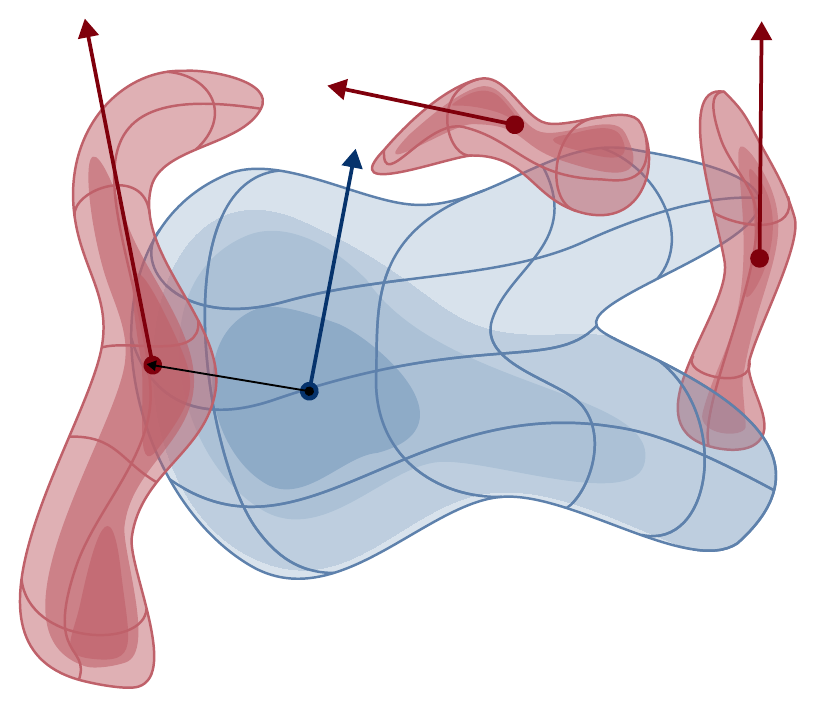}%
    \mylab{-6.25cm}{3.02cm}{$\vect{r}_{u\omega}$}%
    \mylab{-5.35cm}{4.25cm}{$\vect{u}_c$}%
    \mylab{-7.5cm}{4.75cm}{$\vect{\omega}_c$}%
    \caption{Sketch of a segmented region of size $\rmDelta_2$. Red volumes represent enstrophy structures and blue kinetic energy. A detailed description is included in the main text.\label{fig:sketch}}
\end{figure}

\begin{figure}
    \centering\vspace{1em}
    \includegraphics[width=0.99\textwidth]{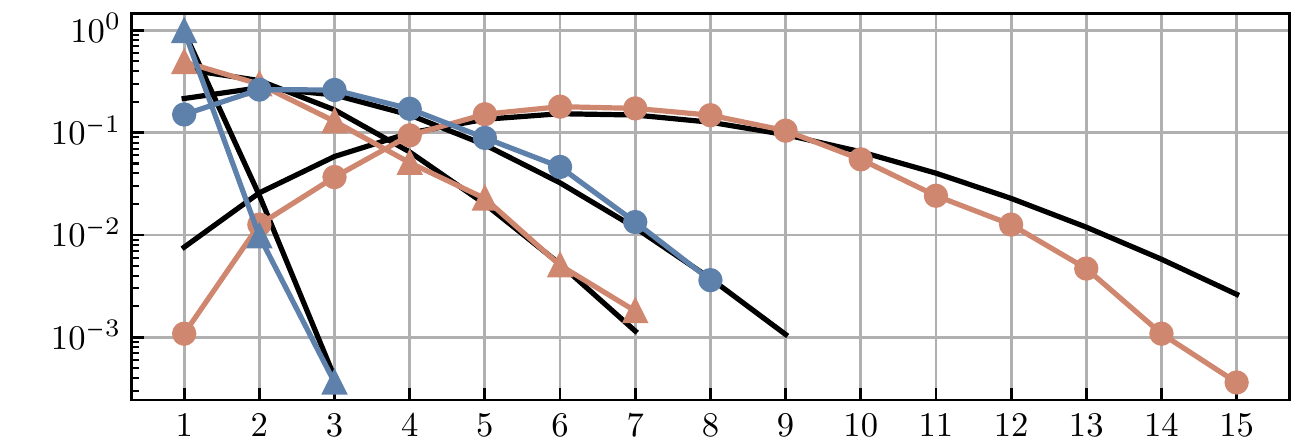}\vspace{2em}
    \includegraphics[width=0.99\textwidth]{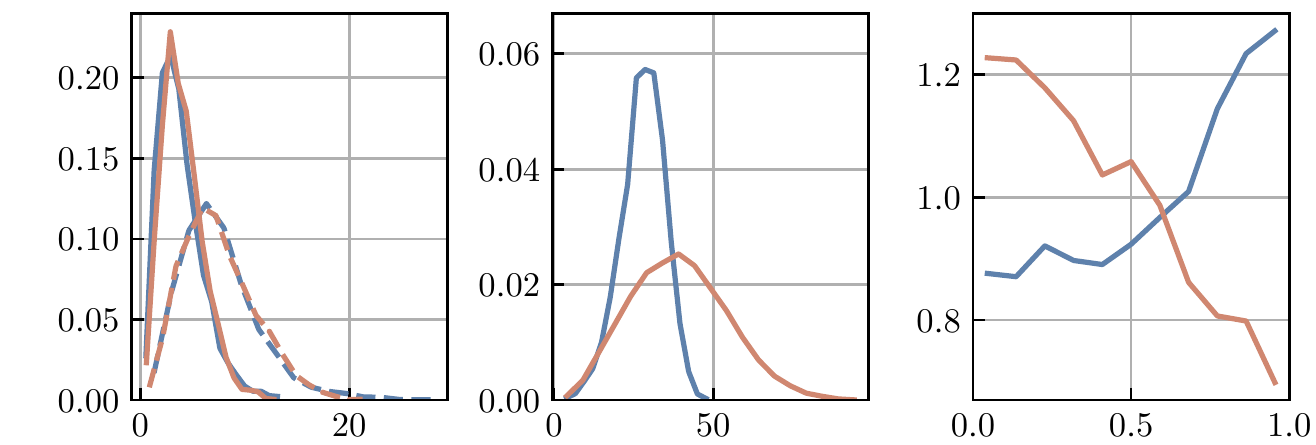}\vspace{2em}
    \mylab{-2.5cm}{-0.3cm}{$\cos{(\theta_{u \omega})}$}%
    \mylab{-4.25cm}{4.3cm}{(d)}%
    \mylab{-6.65cm}{-0.3cm}{$r_{u\omega_\emptyset}/\eta$}%
    \mylab{-8.65cm}{4.3cm}{(c)}%
    \mylab{-10.75cm}{-0.3cm}{$r/\eta$}%
    \mylab{-12.75cm}{4.3cm}{(b)}%
    \mylab{-6.5cm}{4.85cm}{$N$}%
    \mylab{-12.75cm}{9.75cm}{(a)}%
    \mylab{-13.25cm}{7.75cm}{pdf}
    \mylab{-13.25cm}{3.22cm}{pdf}
    \caption{Geometrical statistics of the significant (orange lines) and insignificant (blue lines) regions for the relative $q$-norm for $q$-perturbations of size $\rmDelta_2$. (a) Pdf of the count of kinetic energy structures (triangles) and enstrophy structures (circles). The black lines are Poisson distributions. (b) Pdfs of the distance $r_{ou}$ (solid lines) and the distance $r_{u\omega_0}$ for the strongest enstrophy object (dashed lines). The distances of the significant objects are divided by four. (c) Pdf of the distance $\vect{r}_{u\omega_\emptyset}$ for the enstrophy objects not in (b). (d) Pdf of the angle between $\vect{u}_c$ and the largest $\vect{\omega}_c$. \label{fig:stasmall}}

\end{figure}

Figure \ref{fig:sketch} shows a sketch of a region of size $\rmDelta_2$, which typically contains a few kinetic energy objects and several enstrophy objects.
The centroids of each segmented object, calculated using kinetic energy or enstrophy as the mass distribution, are represented by dots in the sketch. Additionally, the average velocity vector, $\vect{u}_c$, and average vorticity vector $\vect{\omega}_c$, are calculated for each region and represented by arrows in the figure.
Finally, the vectors $\vect{r}_{u\omega}$ connect the centroids of the largest objects of kinetic energy and enstrophy. With these definitions, we can study several metrics that separate significant from insignificant regions and do not directly depend on the intensity of any marker. 

Figure \ref{fig:stasmall} shows several geometric statistics of the significant and insignificant regions. The pdf of the number of objects is shown in \ref{fig:stasmall}(a). In agreement with the examples in figure \ref{fig:3dexample}, both significant and insignificant regions show many more enstrophy objects than kinetic energy ones. Their probability distributions are compatible with Poisson processes, with the exception of the enstrophy objects in significant regions, which are more concentrated than Poisson. A Poisson process implies that the existence of additional objects is independent of existing ones. Thus, the enstrophy objects in significant regions `see' each other, whereas the rest of the objects do not. Insignificant regions essentially contain only one kinetic energy object, but even in the significant regions, the largest kinetic energy object contains 90\% of the kinetic energy on average. A similar picture holds for enstrophy structures in the insignificant regions, where the largest structure contains 87\% of the enstrophy on average. This is not the case with the significant regions, were secondary enstrophy structures have volumes comparable to the largest one. The average volume of their secondary structures is 20\%, with the second largest accounting for 42\% of the total.

Figures \ref{fig:stasmall}(b, c) show the distances between various objects within the regions. The objects found within insignificant regions are located close to $\vect{x}_0$, with distances measuring a few Kolmogorov units. In contrast, objects within significant regions are situated farther away, with the enstrophy objects located at distances equivalent to the perturbation radius. The orange lines in Figure \ref{fig:stasmall}(b) depict the distances for significant regions, scaled by a factor of four, highlighting that they are similar to those found in insignificant regions up to a scale factor. However, this similarity is disrupted by the presence of additional enstrophy objects in significant regions, as shown in Figure \ref{fig:stasmall}(c). These objects do not conform to the same scaling factor and, when considering the scaling factor, they are located much farther away in insignificant regions. This suggests that their nature is different and potentially unrelated to the local dynamics of insignificant regions.
On the other hand, all enstrophy objects found within significant regions are similar to one another, both in distances and volumes. Figure \ref{fig:stasmall}(d) shows the probability density function of the cosine of the angle between $\vect{\omega}_{c}$ of the strongest enstrophy structure and $\vect{u}_c$. This angle should not be confused with the point-wise angle of the vorticity and velocity vectors associated with the depletion of nonlinearity \citep{kra:pan:88}, as the vectors here are not collocated. Instead, this measurement helps interpret the nature of the velocity within the object. Vorticity induces a velocity that is orthogonal to it, as per the Biot-Savart law. This scenario is consistent with the significant regions, where the pdf leans towards a null cosine between the two vectors. In contrast, insignificant regions display a preference for parallel velocity and vorticity, indicating that most of the velocity is generated by the flow outside the region, rather than by the local vorticity. Similar insight is provided by the intersection between enstrophy and kinetic energy objects within each region, as they only share 11\% of their volume in the significant regions compared 47\% in the insignificant ones.

\begin{figure}
    \centering\vspace{1em}
    \includegraphics[width=0.99\textwidth]{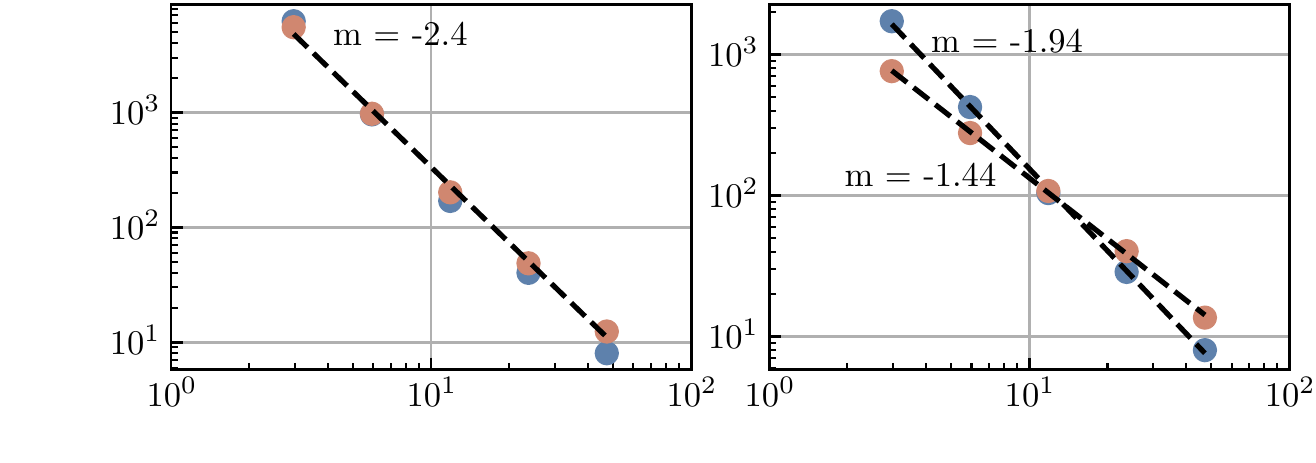}%
    \mylab{-3.25cm}{0.2cm}{$r/\eta$}%
    \mylab{-9.25cm}{0.2cm}{$r/\eta$}%
    \mylab{-12.5cm}{2.9cm}{N}%
    \mylab{-12.2cm}{4.75cm}{(a)}%
    \mylab{-6.15cm}{4.75cm}{(b)}%
    \caption{Average box-counting for insignificant (blue) and significant (orange) regions. Dashed lines are the fractal dimensions from a least squares fit. (a) Kinetic energy regions. (b) Enstrophy regions.\label{fig:fractal}}
\end{figure}


Figures \ref{fig:3dexample}(a, b) suggest that significant regions are inherently more `complex' than insignificant regions.
We quantify this complexity by computing the fractal or box-counting dimension of the kinetic energy and enstrophy segmentations, which is shown in figure \ref{fig:fractal} \citep{moi:jim:04}. There is little difference between the fractal dimension of kinetic energy objects for both sets, with a value close to $D\approx 2.4$. This value falls between a surface and a volume and can be thought as a `thick', complex pancake, as shown in figures \ref{fig:3dexample}(a,b). These figures also give an intuition for the types of structures associated to the values of the fractal dimension obtained from the enstrophy objects. The enstrophy objects inside significant regions have a fractal dimension of $D \approx 1.4$, between a thread and a surface, and associated with vortex clusters 
\citep{moi:jim:04,ala:jim:zan:mos:06}. In contrast to those, insignificant regions contain enstrophy objects with a fractal dimension of $D \approx 1.9$, much closer to vortex sheets (see figure \ref{fig:3dexample}(b), for example).

\section{From structures to dynamical relevance\label{sec:assimilation}}

\begin{figure}
    \includegraphics[width=0.6\textwidth]{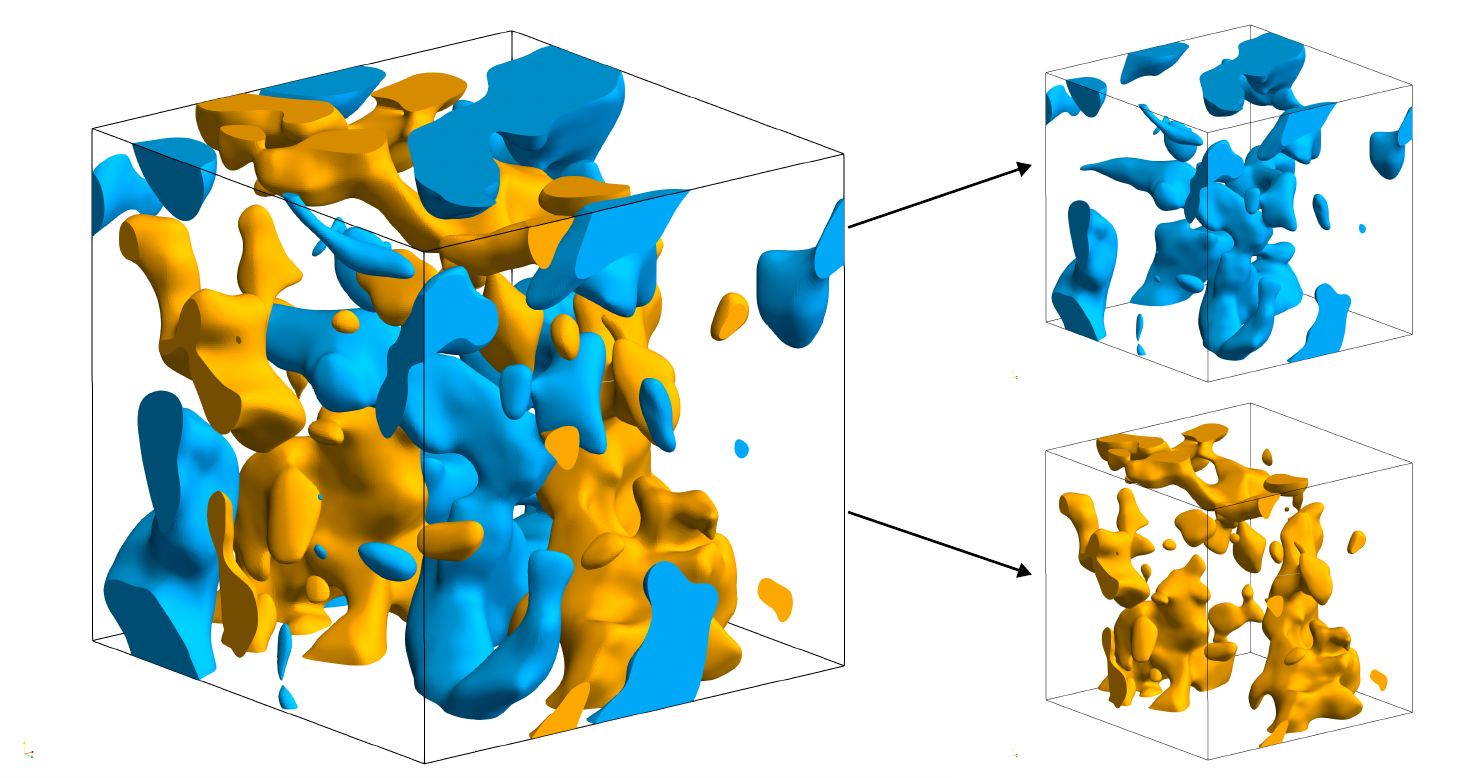}%
    \includegraphics[width=0.39\textwidth]{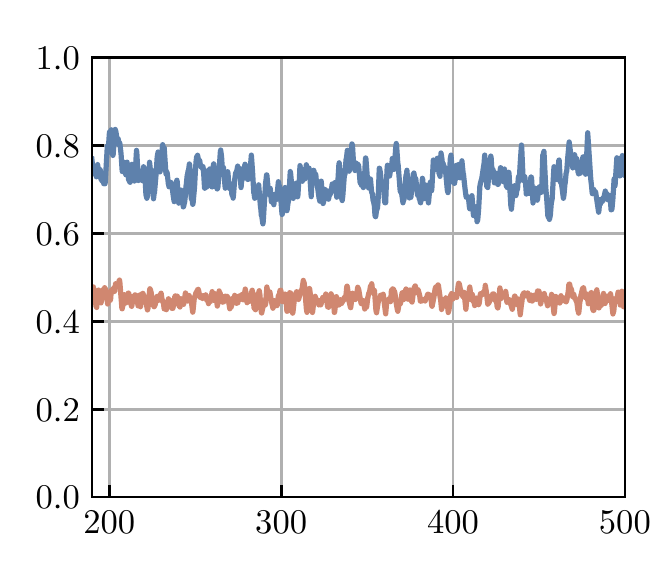}%
    \mylab{-2.75cm}{0.0cm}{$tu^\prime/L_\mathcal{E}$}%
    \mylab{-5.25cm}{2.375cm}{$e$}%
    \mylab{-12.5cm}{0.0cm}{`{A}'}%
    \mylab{-7.5cm}{2.1cm}{`{B}'}%
    \mylab{-7.5cm}{0.0cm}{`{C}'}%
    \mylab{-13cm}{4.5cm}{$(a)$}%
    \mylab{-5.25cm}{4.5cm}{$(b)$}%
    \caption{(a) Sketch of the assimilation scheme. The reference simulation is on the left, and the 10\% sub-volume with the most negative $Q$ is in orange, while its converse is in blue. (b) Relative error of the assimilated simulations. `{B}' is in blue and `{C}' in orange. \label{fig:assimilation}}
\end{figure}

We have identified, classified and analysed significant an insignificant regions. In this last section we show an example of why knowing their properties gives us relevant information about the turbulent flow. For this task we do the following experiment. Consider a simulation of HIT forced with constant energy input, as defined in \S\ref{sec:experiments}. This simulation is used as a reference and labelled as `{A}'. Consider two other simulations, `{B}' and `{C}', which have no forcing other than the condition that some physical subdomain is assimilated at each time step from {A}. This type of experiment is similar to the assimilation experiments of \cite{wan:tam:22} or the nudging in physical space of \cite{di:maz:bif:20}.

Figure \ref{fig:assimilation}(a) shows a sketch of the three simulations run. The subdomains highlighted in orange and blue are assimilated to the simulations `{C}' and `{B}' respectively. First, we compute $Q$ for the filtered velocity field, using $g(\boldsymbol{x}, \rmDelta_2)$ as a filter. We saw in figure \ref{fig:qrcond} that significance is associated to perturbation-scale dissipation dominating over perturbation-scale enstrophy, while the opposite can be said about insignificance. Both states can be identified by the sign of $Q$. Thus, we choose two constant thresholds for the second invariant such that, on average, a constant fraction of the volume of the box is below and above each of them respectively. These are the subvolumes represented in figure \ref{fig:assimilation}(a) for a given instant in blue and orange colours. The former contains the 10\% most positive $Q$ and the latter the 10\% most negative. Assimilation is achieved by copying the value of the three velocity components within the subvolume to the subjugate simulations, correcting for continuity.
The main result of this experiment is shown in figure \ref{fig:assimilation}(b), which shows the relative assimilation error,
\begin{equation}
    e_\mathrm{B}^2(t) = \dfrac{\int_\Omega(u_\mathrm{B}(t)-u_\mathrm{A}(t))^2\dd\vect{x}}{\int_\Omega u_\mathrm{A}^2(t)\dd\vect{x}},
\end{equation}
and equivalently for `C'. The assimilated simulations experience a transient for the first $tu^\prime/L_\mathcal{E} \approx 5$, but remain stable after that, with the assimilation error oscillating with periods comparable with the oscillation of the dissipation of `A'. The simulation `C' has almost 60\% less error than `B', despite both receiving from `A' the same total fraction of flow-field volume. Both simulations receive from `A' approximately the same fraction of the total energy, which is approximately equal to the fraction of volume.
We saw in figure \ref{fig:physper} that perturbation growth is associated to the spreading of the perturbation, rather than local magnitude growth. It thus seem reasonable that coherent vortices take longer to spread the perturbation than regions of strong strain, resulting in smaller assimilation errors when the latter are assimilated. 

\section{Discussion and Conclusions}\label{sec:conclusions}

In this work we have extended the Monte-Carlo ensembles of \cite{jim:18a} to three-dimensional
decaying isotropic turbulence at $\Rey_\lambda \approx 190$. Owing to the multi-scale nature 
of fully developed turbulence, we have tested the effect of different perturbation sizes, as
well as of different types of perturbations. The evolution of the perturbations is tracked in time,
and the properties of the flow where the perturbations are introduced are related to their
growth using a classification algorithm.

The first part of the analysis deals with the properties of the perturbations that either cancel
the velocity or the vorticity within a region. We found that perturbations typically grow as a function of time, the sole exception being weak and small perturbations of sizes in the dissipative range. 
We observed two time scales in the evolution of perturbations in the fluid flow. The first, fast time scale is associated with the growth of small scales within the perturbed region, which is independent of the perturbation size. The second, slower time scale corresponds to the growth of the perturbed region itself, which weakly depends on the perturbation size. Our findings show that the time at which the growth rate of the perturbations becomes independent of the initial conditions is of the order of one large-scale turnover. Furthermore, we found that perturbations grow by increasing in size rather than intensity, leading to small perturbations growing faster but never overtaking larger ones. On average, the diameter of the perturbations was found to be almost linear with time. This type of growth is much faster than that of a passive scalar, highlighting the importance of the active nature of perturbations in the flow for both their spatial growth and intensity.
The 5\% of the perturbations which grow the most or the least are classified as significant or insignificant respectively. 
Our findings indicate that significant perturbations grow faster than insignificant ones.
Their spectra is different from the insignificant perturbations in the large scales, which grow for the significant perturbations but are approximately constant for the insignificant ones. This is due to significant perturbations displacing or deforming large scale structures from their expected evolution, while insignificant perturbations affect mostly scales smaller than themselves.

In the second part of our study, we examined the regions of the fluid flow where the perturbations are introduced. We investigated which properties of the flow field can be coarse-grained to the size of the perturbations and still act as reasonable predictors for future significance. We found that absolute norms, i.e. which perturbation is the strongest after some time, are strongly conditioned by the initial perturbations. Perturbations that modify a lot of kinetic energy initially are the most significant in terms of total kinetic energy error after a while, and a similar statement holds for the enstrophy. We found that the amplifications, i.e. which perturbations grow the most, depend on the magnitude of the local gradients. Of these, the markers based on small-scale gradients are better than the ones based on perturbation-scale ones. The simplest interpretation is that the contents of the region are more important for significance than the `inertial' gradients, with the exception of the inertial strain magnitude. As a result, isolating the significance of inertial scales may require the use of strategies such as wavelets to constrain the support of perturbations in scale space.

Besides the magnitude of observables, we studied the structure of the flow in the significant and insignificant regions. Their structure is studied both at the scale of the perturbations, and of the smaller scales contained in them.
We found the structure of the flow to have a weak dependence on the significance, with only a few important differences. When compared to a random region, significant ones have larger tendency to have vorticity aligned with the most stretching eigenvector of the strain, exhibit larger perturbation-scale strain than vorticity, and have a complex structure that resembles vortex clusters. In contrast, insignificant ones favour vorticity alignment with the most compressive eigenvector, have larger perturbation-scale vorticity than strain, and have a simpler structure reminiscent of vortex sheets. The association of strong or weak enstrophy to vortex clusters or sheets was already observed in \cite{jim:wra:saf:rog:93}. \cite{gop:tai:21} found that strain dominated regions influence the rest of the flow field using a completely different method based on graph theory, which is consistent with our observations. The different fractal dimension of the regions is particularly interesting because it does not depend on any intensity and is purely geometrical. Aside from their significance, the only `global' information the fractal dimension has from the flow is the scale, which is, in principle, equal for both significant regions and insignificant ones. However, the local dissipation is different for both sets of perturbations. A reasonable model is that the local dissipation results in a larger local Reynolds number \citep{kol:61}, which in turn results in a more complex structure.

Finally, we show with a quick assimilation experiment that significant and insignificant regions are related to the flow dynamics. The inertial-scale strain of significant regions can amplify the perturbations and spread them across the flow, whereas the inertial-scale vorticity cannot do it so efficiently. Thus, when a simulation receives from another information about significant regions, it achieves a higher degree of synchronisation with the latter than when the same volume of the flow is received from insignificant regions. Thus, significant regions are more relevant for tasks such as assimilation and control, and algorithms can benefit from this knowledge. For example, the covariance matrices of Kalman filters could be adjusted depending on the significance of the flow. These tasks do not require to compute the prohibitively expensive significance directly. Instead, significance can be estimated from the properties of the significant regions gathered in this paper and still obtain promising results, as in the present work.

\backsection[Funding]{ This work was supported by the European Research Council under the Coturb and CausT grants, ERC-2014.AdG-669505 and ERC-2021.AdG-101018287. }

\backsection[Declaration of interests]{The authors report no conflict of interest.}

\appendix

\section{Statistics for the vorticity perturbations}\label{app:op}

\begin{figure}
    \centering
    \includegraphics[width=0.99\textwidth]{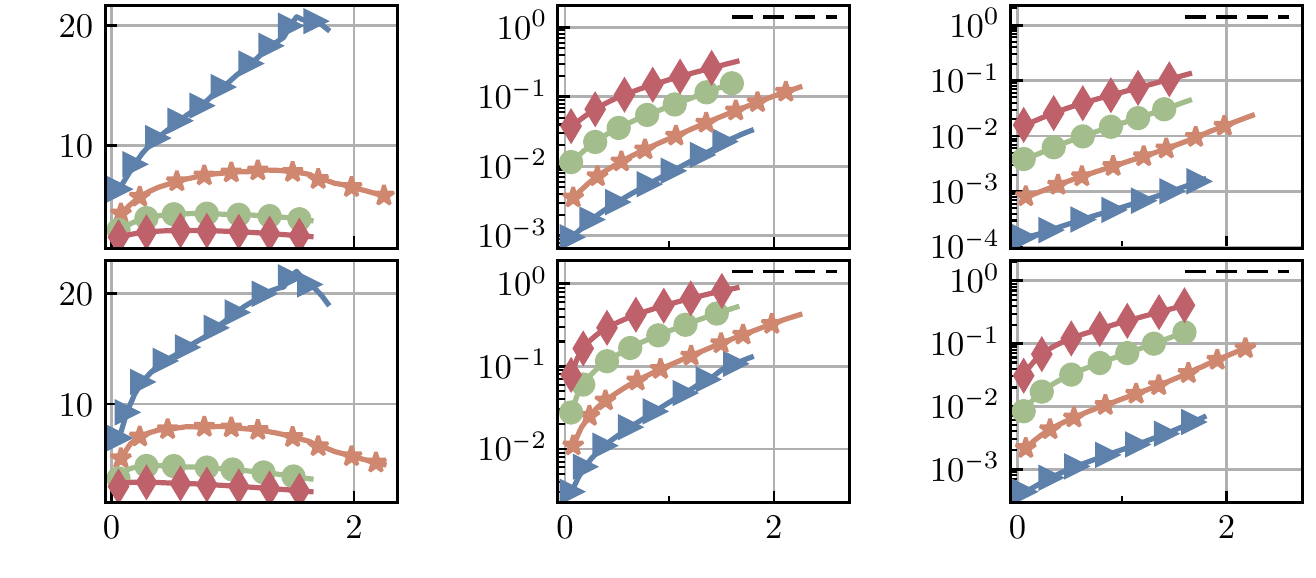}%
    \mylab{-0.145\textwidth}{0pt}{$\tau$}%
    \mylab{-0.478\textwidth}{0pt}{$\tau$}%
    \mylab{-0.811\textwidth}{0pt}{$\tau$}%
    \mylab{-4.6cm}{2.0cm}{$P_\omega^{5}$}%
    \mylab{-4.6cm}{4.5cm}{$P_q^{5}$}%
    \mylab{-9.25cm}{2.0cm}{$P_\omega^{95}$}%
    \mylab{-9.25cm}{4.5cm}{$P_q^{95}$}%
    \mylab{-13.3cm}{2.0cm}{$\mathcal{R}_{\omega}$}%
    \mylab{-13.3cm}{4.5cm}{$\mathcal{R}_{q}$}%
    \mylab{-13.2cm}{5.75cm}{(a)}%
    \mylab{-13.2cm}{3cm}{(b)}%
    \mylab{-9.2cm}{5.75cm}{(c)}%
    \mylab{-9.2cm}{3cm}{(d)}%
    \mylab{-4.6cm}{5.75cm}{(e)}%
    \mylab{-4.6cm}{3cm}{(f)}%
    
    \caption{Statistics of the growth of the 95th ($P^{95}$) and the 5th ($P^{5}$) percentiles of  $\psi'_{q}$ (a,c,e) and $\psi'_{\omega}$ (b,d,f) for enstrophy perturbations. (a,b) Significance ratio, defined as the ratio between the percentiles. (c,d) $P^{95}$. (e,f) $P^{5}$.\label{fig:perto}}
\end{figure}

\begin{figure}
    \centering
    \includegraphics[width=0.99\textwidth]{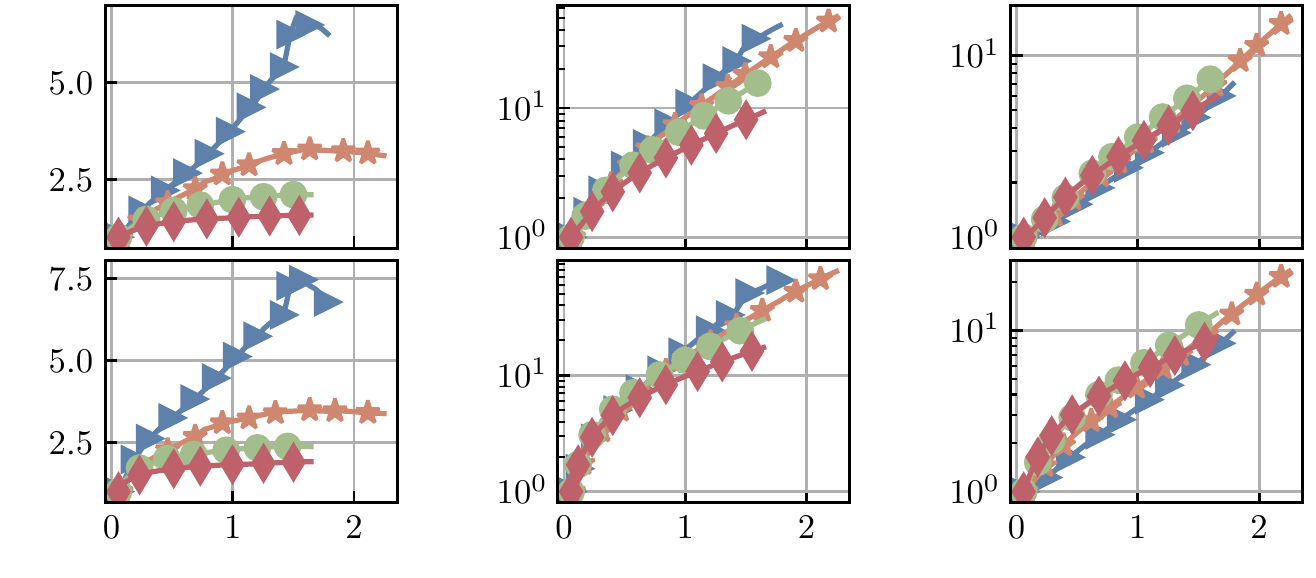}%
    \mylab{-0.145\textwidth}{0pt}{$\tau$}%
    \mylab{-0.478\textwidth}{0pt}{$\tau$}%
    \mylab{-0.811\textwidth}{0pt}{$\tau$}%
    \mylab{-4.6cm}{2.0cm}{$\tilde P_{\omega}^{5}$}%
    \mylab{-4.6cm}{4.5cm}{$\tilde P_{q}^{5}$}%
    \mylab{-9.25cm}{2.0cm}{$\tilde P_{\omega}^{95}$}%
    \mylab{-9.25cm}{4.5cm}{$\tilde P_{q}^{95}$}%
    \mylab{-13.3cm}{2.0cm}{$\tilde R_{{\omega}}^{5}$}%
    \mylab{-13.3cm}{4.5cm}{$\tilde R_{{q}}^{5}$}%
    \mylab{-13.2cm}{5.75cm}{(a)}%
    \mylab{-13.2cm}{3cm}{(b)}%
    \mylab{-9.2cm}{5.75cm}{(c)}%
    \mylab{-9.2cm}{3cm}{(d)}%
    \mylab{-4.6cm}{5.75cm}{(e)}%
    \mylab{-4.6cm}{3cm}{(f)}%
    
    \caption{Statistics of the growth of the 95th ($P^{95}$) and the 5th ($P^{5}$) percentiles of  $\tilde \psi'_{q}$ (a,c,e) and $\tilde \psi'_{\omega}$ (b,d,f) for enstrophy perturbations. (a,b) Significance ratio, defined as the ratio between the percentiles. (c,d) $P^{95}$. (e,f) $P^{5}$.\label{fig:pertor}}
\end{figure}

Figures \ref{fig:perto} and \ref{fig:pertor} are analogous to figure \ref{fig:pertu} but are made for vorticity perturbations. 
Most of the conclusions extracted from the velocity perturbations carry over to them, although some differences exist. For example, the significance of the absolute magnitude is dominated initially by the faster dynamics associated to the vorticity, but after the transient, their growth rate is comparable to that of velocity perturbations, and both types of perturbations peak at comparable time delays.

Comparing figures \ref{fig:pertu}(a) and \ref{fig:perto}(a), shows that the values of significance are about twofold for the vorticity perturbations, but, at the same time, figures \ref{fig:pertur}(g) and \ref{fig:pertor}(a) show that the `relative' significance has similar values. Thus, higher values of the `absolute' significance can be traced to the differences in the initial conditions of the perturbations. Common turbulence knowledge explains this difference by the much higher intermittentcy of the vorticity fields, compared with the velocity ones. Not obvious is the behaviour of $\rmDelta_0$, whose relative velocity significance $\mathcal{\tilde R}_{q}$ is larger for velocity than for vorticity perturbations. The difference lies in the different behaviour of the $P^5$ velocity perturbations, which have a considerable delay in their growth, or even contract for some time. On the other hand, vorticity perturbations always grow, as shown in \ref{fig:pertor}(e). A possible explanation is that while $\rmDelta_0 \approx 15\eta$ is a small size for velocity structures, is of the order of a structure for vorticity ones. While a velocity perturbation of that size at most harms a piece of a larger flow structure, a vorticity perturbation is more likely to affect a complete vorticity flow pattern. Most of the other remarks about the growth rates in figure \ref{fig:pertur} carry over figure \ref{fig:pertor}, even if the perturbations introduced are radically different.

\section{Equations for the growth of the perturbation\label{app:a}}
Introducing the definition of the perturbed flow field \eqref{eq:ploff0} into the N--S equations (\ref{eq:N1}, \ref{eq:N2}) gives,
\begin{gather}
    \partial_t\left(u_i + \pert_i\right) + \left(u_j + \pert_j\right)\partial_j\left(u_i + \pert_i\right) = -\partial_i \tilde p + \nu\partial_{jj}\left(u_i + \pert_i\right)\label{eq:N1tot}\\
    \partial_i \left(u_i + \pert_i\right)= 0\label{eq:N2tot}.
\end{gather}
An evolution equation for the perturbation can be obtained by subtracting from (\ref{eq:N1tot}, \ref{eq:N2tot}) the unperturbed equations,
\begin{gather}
    D_t\pert_i = -\partial_i \breve p - \pert_j\partial_ju_i - \partial_j\left(\pert_i\pert_j\right) + \nu\partial_{jj}\pert_i\label{eq:N1per}\\
    \partial_i \pert_i = 0\label{eq:N2per};
\end{gather}
where $D_t \equiv \partial_t + u_j\partial_j$ is the material derivative and $\breve p$ is a pseudopressure that enforces \eqref{eq:N2per} in \eqref{eq:N1per}. Multiplying \eqref{eq:N1per} by $\pert$,
\begin{equation}
    \frac{1}{2}D_t \pert^2_i = -\partial_j \left[\pert_j\breve p + \frac{1}{2}\pert_j\pert^2_i - 2\nu \pert_iS_{ij} \right] - 2\nu S_{ij}^\pert S_{ij}^\pert - \pert_iS_{ij}\pert_j, \label{eq:perg}
\end{equation}
where $S_{ij}^\pert$ is the rate-of-strain of the perturbation field.
Integration of \eqref{eq:perg} over the whole volume gives,
\begin{equation}
    \frac{1}{2}D_t \pert^\prime_q = -2\nu\!\!\int_\Sigma S_{ij}^\pert S_{ij}^\pert - \int_\Sigma \pert_iS_{ij}\pert_j.
\end{equation}
The only term responsible for the production of error is $\pert_iS_{ij}\pert_j$, which needs to be negative on average for the perturbation to grow. For perturbations that zero the kinetic energy, $\vect\pert_0 \approx - \vect u_0$, which suggests that the term $u_iS_{ij}u_j$ should be related to the significance. The terms under the divergence operator in the right hand sign of \eqref{eq:perg} do not produce net growth but can spread the perturbation through the flow field \citep{tsi:01}.


\section{Relation between the different norms\label{app:b}}
\begin{figure}
    \includegraphics[width=0.99\textwidth]{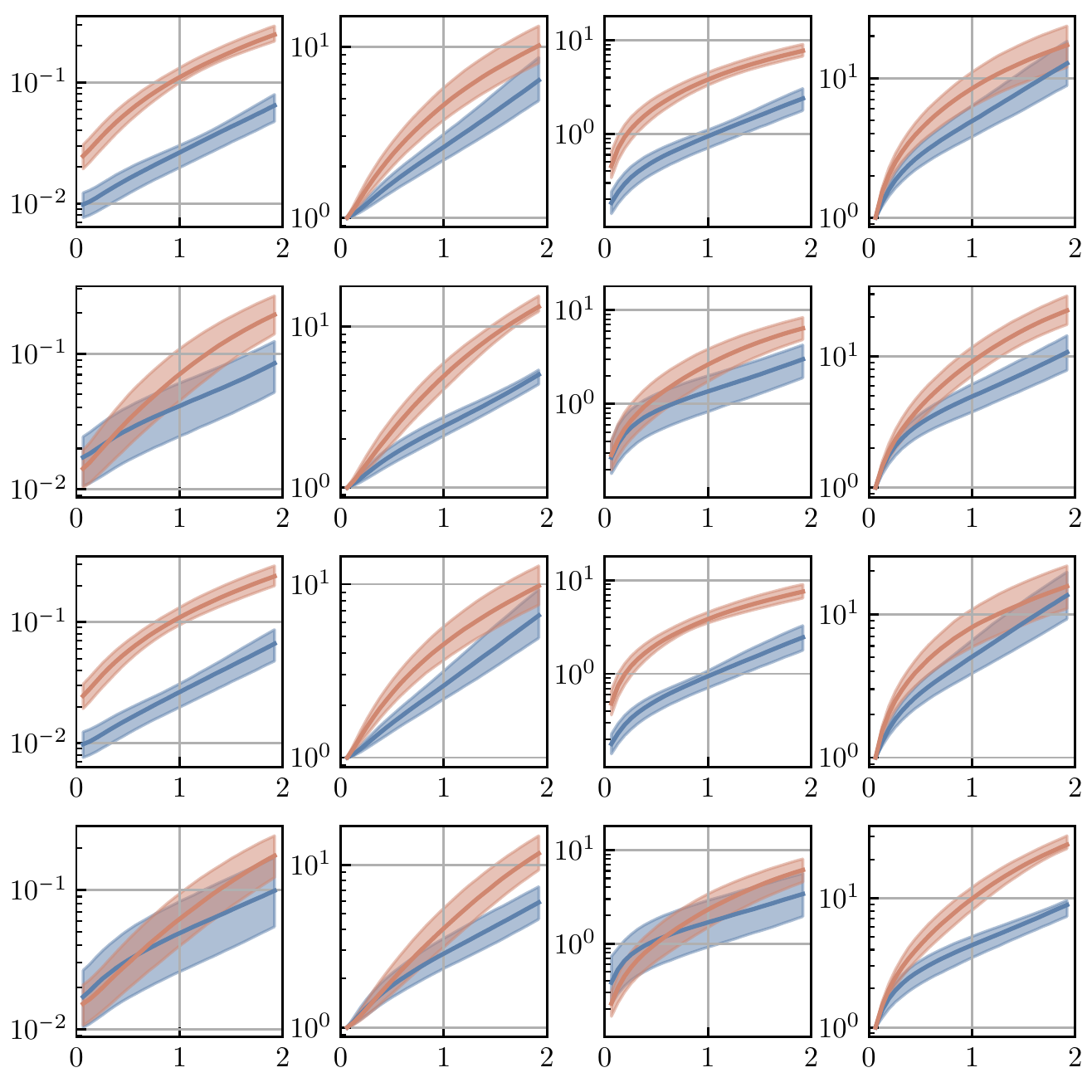}%
    \mylab{-13.cm}{13.15cm}{$(a)$}%
    \mylab{-9.75cm}{13.15cm}{$(b)$}%
    \mylab{-6.5cm}{13.15cm}{$(c)$}%
    \mylab{-3.25cm}{13.15cm}{$(d)$}%
    \mylab{-13.cm}{9.85cm}{$(e)$}%
    \mylab{-9.75cm}{9.85cm}{$(f)$}%
    \mylab{-6.5cm}{9.85cm}{$(g)$}%
    \mylab{-3.25cm}{9.85cm}{$(h)$}%
    \mylab{-13.cm}{6.55cm}{$(i)$}%
    \mylab{-9.75cm}{6.55cm}{$(j)$}%
    \mylab{-6.5cm}{6.55cm}{$(k)$}%
    \mylab{-3.25cm}{6.55cm}{$(l)$}%
    \mylab{-13.cm}{3.25cm}{$(m)$}%
    \mylab{-9.75cm}{3.25cm}{$(n)$}%
    \mylab{-6.5cm}{3.25cm}{$(o)$}%
    \mylab{-3.25cm}{3.25cm}{$(p)$}%
    \mylab{-11.3cm}{0cm}{$\tau$}%
    \mylab{-8.05cm}{0cm}{$\tau$}%
    \mylab{-4.8cm}{0cm}{$\tau$}%
    \mylab{-1.55cm}{0cm}{$\tau$}%

    \caption{Cross classification of the different norms for $q$-perturbations at size $\rmDelta_2$. From left to right, the magnitude represented is $\psi^\prime_q$, $\tilde \psi_{q}^\prime$, $\psi_\omega^\prime$ and $\tilde \psi_{\omega}^\prime$ respectively. From top to bottom, the norm used to classify significance changes in the same order. Significant perturbations are in orange and insignificant ones in blue. Solid lines are the median, and the shaded contour contains 80\% of the probability mass at each time. \label{fig:crosscla}} 
\end{figure}
\begin{figure}
    \includegraphics[width=0.99\textwidth]{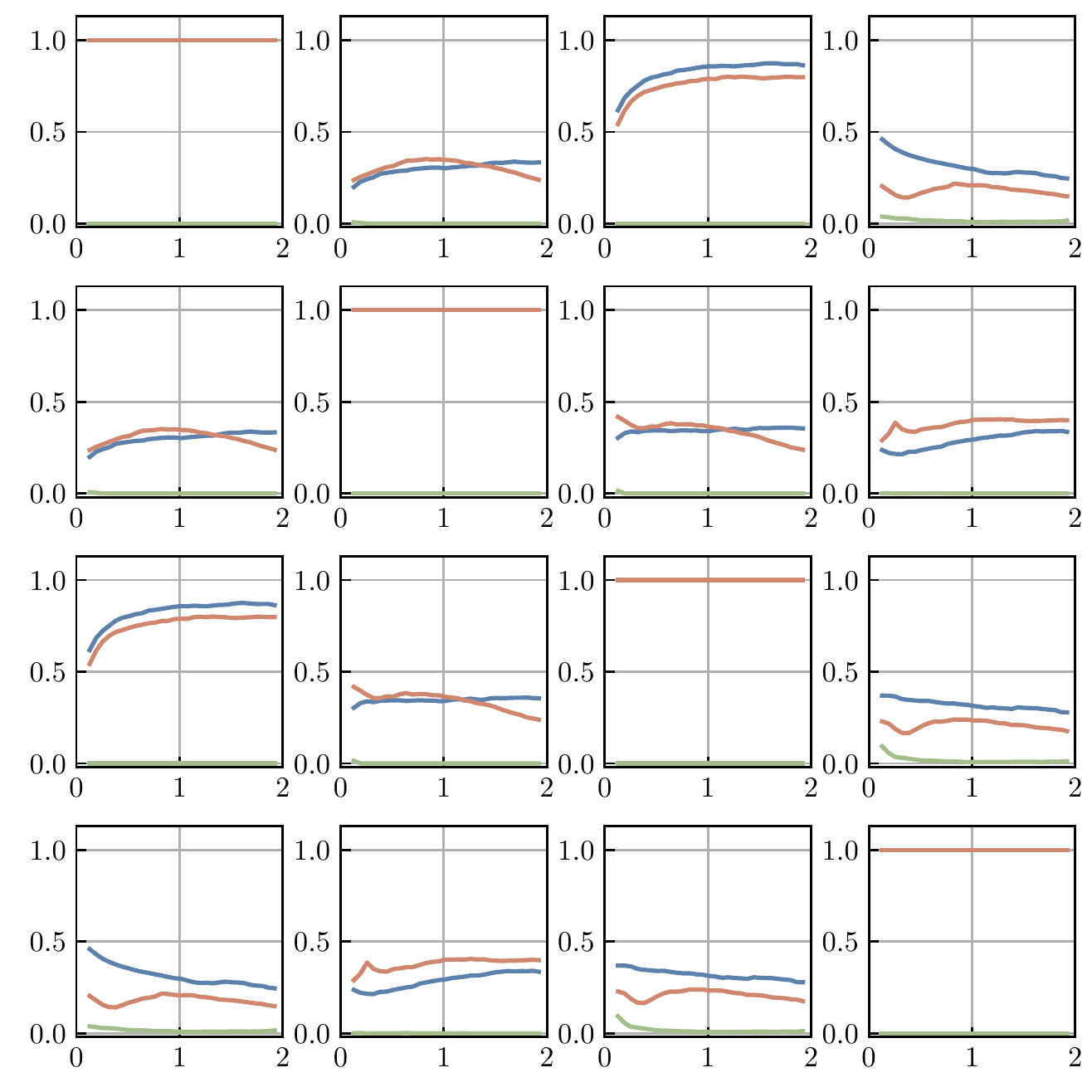}%
    \mylab{-13.cm}{13.15cm}{$(a)$}%
    \mylab{-9.75cm}{13.15cm}{$(b)$}%
    \mylab{-6.5cm}{13.15cm}{$(c)$}%
    \mylab{-3.25cm}{13.15cm}{$(d)$}%
    \mylab{-13.cm}{9.85cm}{$(e)$}%
    \mylab{-9.75cm}{9.85cm}{$(f)$}%
    \mylab{-6.5cm}{9.85cm}{$(g)$}%
    \mylab{-3.25cm}{9.85cm}{$(h)$}%
    \mylab{-13.cm}{6.55cm}{$(i)$}%
    \mylab{-9.75cm}{6.55cm}{$(j)$}%
    \mylab{-6.5cm}{6.55cm}{$(k)$}%
    \mylab{-3.25cm}{6.55cm}{$(l)$}%
    \mylab{-13.cm}{3.25cm}{$(m)$}%
    \mylab{-9.75cm}{3.25cm}{$(n)$}%
    \mylab{-6.5cm}{3.25cm}{$(o)$}%
    \mylab{-3.25cm}{3.25cm}{$(p)$}%
    \mylab{-11.3cm}{0cm}{$\tau$}%
    \mylab{-8.05cm}{0cm}{$\tau$}%
    \mylab{-4.8cm}{0cm}{$\tau$}%
    \mylab{-1.55cm}{0cm}{$\tau$}%
    \caption{Intersection ratios for the cross-classifications of figure \ref{fig:crosscla}. Orange lines are the significant-significant intersection, blue lines are the insignificant-insignificant intersection, and green lines are the sum of the insignificant-significant intersection and its converse.\label{fig:crossinte}}
\end{figure}

In this appendix we show the relation between the four different norms with more detail than in the main manuscript. Figure \ref{fig:crosscla} shows significant perturbations and insignificant ones classified under one norm represented using a different one. Thus, plots in the diagonal are represented in the same norm used to classify them, and provide no additional information. More interesting are the off-diagonal plots. The two absolute norms separate each other sets very well, as seen in the plots (c) and (i) of the 4-by-4 tiling. A similar story holds for the two relative norms among themselves. The classification using an absolute norm stays relevant from the point of view of the associated relative norm, albeit at an earlier time, as shown in plots (b) and (l). This implies that relative significance precedes absolute one. The same effect can be observed at later times in figure \ref{fig:crosscla}(e). For $q$-norms, it is able to fully separate significants from insignificants by the end of our time window. The worst cross classifier is the relative $\omega$-norm, which is almost unrelated to either absolute norm, and we recall that it was almost impossible to classify with any flow marker. The other three norms have reasonable agreement among themselves, which show why similar results are obtained by looking at either of the three.

That significant perturbations and insignificant ones are separated from each other in figure \ref{fig:crosscla} does not imply that they are equivalent to their definitions under other norms; only that their classifications are not too far appart. Figure \ref{fig:crossinte} shows the difference between classifications by means of the intersection of the significant and insignificant sets under pairs of norms. Only the upper triangular plots need to be considered, as the tiling is symmetric respect to the diagonal and the latter only shows that the self intersection is always one. The green line represents the misclassification, and contains the sum of significant events being classified as insignificant and its converse. In agreement with the previous figure, $\tilde \psi_\omega^\prime$ is the only norm giving appreciable misclassification, specially at earlier times. 
The conclusion is that times of the order of one turnover are probably too large for a metric that measures the amplification of gradients, which have a much faster time scale.

\begin{figure}
    \centering\vspace{1em}
    \includegraphics[width=0.99\textwidth]{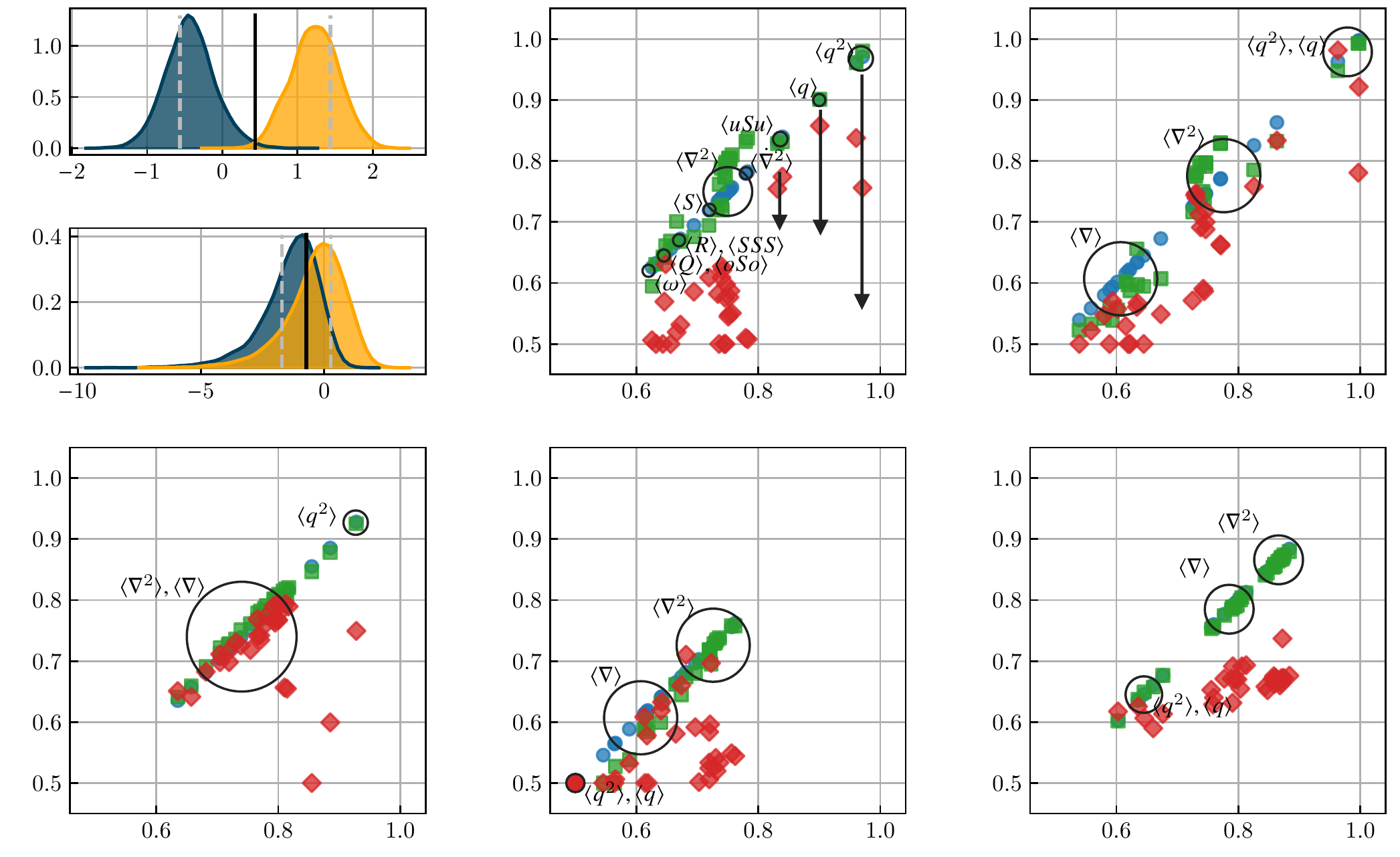}%
    \mylab{-2.25cm}{0pt}{$\mathcal{M}$}%
    \mylab{-6.75cm}{0pt}{$\mathcal{M}$}%
    \mylab{-11.5cm}{0pt}{$\mathcal{M}$}%
    \mylab{-4.4cm}{2.3cm}{${\mathcal{M}}$}%
    \mylab{-9.0cm}{2.3cm}{${\mathcal{M}}$}%
    \mylab{-13.6cm}{2.3cm}{${\mathcal{M}}$}%
    \mylab{-2.25cm}{4.2cm}{$\mathcal{M}$}%
    \mylab{-6.75cm}{4.2cm}{$\mathcal{M}$}%
    \mylab{-4.4cm}{6.5cm}{${\mathcal{M}}$}%
    \mylab{-9.0cm}{6.5cm}{${\mathcal{M}}$}%
    \mylab{-11.0cm}{4.05cm}{${\log\so}$}%
    \mylab{-11cm}{6.2cm}{${\log\sud}$}%
    \mylab{-13.55cm}{7.25cm}{{pdf}}%
    \mylab{-13.55cm}{5.0cm}{{pdf}}%
    \mylab{-13.7cm}{7.9cm}{(a)}%
    \mylab{-13.7cm}{5.75cm}{(b)}%
    \mylab{-9.15cm}{7.9cm}{(c)}%
    \mylab{-4.5cm}{7.9cm}{(d)}%
    \mylab{-13.7cm}{3.85cm}{(e)}%
    \mylab{-9.15cm}{3.85cm}{(f)}%
    \mylab{-4.5cm}{3.85cm}{(g)}%

    \caption{Classification of significant/insignificant regions. (a) Example of a good classifier ($\log\langle{q^2}\rangle$ for $q$-perturbations at size $\rmDelta_3$) and the norm $\mathcal{R}_\omega^{5}$. (b) Example of a bad classifier for the same case ($\log\so$) (c-g) Classification score for different perturbations and significance criteria. The abscissa is always the score associated to $\mathcal{R}_q^{5}$. Colours are the classifier of the ordinate, (blue) $\mathcal{R}_q^{5}$, (green) $\mathcal{R}_\omega^{5}$ and (red) $\mathcal{\tilde R}_\omega$. (c) $q$-perturbations at $\rmDelta_2$. (d) Classification of $q$-perturbations at $\rmDelta_3$. (e) Classification of $q$-perturbations at $\rmDelta_0$. (f) Classification of $\omega$-perturbations at $\rmDelta_2$. (g) Classification of $\omega$-perturbations at $\rmDelta_0$.\label{fig:class1da}}
\end{figure}

Finally, figure \ref{fig:class1da} shows the classification scores for both enstrophy-based norms, and it should be compared with figure \ref{fig:class1d}. First, the green squares represent the absolute $\omega$-norm $\mathcal{R}_\omega$
Essentially, classification by the gradient markers improves with respect to the absolute $q$-norm, and classification by the kinetic energy worsens slightly, although the effect is marginal. The most notable effect is the increase in the effectiveness of the indicators of small-scale gradients ($\langle\nabla^2\rangle, \langle\dot\nabla^2\rangle$). Small perturbations ($\rmDelta_0$) have almost identical scores for both absolute norms, probably because they contain essentially one scale.
Lastly, the amplification of gradients, shown in red diamonds, is notoriously hard to classify, and almost every marker has low score, probably because $\mathcal{\tilde R}_\omega$ evolves at a faster time scale.

  

\bibliographystyle{jfm}

\bibliography{jfm}




\end{document}